\documentclass[twocolumn,traditabstract]{aa}
\usepackage{graphicx,amsmath}
\usepackage{hyperref}
\usepackage{natbib}
\usepackage{stfloats}
\usepackage{multirow}
\usepackage{color}
\usepackage{txfonts}
\usepackage{upgreek}
\usepackage{url}
\usepackage{epsfig}
\usepackage{epsf}
\usepackage{float}
\usepackage{slashbox}
\usepackage[outercaption]{sidecap}
\usepackage{fancyhdr}
\usepackage{afterpage}
\usepackage{soul}
\usepackage{booktabs,caption}
\usepackage[flushleft]{threeparttable}
\usepackage[export]{adjustbox}
\usepackage{subfig}

\newcommand{\planck}{\textit{Planck}}
\newcommand{\herschel}{\textit{Herschel}}

\newcommand{\polang}{\psi}
\newcommand{\bang}{\psi_{B}}
\newcommand{\snr}{S/N}
\newcommand{\mn}{North-Main}
\newcommand{\valeast}{3^{\circ}}
\newcommand{\valwest}{-2.6^{\circ}}
\newcommand{\stdeast}{5.4^{\circ}}
\newcommand{\stdwest}{5.3^{\circ}}

\newcommand{\mx}{\mathbf{x}}
\newcommand{\mxl}{\mathbf{x}+\mathbf{l}_i} 
\newcommand{\co}{$^{12}\mathrm{CO}$}
\newcommand{\coo}{$^{13}\mathrm{CO}$}

\title{Large-scale magnetic field in the Monoceros OB-1 East molecular cloud}
\titlerunning{Magnetic field in Mon OB-1 East}
\author{D. Alina\inst{1,3} 
\and J. Montillaud\inst{2} 
\and Y. Hu\inst{4,5}
\and A. Lazarian\inst{5,6}
\and I. Ristorcelli\inst{3}
\and E. Abdikamalov\inst{1,7} 
\and S. Sagynbayeva\inst{1}
\and M. Juvela\inst{8}
\and T. Liu \inst{9}
\and J.-S. Carri\`{e}re\inst{3}}
\authorrunning{D. Alina \and J. Montillaud \and Y. Hu \and A. Lazarian}
\institute{
\inst{1}{Department  of  Physics,  School  of  Sciences  and  Humanities, Nazarbayev University, Nur-Sultan 010000, Kazakhstan}\\
\inst{2}{Institut UTINAM - UMR 6213 - CNRS - Universit\'{e} Bourgogne Franche Comt\'{e}, France, OSU THETA, 41bis avenue de l'Observatoire, 25000 Besan\c{c}on, France}\\
\inst{3}{IRAP, Universit\'{e} de Toulouse CNRS, UPS, CNES, F-31400 Toulouse, France} \\
\inst{4}{Department of Physics, University of Wisconsin-Madison, Madison, WI 53706, USA} \\
\inst{5}{Department of Astronomy, University of Wisconsin-Madison, Madison, WI 53706, USA}\\
\inst{6}{Center for Computation Astrophysics, Flatiron Institute, 162 5th Ave, New York, NY 10010} \\
\inst{7}{Energetic Cosmos Laboratory, Nazarbayev University, Nur-Sultan 010000, Kazakhstan} \\
\inst{8}{Department of Physics, PO Box 64, University of Helsinki, 00014, Helsinki, Finland}\\
\inst{9}{Shanghai Astronomical Observatory, Chinese Academy of Sciences, 80 Nandan Road, Shanghai 200030, People's Republic of China} 
}
\date{Received date /
Accepted date }
\begin{document}

\abstract
{
The role of large-scale magnetic fields in the evolution of star forming regions remains elusive. Its understanding requires observational characterisation of well-constrained molecular clouds. 
The Monoceros OB1 molecular cloud is a large complex containing several structures which were shown to be in an active interaction and to have a rich star formation history. However, magnetic fields in this region have only been studied on small scales.}
{We study the large-scale magnetic field structure and its interplay with the gas dynamics in the Monoceros OB1 East molecular cloud.}
{We combine observations of dust polarised emission  from the {\planck} telescope and CO molecular line emission observations from the Taeduk Radio Astronomy Observatory 14-metre telescope. We calculate the strength of the plane-of-the-sky magnetic field using a modified Chandrasekhar-Fermi method and estimate mass over flux ratios in different regions of the cloud. We use the comparison of the velocity and intensity gradients of the molecular line observations with the polarimetric observations to trace dynamically active regions.} 
{The molecular complex shows an ordered large-scale plane-of-the-sky magnetic field structure. 
In the Northern part, it is mostly orientated along the filamentary structures while the Southern part shows at least two regions with distinct magnetic field orientations. 
Our analysis reveals a shock region in the Northern part right in-between two filamentary clouds which were previously suggested to be in collision.
The magnetic properties of the North-Main and North-Eastern filaments suggest that these filaments once formed one, and the magnetic field evolved together with the material and did not undergo major changes during the evolution of the cloud.
In the Southern part, we find that either the magnetic field guides the accretion of interstellar matter towards the cloud or it is dragged by the matter falling towards the main cloud.}
{The large-scale magnetic field in Monoceros OB-1 East molecular cloud is tightly connected to the global structure of the complex. In the Northern part, it seems to have a dynamically important role by possibly providing support against gravity in the direction perpendicular to the field and to the filament. In the Southern part, it is probably the most influencing factor which governs the morphological structure, guiding possible gas inflow.  
A study of the whole Monoceros OB-1 molecular complex at large scales is necessary in order to form a global picture of the formation and evolution of the Monoceros OB1 East cloud and the role of the magnetic field in this process.}
\maketitle

\section{Introduction}
\label{sec:introduction}
Magnetic fields are one of the key factors that regulate dynamical processes in molecular clouds along with gravity and turbulence. 
Studies of the relative orientation between filamentary molecular clouds and interstellar magnetic fields traced by polarimetric observations of dust emission are one of the main tools to probe how the magnetic field affects the evolution of the interstellar medium (ISM) and the formation of dense structures. 
\cite{planck2015-XXXV} showed that in nearby molecular clouds the relative orientation changes from parallel to perpendicular with increasing column density. The latter effect can be understood on the basis of MHD turbulence properties \citep{xu2019} as well as on their inner morphology and evolutionary stage \citep{liu2018,doi2020,soam2019,malinen2016,alina2019}. \cite{alina2019} suggested that the high density contrast filaments could be those where self-gravity takes over, and the magnetic field turns out to be perpendicular to the over-densities. 
MHD simulations reveal that the formation of structures within molecular clouds is highly affected by the magnetic field while the magnetic field strength and structure are both affected by the turbulent motions of matter \citep{andre2014,li2014,hennebelle2013,federrath2016}. 

Polarimetric observations of the interstellar dust allows us to trace indirectly the magnetic field orientation, but it suffers from the signal integration along the line of sight and from the ambiguity coming from the projection of distinct structures onto the plane-of-the-sky (POS), which in this case may seem to be connected. 
If the former constraint can partly be alleviated for observations in the lines of sight out of the Galactic disk and within the optically thin emission assumption, the latter can only be examined using molecular line data. A recently developed technique of velocity gradients (see  \citealt{yuen2017vgt}, \citealt{lazarian2018vgt}, \citealt{2018MNRAS.480.1333H}) provides  a new promising way of using spectroscopic data for sampling magnetic fields in molecular clouds \cite[see][and references therein]{hu2019}.

Connecting information on the dynamics of molecular clouds to their magnetic field structure allows one to better understand how turbulence, magnetic fields, and gravity are regulating dynamical processes in molecular cloud filaments and in star formation process.
To achieve this goal, it is ineluctable to combine polarimetric and spectroscopic data, and recently there has been a growth of synthesized studies. 
A combination of {\planck} and {\herschel} continuum data and {\co~and \coo} emission data was used by \cite{malinen2016} to study the L1642 cloud. This allowed the authors to show a tight connection between the magnetic field structure and the morphology of the cloud and to confirm the connection of striations to the clumps embedded in the cloud.
\cite{heyer2020} also used the {\planck} polarisation data and {\co~and \coo} emission data to study the role of the magnetic field in structuring the Taurus molecular cloud. Their analysis of variations in the relative orientation between the magnetic field and the gradients of surface brightness suggested a presence of local variations of the Alfv\'{e}nic Mach number at different layers of the cloud. 
\cite{fissel2019} compared the orientation of the structures traced by {\coo} and C$^{18}$O data with the magnetic field derived from the BLASTPol balloon experiment polarimetric data in the Vela C molecular cloud to infer the density threshold of the transition between parallel and perpendicular relative orientations.
It seems us necessary to widen the scope of combined analyses of polarimetric and spectroscopic data to connect the magnetic field structure to the dynamical processes in molecular clouds.

The Monoceros OB-1 molecular complex is well suited for such studies. 
It is located near the anti-centre direction, and the Eastern part of the complex has a latitude range sufficient to avoid confusion with the background material of the Perseus galactic arm. It also seems to be free of any significant amount of foreground material, according to, for example, the 3D extinction map by \citet{green2019}. Thanks to its intermediate distance of 723 pc \citep{cantat-gaudin2018}, it is possible to map the entire complex ($\sim 60$ pc, i.e. $~5$ degrees) at millimetre wavelengths with a spatial resolution of the order of the core scales ($\sim 0.15$ pc, i.e. 45\arcsec) and it has been the target of a wealth of observations confirming an active star forming process. 
In this study, we focus on the Eastern part of the complex. It hosts NGC 2264, a $\sim 5$ Myr old open cluster which contains over 1000 stars, including not only several bright O and B stars, but also many T Tauri stars along with signs of on-going star formation activity such as young stellar objects (YSOs) and dense cores \citep{park2002,dahm2005, chen2007, gregorio2003, wolf-chase2003}.
\cite{rapson2014} studied the Monoceros OB1 East cloud using \textit{Spitzer} data and showed that the NGC 2264 and the remainder of the cloud are significantly different regarding the star formation activity. The open cluster is undergoing an active star formation process, and its natal gas around the overdensity of YSOs has already been dispersed. 
The remainder of the cloud has a dispersed population of old stars along with signs of a recent starburst activity, and it is globally quiescent.
The dynamics of the filamentary structure of G202.3+2.5, which is located in the Northern part of the Monoceros OB-1 East molecular cloud, was extensively studied by \cite{montillaud2019a,montillaud2019b} using observations of the TRAO (Taeduk Radio Astronomy Observatory) 14-metre and IRAM (Institut de Radio Astronomie Millim\'{e}trique) 30-metre telescopes.
They showed that the cloud is likely experiencing a collision of two filaments, and the junction region exhibits signs of intense star formation activity. 
In contrast, the magnetic field in this region remains poorly known. \citet{dotson2010} report polarimetric observations at 350 $\mu$m toward three targets in Monoceros OB-1. However those measurements have a very limited extent of approximately 2\arcmin, focused on selected dense cores.

The aim of this paper is to study the large-scale magnetic field of the Monoceros OB-1 East molecular cloud and to investigate how it interlaces with the global dynamics.
To do so, we use the {\planck}\footnote{\url{http://www.esa.int/Planck} is an ESA mission with participation of NASA and Canada.} telescope polarimetric observations of the interstellar dust emission and the new TRAO 14-metre telescope observations of {\co} and {\coo} (J=1-0) emission.
We use the novel technique of comparison of the spectroscopic velocity and intensity gradients with the magnetic field orientation \citep{yuen2017vgt,lazarian2018vgt} in order to trace the dynamical processes within MHD turbulence. 

The paper is organised as follows: we describe the data and the methods in Sect.~\ref{sec:data} and \ref{sec:methods} respectively; we present and discuss the results in Sect.~\ref{sec:results} and \ref{sec:discussion}; and we propose our vision of the interplay between the magnetic fields and the evolution of the cloud in Sect.~\ref{sec:conclusion}.

\section{Data used}
\label{sec:data}
We combine {\planck} 353 GHz intensity and polarisation data as well as the derived column density map, and the TRAO 14-metre telescope observations of the $^{12}$CO and $^{13}$CO emission that we describe below.
\subsection{Continuum observations and data}
\label{sec:continuum}

\subsubsection*{{\planck} data}

To trace the magnetic field orientation in the POS, we use the 353 GHz polarised channel data from the {\planck} PR3 release \citep{planck2018-i}. 
It provides the Stokes $I$, $Q$, $U$ maps and the corresponding noise variances.
To increase the signal-to-noise ratio ($\snr$), we smooth all maps from nominal angular resolution ($5\arcmin$) up to a resolution of $7 \arcmin$ using a Gaussian kernel.
We follow the procedure described in \cite{planck2014-xix}, where the local polarisation reference frame rotation is carefully taken into account.

\begin{figure}
\begin{center}
\includegraphics[width =  0.45\textwidth]{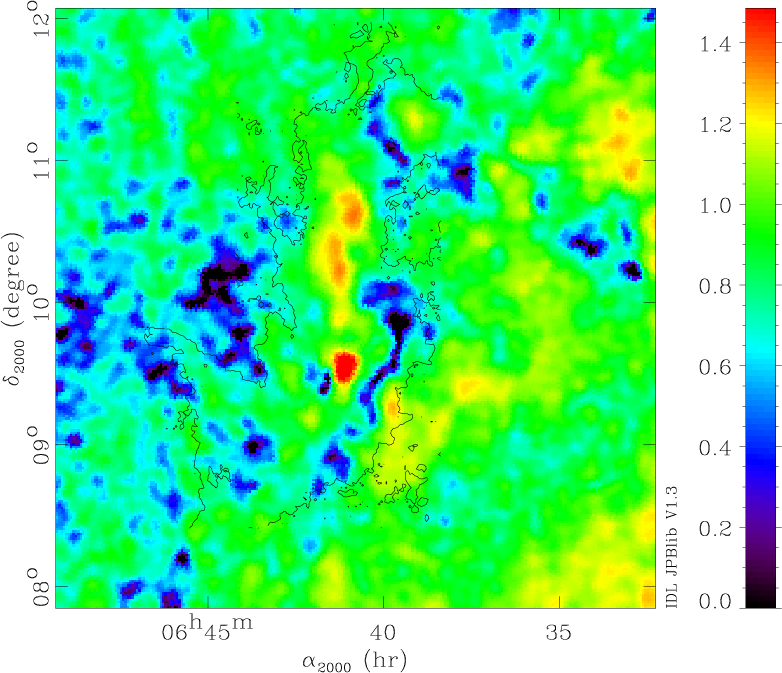} \\
\includegraphics[width =  0.45\textwidth]{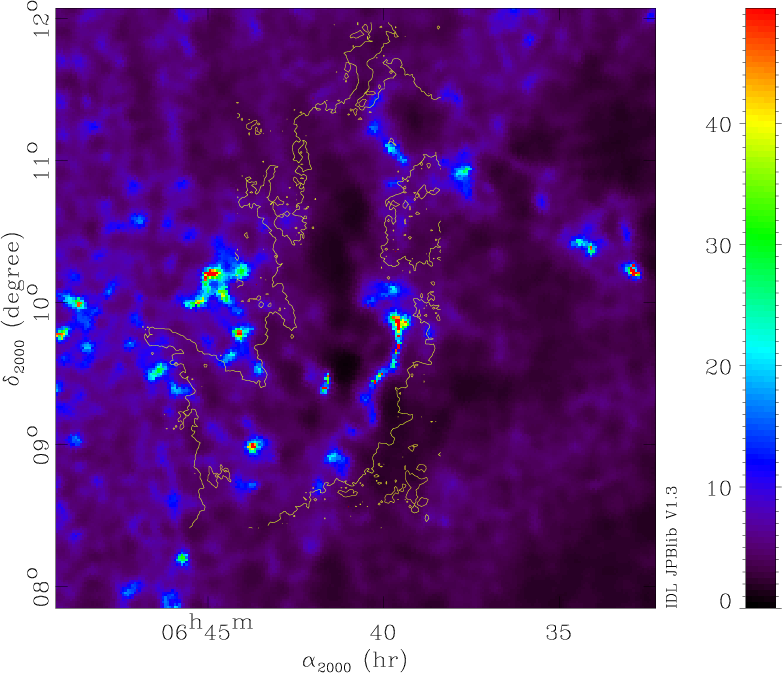} 
\caption{ {\bf Top}: Signal-to-noise ratio of the polarisation. The colour scale shows $log_{10} (p/\sigma_p)$. {\bf Bottom}: Dispersion of the polarisation angle $\sigma_{\polang}$ in degrees. Both are estimated using Bayesian analysis of the {\planck} data. Contours correspond to the value of $2$ K km s$^{-1}$ of the $^{12}$CO integrated intensity, represented in the right panel of Fig.~\ref{fig:pol_map}. }
\label{fig:snrp}
\end{center}
\end{figure}

The polarisation fraction ($p$) and polarisation angle ($\polang$) are calculated from the measured Stokes intensity $I$ and linear polarisation parameters ($Q,\,U$) as follows:
\begin{equation}
p = \frac{\sqrt{Q^2+U^2}}{I} ;
\end{equation}
\begin{equation}
\polang = 0.5 \mathrm{atan} (-U, Q) \, ,
\label{eq:polang}
\end{equation}
where the two-argument function $\mathrm{atan}$ is used to account for the $\pi$-periodicity.
The POS magnetic field angle $\bang$ is obtained via rotation of $\polang$ by $\pi/2$: 
\begin{equation}
\bang = \polang + \pi/2 \, .
\label{eq:bang}
\end{equation}
The {\planck} data are in the COSMO convention adopted from \cite{zaldarriaga1998}, this is the reason for taking the negative value of $U$. Both angles are defined in the range from $-90^{\circ}$ to $90^{\circ}$ in Eq.~\ref{eq:polang}, and are counted positively from Galactic North to East according to the IAU convention. 
The maps were extracted from the all sky maps in HEALPix format provided by the {\planck} Legacy Archive\footnote{\url{pla.esac.esa.int}} and brought to the Equatorial coordinates system. \\
The polarisation fraction and angles are biased due to the non-linearity of the equations above and to the presence of noise in the data \citep{serkowski1958,quinn2012,montier1}, especially at low signal-to-noise ratios ($\snr$).
We estimate the $\snr$ of the polarisation parameters by calculating the classical estimates of the uncertainties \citep{montier1}, which takes into account the full noise variance matrix. 
This allowed us to reveal that the uncertainty of the angle in some regions of the map can be high up to $20^{\circ}$ even for reasonable, larger than three, $\snr$ of the polarisation fraction.
It is worth noting that the $\snr$ is overestimated as the classical estimate fails at low true 
$\snr$s \citep{Montier2}. 
Thus,  to increase the reliability of the data and of the {$\snr$} estimation, we compute the Bayesian estimates of $p, \, \polang$, and the corresponding uncertainties by performing Monte-Carlo simulations to build posterior probability density functions (PDF) as described in \cite{planck2014-xix}. 
In what follows, $p$ and $\polang$ stand for the mean posterior Bayesian estimates of polarisation fraction and angle, while $\sigma_p$ and $\sigma_{\polang}$ denote their uncertainties calculated from variances over the PDFs.
The resulting {$\snr$} of $p$ and the uncertainty of $\polang$ are shown in Fig.~\ref{fig:snrp}. 
In this study, we considered only pixels with $\snr(p) \geq 2$ and $\sigma_{\polang} < 10^{\circ}$. 
This selection leads to a distribution of $\sigma_{\polang}$ that peaks around $3.5^{\circ}$ (Fig.~\ref{fig:sigpsi}).

\begin{figure*}[htbp]
\begin{center}
\begin{tabular}{cc}
\includegraphics[height = 10 cm]{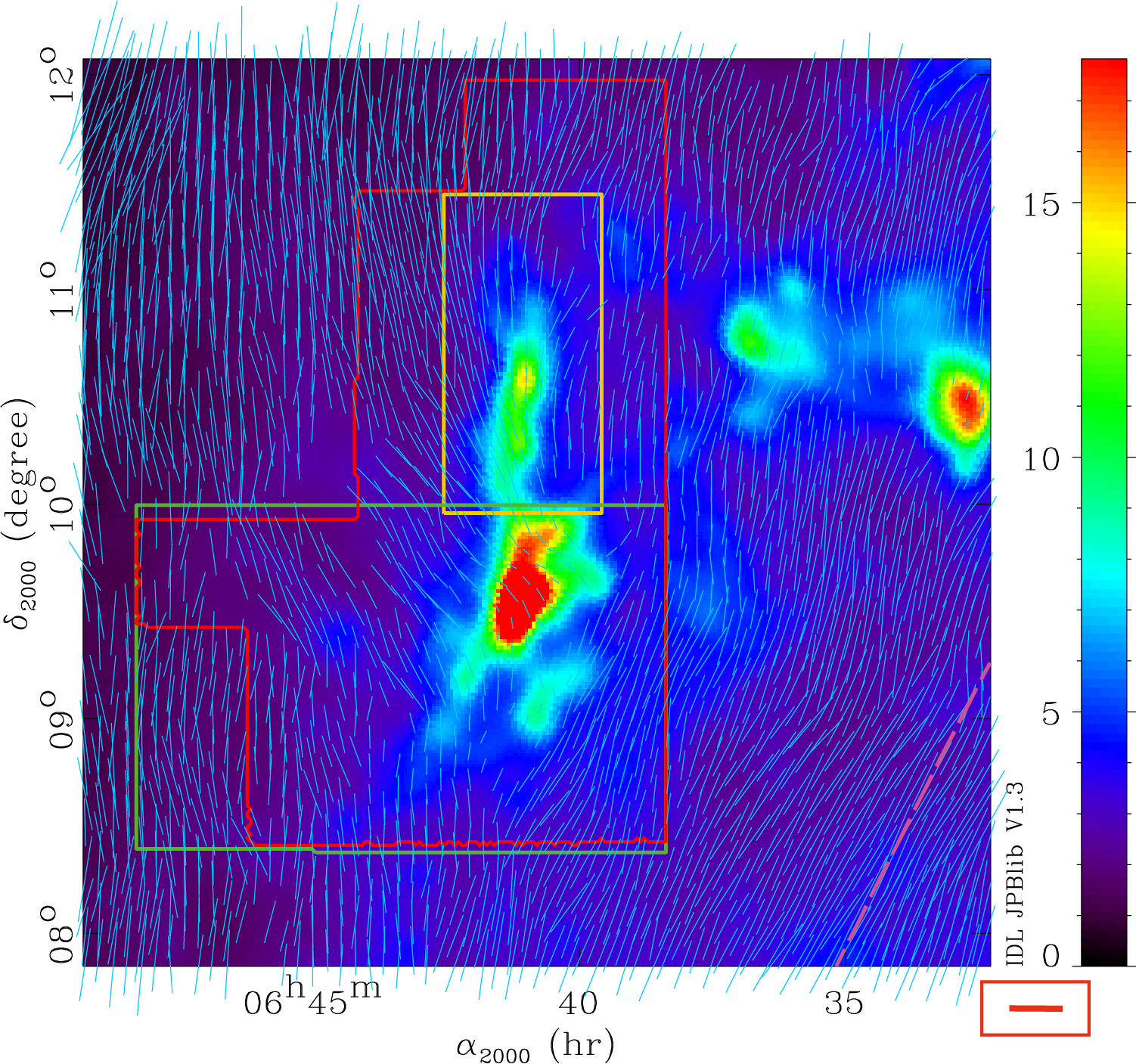} &
\includegraphics[height = 9.55 cm]{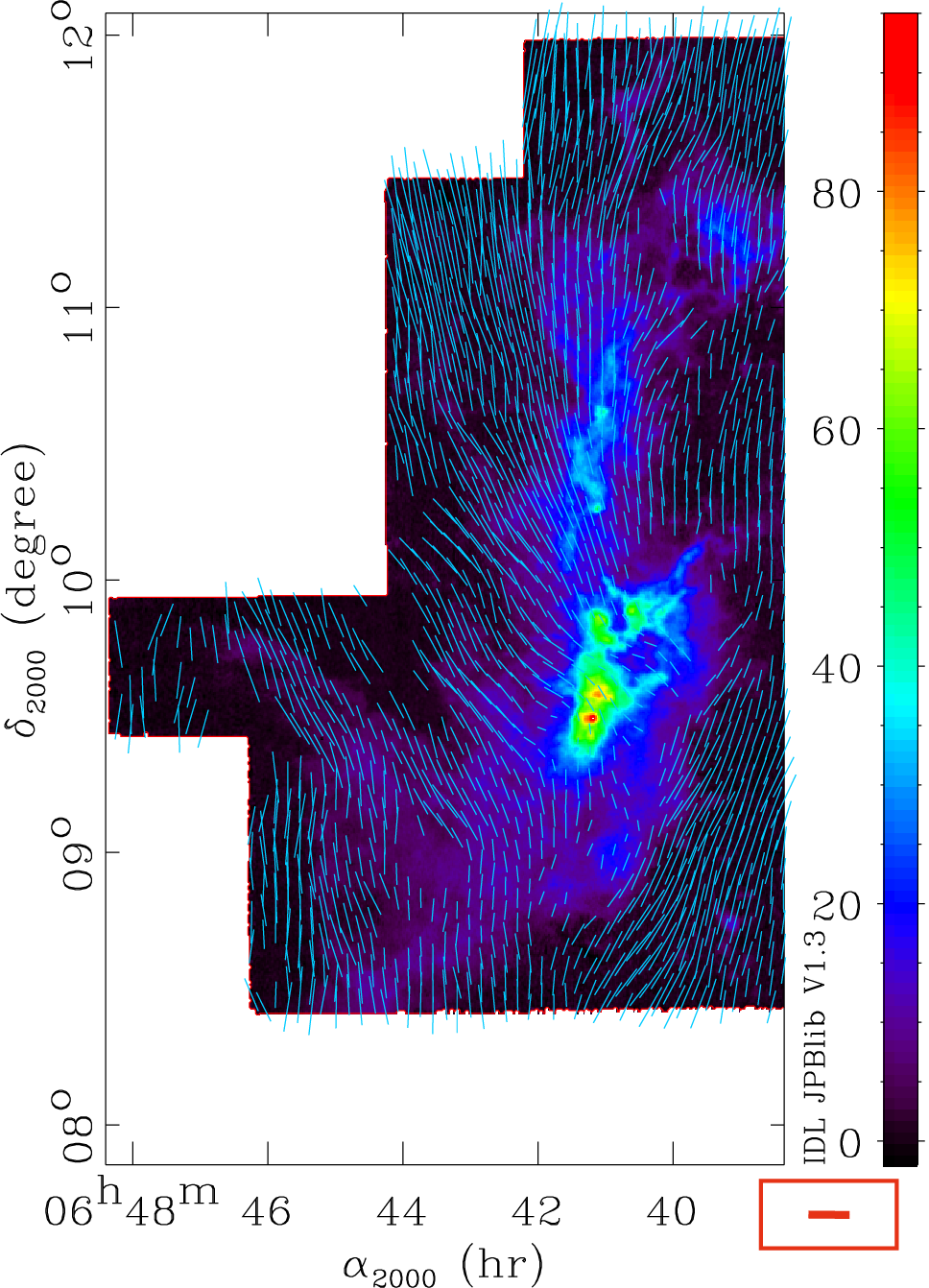}
\end{tabular}
\caption{{\bf Left panel:} {\planck} $353$ GHz intensity map in MJy/sr at the angular resolution of $7\arcmin$ with the POS magnetic field orientation overlaid as blue segments. The length of the segments corresponds to polarisation fraction, with the reference length of $15 \arcmin$ corresponding to $p=0.05$ (shown in the red box in the bottom right corner). Only data satisfying $\snr(p) \geq 2$ and $\sigma_{\polang} < 10^{\circ}$ are used for polarisation studies. Red, yellow, and green lines show contours of the TRAO map used in this analysis, the Northern region (Fig.~\ref{fig:qu_smallmap}), and the Southern region (Fig.~\ref{fig:b_vs_grads_south}) respectively. The pink dashed line shows the isolatitude  $b_{\mathrm{II}} = 0^{\circ}$.\\ {\bf Right panel:} TRAO $^{12}$CO (J=1-0) integrated (between $v_{\rm{lsr}} = -3$ and $17$ km\,s$^{-1}$) $T_a^*$ intensity map in K\,km\,s$^{-1}$ at angular resolution of $47\arcsec$ with the POS magnetic field orientation derived from the {\planck} data overlaid as blue segments. The red contour corresponds to the red contour in the left panel. }
\label{fig:pol_map}
\end{center}
\end{figure*}

\subsubsection*{Column density map}

We also use the {\planck} data to derive a column density map, to address the gravitational stability in Sect.~\ref{sec:strength}. We first calculated the color temperature map  on the basis of the {\planck} 857, 545, and 353 GHz bands and the IRIS 3 THz band, convolved to the same angular resolution ($7\arcmin$). 
The spectral energy distribution for each pixel is fitted by the modified black-body law $B_{\nu}(T) \nu^{\beta}$ using a spectral index $\beta=2$, which corresponds to the value adopted for the {\planck} Cold Clumps in \cite{planck2016-XXVIII}.
The column density map is then calculated using the {\planck} 857 GHz channel flux density and 
\begin{equation}
N_{\mathrm{H}_2} = \frac{I_{\nu} }{ \mu m_{\mathrm{H}}\,  B_{\nu}(T) \,  \kappa_{\nu}} \, ,
\label{eq:nh}
\end{equation}
where the dust opacity $\kappa_{\nu}$ is taken to be  $0.1\,(\nu/1\,\rm{THz})^{\beta}\,
\rm{cm^{2}g^{-1}}$ according to \cite{beckwith1990}, and $\mu=2.8$ amu is the mean molecular weight per H$_2$ molecule. 
\begin{figure}
    \centering
    \includegraphics[width = 0.45 \textwidth]{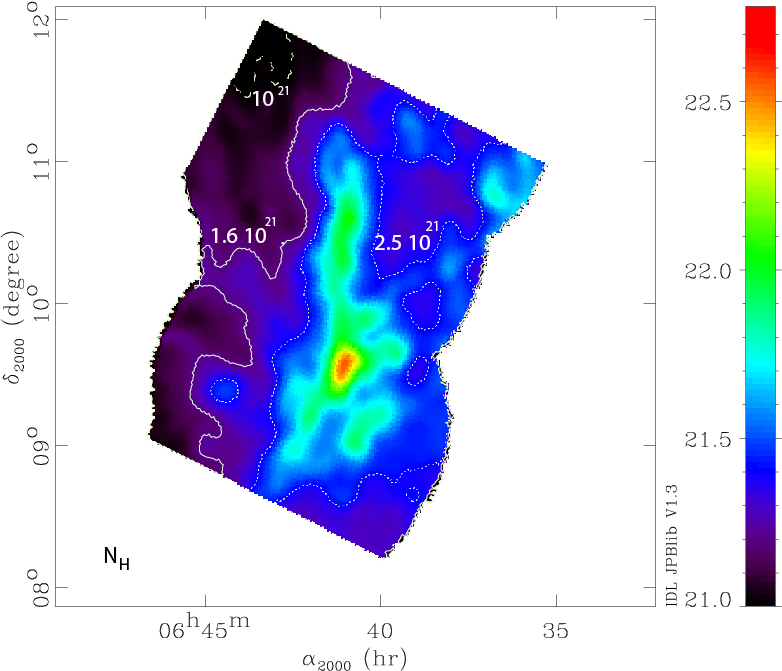}
    \caption{Column density ($N_H$) map derived using the {\planck} and IRIS data, in the logarithmic scale and in cm$^{-2}$. The contours correspond to $1.26 \times 10^{21}, \, 1.78 \times 10^{21}$, and $3.16 \times 10^{21}$ cm$^{-2}$.}
    \label{fig:nhmap}
\end{figure}
The resulting $N_H$ map is represented in Fig.~\ref{fig:nhmap}.

\subsection{Molecular line observations and data}
\label{sec:spectro}
The Monoceros OB 1 molecular complex was observed with the TRAO 14-meter telescope \citep{jeong2019} as part of the COMMON large programme (COMplete view of the MONoceros OB1 molecular complex, P.I.: J. Montillaud), from March till May 2019, and from November 2019 till April 2020. The $^{12}$CO and $^{13}$CO (J=1-0) rotational lines at 115.271 and 110.201 GHz, respectively, were detected with the SEQUOIA-TRAO frontend, a $4 \times 4$ multi-beam receiver, with a spectral resolution of $\sim 0.04$ km\,s$^{-1}$ and a beam size of 47\arcsec. The observations were conducted in the on-the-fly mode, and the data were reduced with the otftool-TRAO software to produce maps with 20{\arcsec} cells. After smoothing the spectra to an effective resolution of 0.2 km\,s$^{-1}$, the achieved sensitivity is rms($T_a^*$)$\approx 0.35$ K and $\approx 0.15$ K for $^{12}$CO and $^{13}$CO, respectively. The dataset is presented in-depth by Montillaud et al. (in prep.). In the present paper, we use the 6 deg$^2$ covering the Eastern part of the complex, as presented in the right panel of Fig.~\ref{fig:pol_map}. 
The Monoceros OB-1 complex is detected in the velocity range from $-3$ to $17$ km s$^{-1}$, and the analysis of the TRAO data in this paper is restricted to these limits.

\section{Methods}
\label{sec:methods}

\subsection{Velocity coherent structures}
\label{sec:vcs}
The TRAO $^{13}$CO data were analysed to identify velocity coherent structures (VCS). We adapted the method presented in \citet{montillaud2019b}, which in turn is adapted from \citet{hacar2013}. In short, we start by smoothing each velocity channel with a Gaussian kernel of full width half maximum (FWHM) of 100\arcsec, to improve the {$\snr$}. Each pixel is then fitted with up to three Gaussian components, and the central velocities of the components peaking at values with $\snr>12$ are used to build a cube of discrete points. A friends-of-friends algorithm is then used to connect the most related points. Two points are considered to be "friends" when they are within a $5 \times 5 \times 5$ box, where pixels are 20\arcsec\, wide and the channel width is 0.2 km\,s$^{-1}$. These values are found to provide VCSs small enough to disentangle the main parts of the clouds, and large enough to enable an analysis of the magnetic field properties.

\subsection{Determination of the magnetic field strength}
\label{sec:dcf}

\citeauthor{chandrasekhar1953} (\citeyear{chandrasekhar1953}) and \citet{davis1951} led a pioneering work on the determination of the magnetic field strength from the dispersion of polarisation angles in the Galactic plane, which we referred to as DCF hereafter. Further, the principle of linking the magnetic field strength to the dispersion of observed angles has been used to study the balance between magnetic and turbulent energies in molecular clouds. Among many modifications to the method that were proposed to resolve issues related to observational constraints or to the underlying assumptions on the nature of turbulence, we choose the approach proposed by \cite{houde2009}. It assumes the turbulence to be isotropic and homogeneous which is not the case in molecular clouds where gradients of the MHD turbulent velocities are dependent on the magnetic field orientation. However, this method accounts for the effects of beam dilution and integration through the thickness of the cloud which are the main sources of overestimation of the magnetic field strength using the DCF method \citep{crutcher2012}. 
The method consists in evaluating the dispersion function of polarisation angles at different lags, or distance, $l$, expressed by $\mathcal{C} =$ $1 - \cos \langle \Delta \polang (l) \rangle$, that makes it possible to determine the turbulent ($B_t$) to ordered ($B_0$) magnetic field strength ratio through the following equation:
\begin{eqnarray}
\mathcal{C} = 1 - \langle \cos [ \Delta \polang(l)] \rangle \simeq  \sqrt{2 \pi} \frac{\langle B_t^2 \rangle}{\langle B_0^2 \rangle} \left[ \frac{\delta^3}{(\delta^2+2 R^2) L}\right] \nonumber \\
\hspace{0.5 cm} \times  \left[ 1 - e^{-l^2/2(\delta^2+2R^2)} \right] + a l^2\, ,
\label{eq:houde53}
\end{eqnarray}
where $\delta$ is the turbulent correlation length, $R$ is the telescope beam radius, $L$ is the thickness of the cloud, and $a$ is a parameter defining the large-scale component. The right-hand term of the equation is calculated from the {\planck} data.
For each lag $l$, we divide data into distance bins where the bin width is constant and is equal to $3 \arcmin$. Note that only unique pairs of points are considered. For each pair $i,j$, the difference $\Delta \psi_{ij}$ is given through the corresponding Stokes $Q$ and $U$ parameters:
\begin{equation}
\Delta \polang_{ij} = 0.5 \mathrm{atan} (Q_j U_i - Q_i U_j, Q_i Q_j + U_i U_j) \, .
\end{equation}
The PDF of the uncertainties on $\polang$ in individual pixels are shown in Fig.~\ref{fig:sigpsi} and are generally less than $5^{\circ}$.
We estimate the beam radius with $FWHM = 7 \arcmin$, and we assume the thickness of the cloud along the LOS in a given region to be equal to the region's width in the POS. 

The fit to the dispersion function provides $\langle B_t^2 \rangle / \langle B_0^2 \rangle$, $a$ and $\delta$.
The strength of the POS component of the large-scale magnetic field is determined using the DCF formula \citep{chandrasekhar1953}:
\begin{equation}
B_0 = \sqrt{4 \pi \rho}\,  \sigma(v) \sqrt {\frac{\langle B_0^2 \rangle}{\langle B_t^2 \rangle}}\, ,
\end{equation}
where $B_0$ is the value we are searching for, while $\langle B_0^2 \rangle / \langle B_t^2 \rangle$ is the fitted parameter, $\sigma(v)$ is the LOS velocity dispersion, and $\rho$ is the mass density.
We obtain $\sigma(v)$ from the TRAO $^{13}$CO emission maps degraded to the same spatial resolution as the {\planck} data. The density $\rho$ is evaluated for the volume density $n = 10^2$ cm$^{-3}$, which is motivated, first, by N$_2$H$^+$ emission being mostly limited to the cores of the junction region in the IRAM observations of \citet{montillaud2019b}. Second, as we will see in Sect.~\ref{sec:dcf}, the magnetic field orientation derived from the {\planck} polarisation data and from the gradient technique applied to the TRAO spectroscopic data shows a better agreement for the $^{12}$CO emission, which suggests that we also detect the magnetic field in the extended large-scale structure, from which we derive the velocity coherent structures (described in Sect.~\ref{sec:vcs}). It is worth noting that in a recent study \citet{evans2020} argued that most of the $^{13}$CO emission in molecular clouds could arise from the gas at the density around $10^2$ cm$^{-3}$. 
We also make a tentative estimation of the volume density by assuming the depth of the cloud $L$ to be equal to its width (approximate angular widths are taken to be $30\arcmin,\,15\arcmin,\,15\arcmin$ for the three VCSs described in Sect.~\ref{sec:strength}) and using $n({\rm H}_2) = N_{{\rm H}_2}/L$ as in \cite{liu2018b}. The obtained values range between $\simeq 10^2$ and $\simeq 2\times 10^2$ cm$^{-3}$ and  are of the adopted order of magnitude.

\subsection{Velocity and intensity gradient techniques}
\label{sec:vgt}

The velocity gradient technique  \citep[VGT; ][]{gonzales-casanova2017, yuen2017, lazarian2018} is a new method to trace the magnetic fields.
It employs the anisotropic properties of MHD turbulence \citep{GS95} and the theory of turbulent reconnection \citep{LV99}, i.e., the fact that turbulent eddies are elongating along with the local magnetic fields for both subsonic and supersonic cases.  For subsonic motions, the velocity and density fluctuations exhibit similar statistical properties so that the velocity gradient (called VGs hereafter) and density gradient (or intensity gradients, called IGs hereafter) both are perpendicular to the magnetic field.
This is the basis on which the intensity gradient technique (IGT)\footnote{The IGT should not be confused by the Histograms of Relative Orientation (HRO) proposed by \citet{soler2013}. While both techniques employ intensity gradients, the IGT uses the set of procedures from the VGT to obtain the orientation of the magnetic field. On the contrary, the HRO gets the magnetic field via polarization measurements and compares those with the intensity gradients. A detailed comparison of IGT and HRO is presented in \citet{2019ApJ...886...17H}.} was developed \citep{yuen2017,2019ApJ...886...17H}. 

A molecular cloud is generally supersonic \citep{zuckerman1974,1999ApJ...525..318P,2007prpl.conf...63B}. When a shock appears, it breaks the anisotropy relation of density field. In this case, the IGs would be parallel to the local magnetic fields, rather then perpendicular, in front of shocks. As for the VGs, the relative orientation with respect to the magnetic fields is still perpendicular \citep{2019ApJ...886...17H}. Therefore, the comparison between the VGs and the IGs, can serve as an indication of regions of shock. Another particular situation for VGs and IGs is the self-gravity. \cite{hu2020} numerically demonstrated that the inflow induced by gravitational collapse could flip the orientation of both IGs and VGs by 90 degrees, from perpendicular to parallel to the magnetic field.  This phenomenon has been observed in the molecular clouds NGC 1333 and Serpens \citep{hu2019}. The comparison of the magnetic field orientations inferred from the interstellar dust polarisation measurements with the orientations of VGs can then reveal regions of gravitational collapse as well as quiescent regions where turbulent motions, thermal pressure, and magnetic support dominate over gravitational energy. 

The applicability of the techniques has been observationally demonstrated by making comparisons with {\planck} 353 GHz polarization and BLASTPol polarization observations \citep{hu2019,2019ApJ...884..137H}. The notion of tracing the local orientation is important as it means that the gradients can map the detailed structure of magnetic field and not only the mean magnetic field orientation\footnote{The orientation of velocity in turbulent eddies with respect to the local orientation of the magnetic field passing through the eddies follows from the theory of turbulent reconnection \citep{LV99} and is well supported by numerical simulations, e.g. Cho \& Vishniac 2000,  \citet{2002ApJ...564..291C}.}. Within this paper, using the spectroscopic data, the technique allows us to trace the magnetic fields by calculating the IGs and VGs from the integrated intensity map (moment-0) and the velocity centroid map (moment-1), respectively \citep{2017ApJ...837L..24Y,2019ApJ...886...17H}. We briefly describe the calculation procedures in the following.

The gradient calculation is performed by convolving individual 2D moment maps with 3 $\times$ 3 Sobel kernels $G_x$ and $G_y$ $$,
	G_x=\begin{pmatrix} 
	-1 & 0 & +1 \\
	-2 & 0 & +2 \\
	-1 & 0 & +1
	\end{pmatrix} \quad,\quad
	G_y=\begin{pmatrix} 
	-1 & -2 & -1 \\
	0 & 0 & 0 \\
	+1 & +2 & +1
	\end{pmatrix}
	$$
as follows:
\begin{equation}
\label{eq:conv}
\begin{aligned}
\bigtriangledown_x f(x,y)=G_x * f(x,y)  \\  
\bigtriangledown_y f(x,y)=G_y * f(x,y)  \\
\psi_{g}(x,y)=\tan^{-1}\left(\frac{\bigtriangledown_y f(x,y)}{\bigtriangledown_x f(x,y)}\right) \, ,
\end{aligned}
\end{equation}
where $f(x,y)$ represents either moment-0 or moment-1 maps, $\bigtriangledown_x f(x,y)$ and $\bigtriangledown_y f(x,y)$ are the $x$ and $y$ components of the gradient respectively, and $*$ denotes the convolution. However, the anisotropy of MHD turbulence concerning the local magnetic field is a statistical concept.  The pixelized raw gradient map $\psi_{g}(x,y)$ is not necessarily required to have any relation to the local magnetic field orientation. The perpendicular relative orientation of gradients and magnetic field only appears when the gradient sampling is enough. The statistical sampling procedure utilizes the sub-block averaging method, which is proposed by \citet{2017ApJ...837L..24Y}. The sub-block averaging method firstly takes all gradient orientations within a sub-block of interest and then plots the corresponding histogram. Because the histogram is close to a Gaussian distribution, the expectation value of the Gaussian distribution reflects the statistically most probable orientation of the gradient. The expectation value of gradients defines the mean gradient orientation for the sub-block and is expected to be perpendicular to the magnetic field. 

The velocity channel gradients (VChGs) \citep{lazarian2018} are calculated similarly to VGs using the thin velocity channel map Ch($x,y$). Ch($x,y$) is defined as:
\begin{equation}
    \begin{aligned}
        \mathrm{Ch}(x,y)=\int_{v_0-\Delta v/2}^{v_0+\Delta v/2}\mathrm{T_R}(x,y,v) dv
    \end{aligned}
    \label{eq2}
\end{equation}
where $\rm T_R$ is the radiation temperature and $v_0$  is the velocity of the averaged emission line maximum. The channel width $\Delta v$ satisfies $\Delta v<\sqrt{(\delta v^2)}$, where $\sqrt{(\delta v^2)}$ is the velocity dispersion. The thin velocity channel maps are used within the assumption that the velocity fluctuations are dominating over density fluctuations due to the velocity caustic effect \citep{LP00}.

The rms noise of the spectroscopic data can in principle affect the calculation of the gradients. 
We represent in Fig.~\ref{fig:rms} the rms noise of {\coo} data which varies from one tile of the mosaic to another between T$\simeq 0.12$K and T$\simeq 0.18$K. It shows that apart from the tile-to-tile rms variations, there is no significant structure (e.g. scanning artefacts) in the noise which may create spurious dominant gradient orientation in the noise-dominated areas. The linear structures at the junction between adjacent tiles are the strongest noise structures in our data set and are located at predictable places. We see in the results presented in Sect.~\ref{sec:results} that no correlations between the observed gradients and the tile overlaps appear, even in the lowest accepted SNR regions.
In addition, the sub-block averaging redresses this issue.
In this work, we apply IG, VG, and VChG techniques to the TRAO data, with sub-block size of 20$\times$20 pixels (one pixel is $20\arcsec$ large) which is the empirical minimum value \citep{2018ApJ...853...96L}, and we also smooth VGs and IGs to $7\arcmin$ resolution of the {\planck} data.
We also tested 30-pixel large box size and observed that the results are stable against averaging.
For a fully noise-dominated sub-region, the corresponding histogram of the gradients would be a uniform distribution, for which sub-block averaging gives a noise-type output. In the case of a spurious signal, the histogram would exhibit a peak in the given sub-block, and the averaging method minimizes such a contribution. The stability of the results at larger block size suggests this effect is not significant since it is less likely that spurious noise features still dominate at larger scales. To further improve the accuracy of the gradients, we make a selection upon the error of Gaussian fitting which is intrinsically based on the noise level of the spectroscopic data. The details of the used thresholds are reported in Sect.~\ref{sec:gradients}.

\section{Results}
\label{sec:results}
Figure~\ref{fig:pol_map} shows the 353 GHz dust emission map and the {\co} integrated intensity map of the Monoceros OB-1 East molecular cloud. It spans over more than 25 pc in the South-North direction and has a complex morphology.
We divide the cloud into two sub-regions: the Northern part, which has a clear filamentary shape, and the Southern part that includes the NGC 2264 open cluster. Its location corresponds to the red color area in the {\planck} intensity map.
In the left panel of Fig.~\ref{fig:pol_map}  the Northern and Southern parts are delimited with yellow and green boxes respectively.
As we shall see, the large-scale magnetic field in the two parts has different properties regarding its structure and its dynamical role.

We observe that the Stokes I map (Fig.~\ref{fig:pol_map}, left panel), the column density map (Fig.~\ref{fig:nhmap}) and the integrated CO gas emission map at the velocities of the studied cloud (Fig.~\ref{fig:pol_map}, right panel) trace the same structure, at column densities larger than $\sim 1.8 \times 10^{21}$ cm$^{-2}$. This threshold is used to define the regions where the background polarised signal can be neglected with respect to the polarisation coming from the Monoceros OB-1 East cloud.

\begin{figure}
  \hfill \begin{minipage}[c]{0.5\linewidth}
    \includegraphics[height = 6.5 cm]{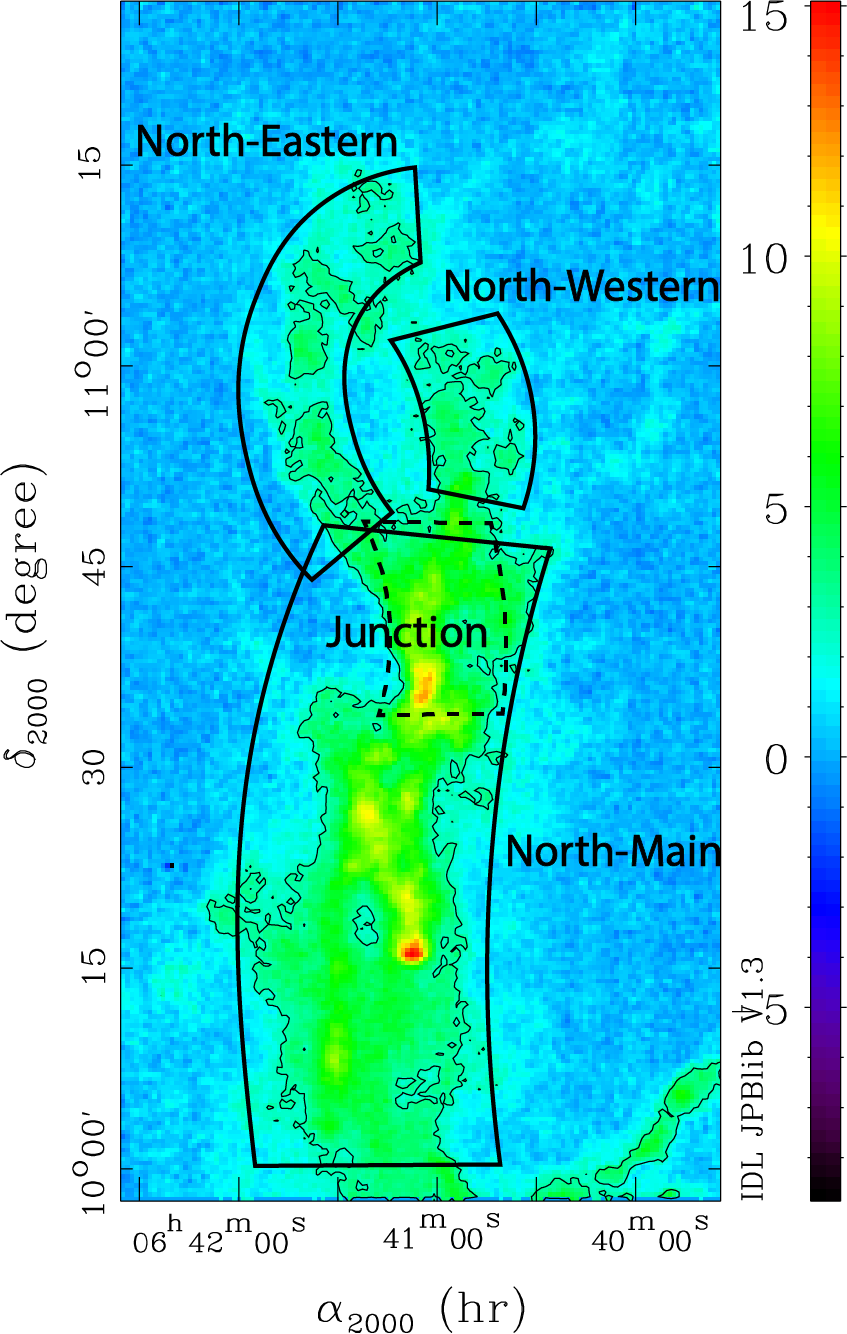}
  \end{minipage}
  \begin{minipage}[c]{0.4\linewidth}
    \caption{The TRAO $^{13}$CO integrated (between $v_{\rm{lsr}} = -3$ and $17$ km s$^{-1}$) $T_a^*$ emission map of the Northern part of the cloud in K\,km\,s$^{-1}$ with the schematic representation of the sub-structures. Contours are taken at $5$ K\,km\,s$^{-1}$.} 
    \label{fig:names}
  \end{minipage}
\end{figure}

In what follows, we adopt the nomenclature introduced by \cite{montillaud2019b} for the sub-structures in the Northern part. There, the {\mn} filament is located in the lower part, connected to two upper filaments that extend slightly to the East and West. They will be called North-Eastern and North-Western respectively, and the three filaments are connected at the junction region. The schematic representation of these structures is shown in Fig.~\ref{fig:names}.

\subsection{Plane-of-the-sky magnetic field structure}
\label{sec:polar_description}
The left panel of Fig.~\ref{fig:pol_map} shows the {\planck} intensity map and the POS magnetic field orientation represented by blue segments. The length of the segments corresponds to the value of $p$ in a given pixel.
The Monoceros OB1 East molecular cloud is located around two degrees above the Galactic plane and we observe the uniformly orientated magnetic field in the South-Western part of the map, which is globally aligned with the Galactic plane, with the $\mathit{b_{\rm II}} = 0^{\circ}$ latitude shown by the pink dashed line. 
Moving from the Galactic Plane towards the centre of the map, the POS magnetic field orientation changes gradually with increasing intensity, then shifts by $90$ degrees from the SE-NW (South-East - North-West) to the SW-NE (South-West - North-East) orientation at the Western border of the main part of the cloud, and the SW-NE orientation prevails throughout the densest part.
The low-intensity Eastern part of the map also shows a uniform orientation of the magnetic field, orientated North, that changes in the brighter region.
We observe that the Northern and Southern parts of the cloud show different POS magnetic field structures.

\begin{figure}[htbp]
\begin{center}
\begin{tabular}{cc}
\includegraphics[height = 6.5 cm]{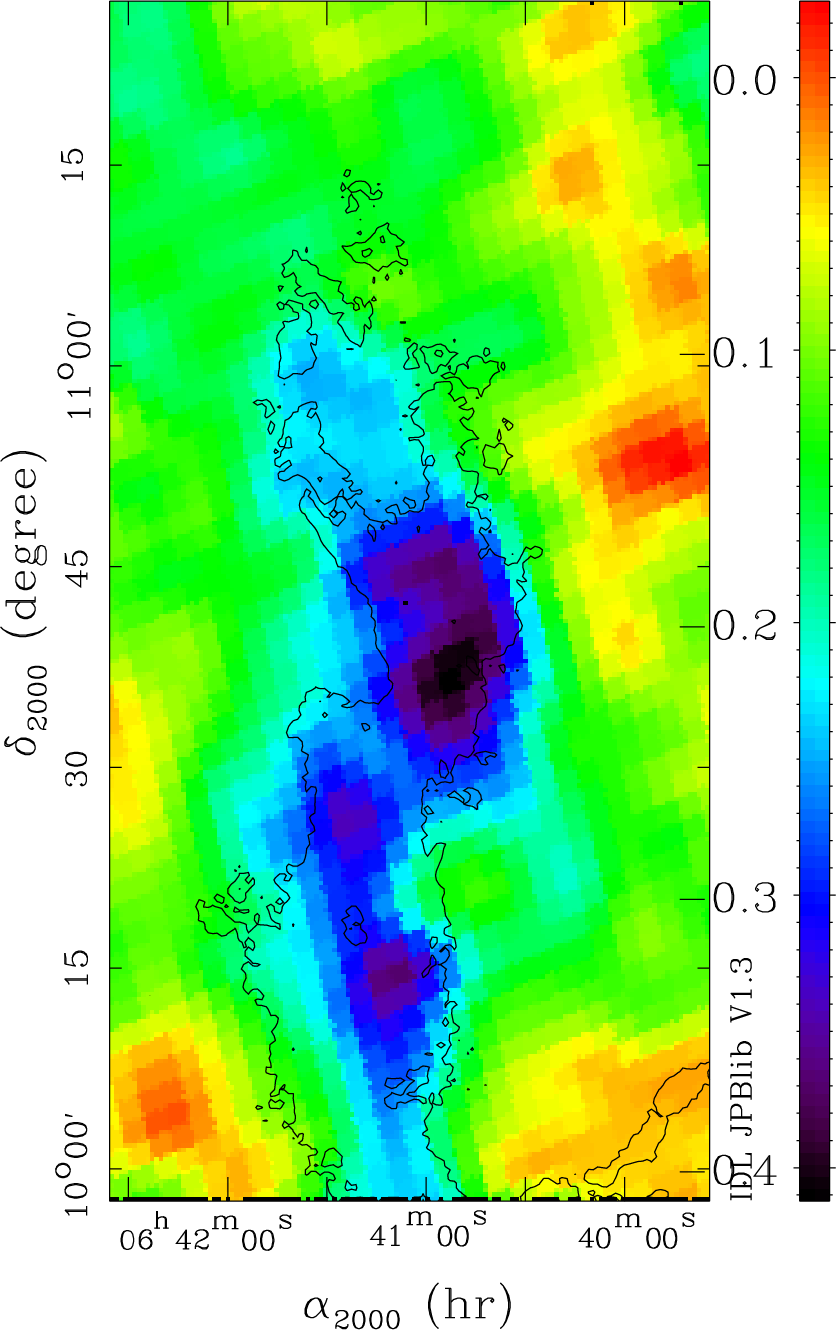} &
\includegraphics[height = 6.5 cm]{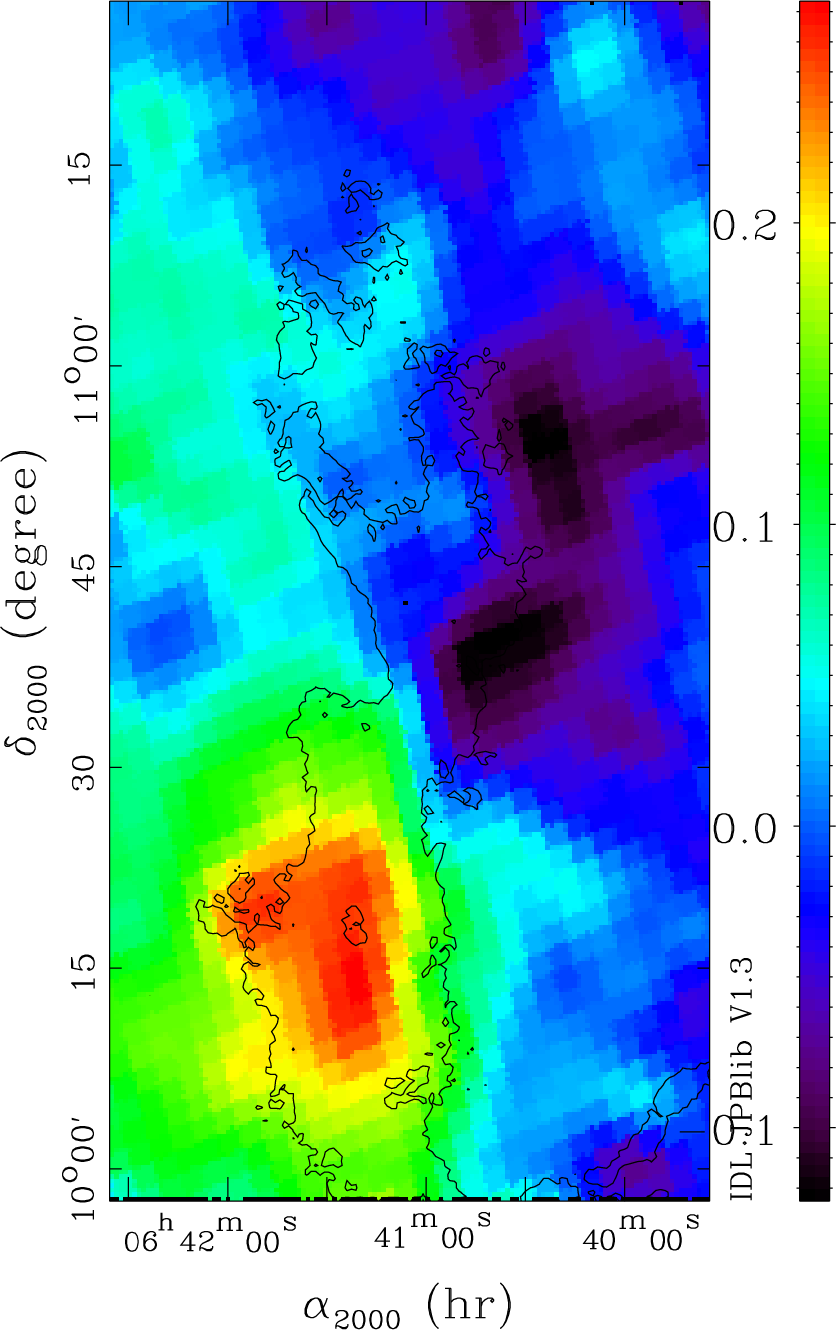} 
\end{tabular}
\caption{{\planck} Stokes $Q$ (left panel) and $U$ (right panel) maps of the Northern part of Monoceros OB1 East, shown in yellow box in Fig.~\ref{fig:pol_map}. Contours correspond to the TRAO $^{13}$CO emission, as shown in Fig.~\ref{fig:names}.}
\label{fig:qu_smallmap}
\end{center}
\end{figure}

The Northern part of the cloud shows a large-scale magnetic field which is mostly orientated in the South-North direction. 
In order to analyse the variations along the cloud, we show in Fig.~\ref{fig:qu_smallmap} the maps of polarised linear intensities in that region, $Q$ and $U$, used for the calculation of $\bang$ (Eq.~\ref{eq:bang}), so that their relative variations partially reflect variations of $\bang$.
The {\mn} filament and the North-Eastern filament, which are proposed by \cite{montillaud2019b} to form one structure, are clearly detected in $Q$ as a strong negative signal in a mostly uniform background with the maximum (in absolute value) in the junction region.  
The $U$ parameter shows a gradient in the direction from South-East to North-West.
Remarkably, the North-Eastern and North-Western filaments have opposite $U$ polarities (positive and negative respectively), which means that the magnetic field orientation is systematically directed North with an inclination to the East and West respectively. 
This is also seen when inspecting the map of the magnetic field angles in Fig.~\ref{fig:pol_map} and when considering the filamentary structures detected in the {\planck} column density map using RHT (see details in App.~\ref{app:rht}). 
The mean magnetic field angles are around $\valeast$ and $\valwest$ with the mean dispersion of $\stdeast$ and $\stdwest$ in the North-Eastern and North-Western filaments respectively. 
The decrease in $Q$ lies between the junction region and the {\mn} filament, and $U$ changes its sign at the southernmost end of the junction region. This {suggests} that the two structures have distinct polarisation properties.

The Southern part of Monoceros OB1 East, shown in the green box in Fig.~\ref{fig:pol_map}, has a clearly different  magnetic field structure. 
In particular, at the Eastern border, we observe a region with a distinct uniform magnetic field orientated roughly perpendicular to the Galactic plane, extending from low-intensity regions towards the brightest part, which includes NGC 2264.
We note that the brightest part seems to be perpendicular to this uniform magnetic field.
The Southernmost end of the region shows a global North-South orientation of the POS magnetic field inferred from the {\planck} data, and the two meet in the brightest part of the cloud (see the right panel of Fig.~\ref{fig:pol_map}). The global SW-NE orientation is also detected in the southernmost Western corner of the Northern part. 

\subsection{The strength of the magnetic field and the effect on filaments stability}
\label{sec:strength}
\begin{figure}[htbp]
\begin{center}
\begin{tabular}{cc}
\includegraphics[width = 0.23\textwidth]{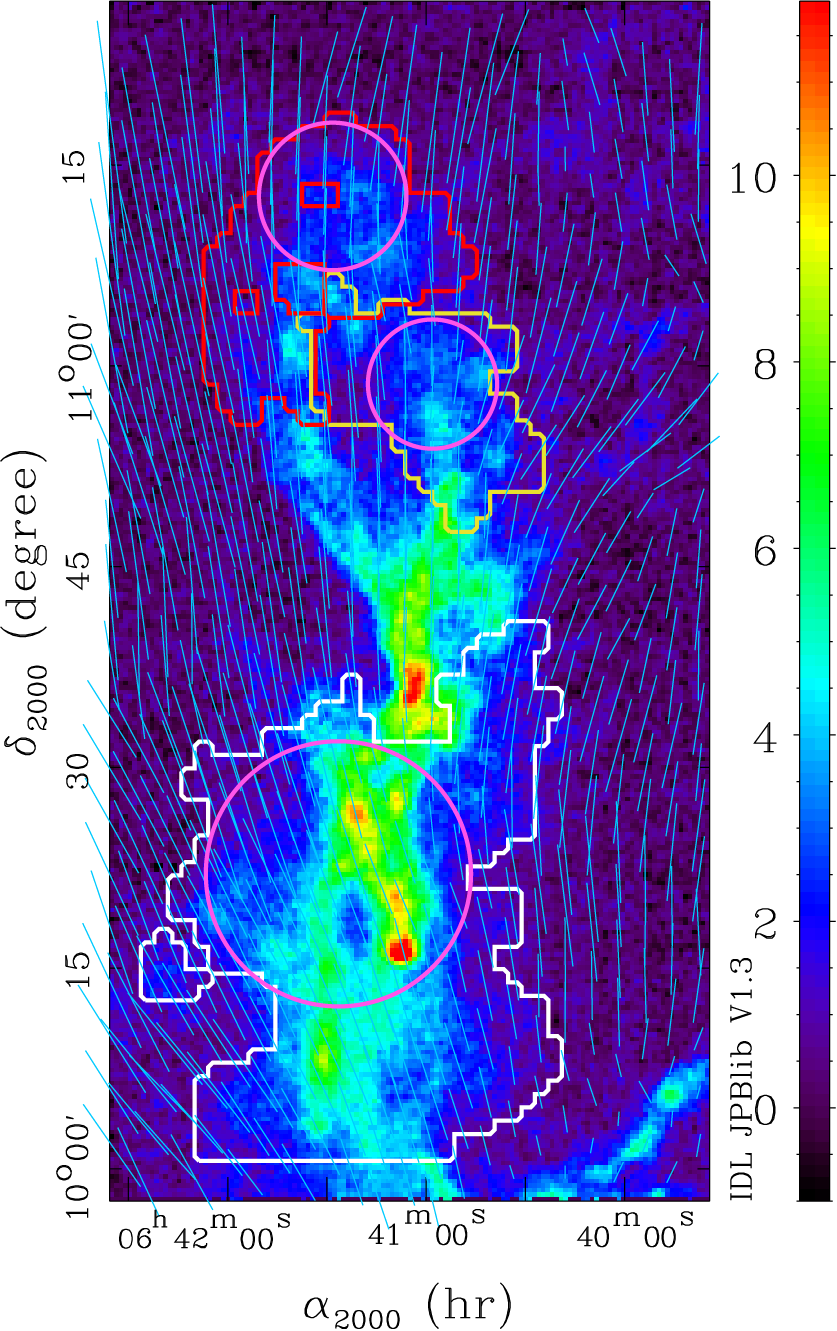}&
\includegraphics[width = 0.23\textwidth]{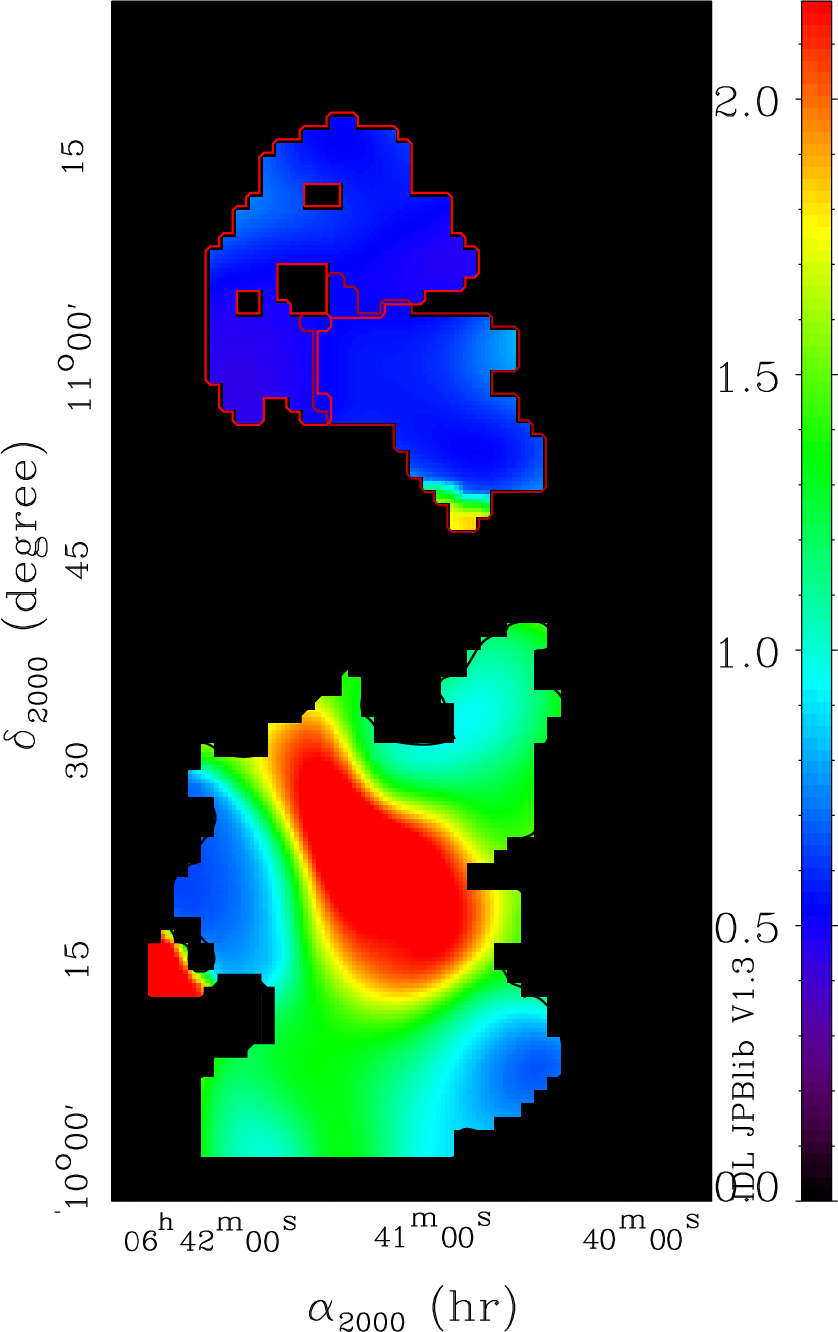} 
\end{tabular}
\caption{\textbf{Left panel:} location of the {\mn} (white contour), North-Western (yellow contour), and North-Eastern (red contour) velocity coherent structures overlaid on the TRAO $^{13}$CO integrated intensity map, in K\,km\,s$^{-1}$. Segments represent the POS magnetic field orientation derived from the {\planck} data, the length of the segments corresponds to the polarisation fraction, with the reference length of $10 \arcmin$ corresponding to $p=0.05$. \textbf{Right panel:} velocity dispersion in the three subregions from the $^{13}$CO emission TRAO data, smoothed to $7'$ resolution, in km s$^{-1}$. 
The purple circles correspond to the largest circles inscribed in the area of each VCS.
}
\label{fig:vcs_map}
\end{center}
\end{figure}

We estimate the POS magnetic field strength according to Sect.~\ref{sec:dcf} in the North-Eastern, North-Western, and {\mn} filaments, in particular, in VCSs that are identified in $^{13}$CO emission (described in Sect.~\ref{sec:vcs}). This allows us to constrain not only the spatial extent of the regions, but also the associated velocity profiles and their dispersions. We require each VCS to contain at least $100$ pixels (after projection onto the {\planck} grid). 
The {\mn} filament contains several VCSs in the junction region. We do not consider that part of the filament to avoid confusion. We also note that the VCS corresponding to the North-Eastern filament covers mostly the Northern part of the filament.
The contours of the VCS and the velocity dispersion map are represented in Fig.~\ref{fig:vcs_map}.
We can see that the three VCS do not spatially overlap, and because each of the structures accounts for most of the observed emission, the use of the {\planck} polarisation data is justified.

\begin{figure}[htbp]
\center
\includegraphics[width = 0.45 \textwidth]{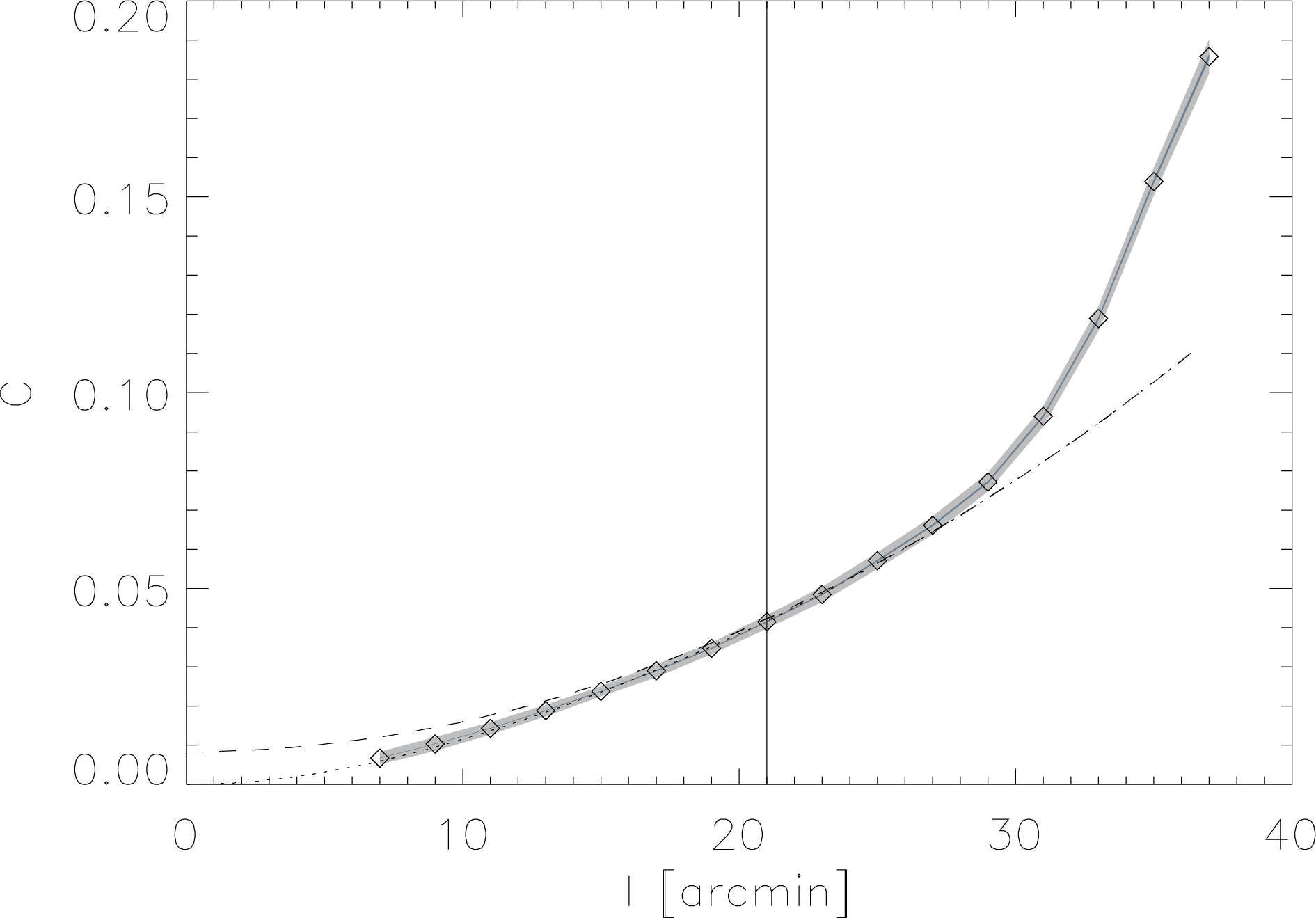} \\
\includegraphics[width = 0.45 \textwidth]{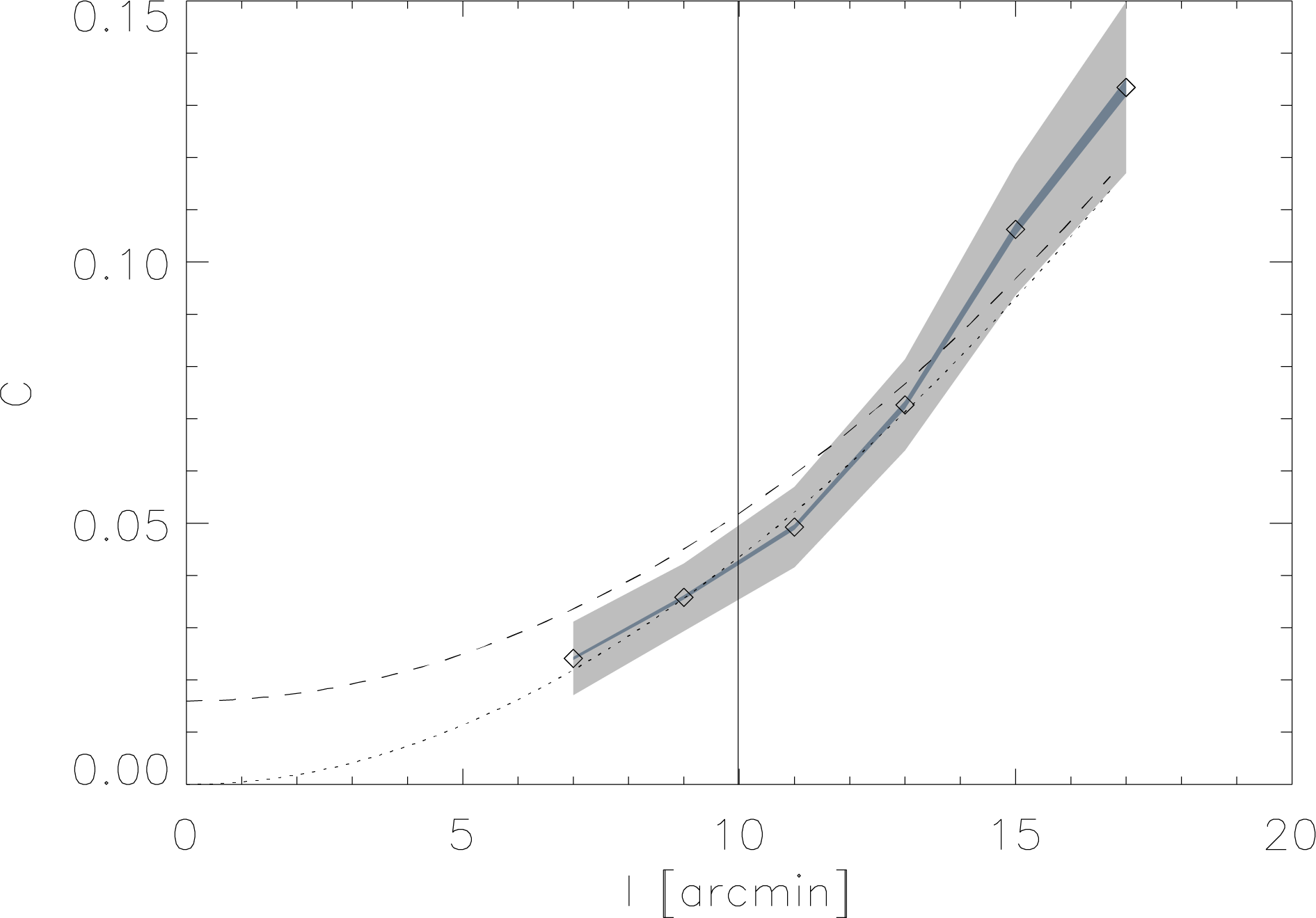} \\
\includegraphics[width = 0.45 \textwidth]{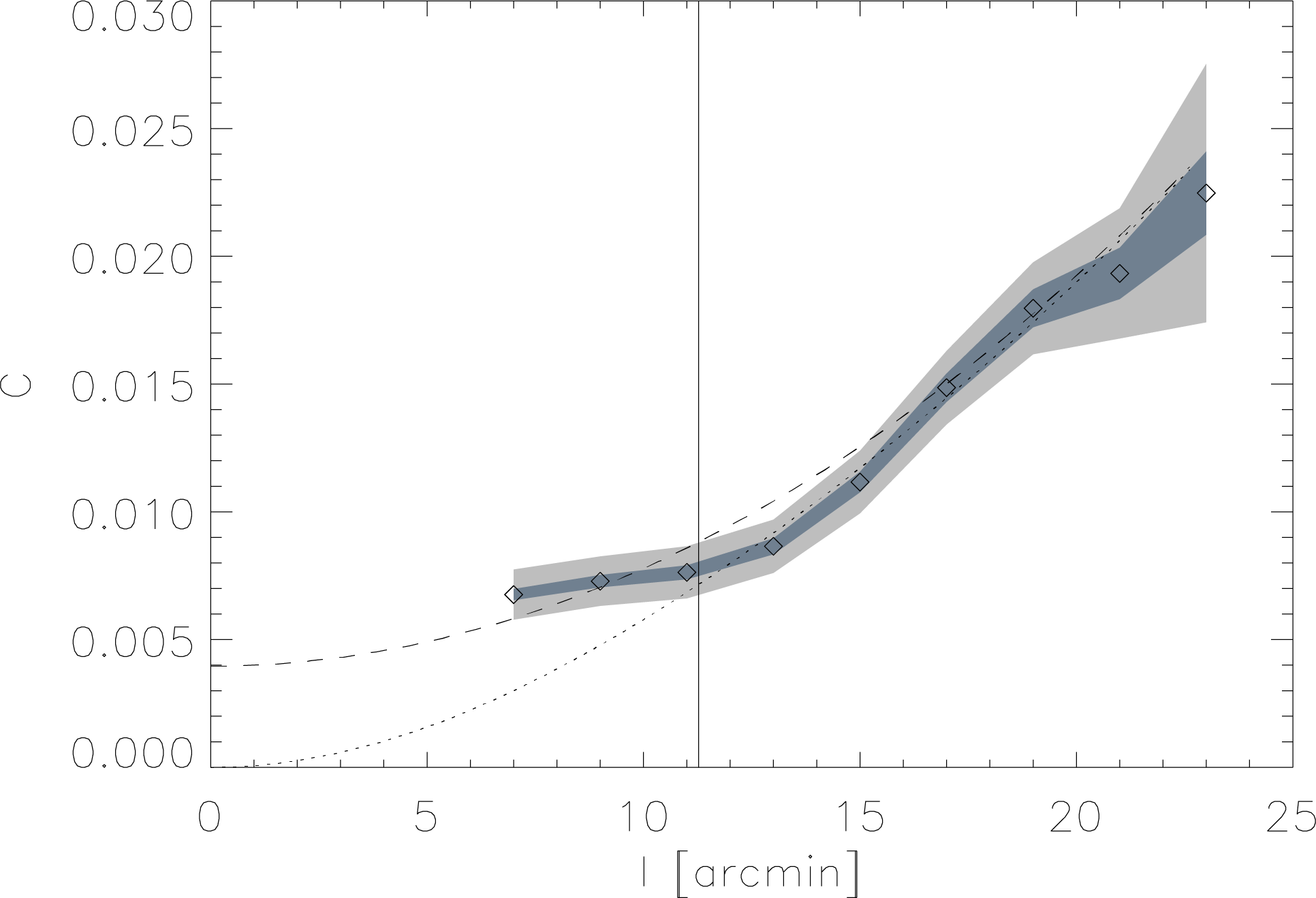} 
\caption{Angular dispersion functions $\mathcal{C}$ ($\mathcal{C} = 1 - \cos(\langle \Delta \polang (l) \rangle)$) and the corresponding fits to the data using Eq.~\ref{eq:houde53} (dotted curves) for the North Main, North Western, and North Eastern regions from top to bottom respectively. The dashed curves show the integrated turbulent component  obtained from Eq.~(\ref{eq:houde53}) without the exponential term. The grey area shows the error estimated using Monte-Carlo simulations that take into account the noise variances of the data and the dark gray area shows the uncertainty due to the number of available pixel pairs for each lag $l$. 
The vertical lines show the diameters of the purple circles from Fig.~\ref{fig:vcs_map}.
}
\label{fig:ang_disp}
\end{figure}

To estimate the ratio of turbulent to ordered magnetic field strengths, we compute the angular dispersion function $\mathcal{C}$ and the corresponding fits as described in Sect.~\ref{sec:dcf}. 
The results are shown in Fig.~\ref{fig:ang_disp}. 
The uncertainties on the measurement of the Stokes $Q$ and $U$ in the {\planck} data propagate non-linearly to $\mathcal{C}$. We use Monte-Carlo simulations that take into account the noise variances of $Q$ and $U$ and their co-variance noise. The range of the uncertainties, within $95\%$ confidence interval, is indicated in dark grey in Fig.~\ref{fig:ang_disp}, except for the {\mn} region where the uncertainties are small and are comparable to the symbol size.
The areas of the VCS having elongated shapes, largest distances have smaller number of pixels pairs. We represent the largest circles inscribed in each VCS in Fig.~\ref{fig:vcs_map} and report the corresponding diameters in Fig.~\ref{fig:ang_disp}. This corresponds to the lags for which the determination of pixel pairs is isotropic. 
We also assess the uncertainty due to the decreasing number of pixel pairs with increasing lag $l$ for each VCS using Monte-Carlo simulations of random angles, and report them as gray shaded areas in Fig.~\ref{fig:ang_disp}.
The fit was applied to intermediate lags, because, at low separations between data points, the assumption of the Gaussian turbulent auto-correlation function is not relevant \citep{houde2009}, while at large separations one might be tracing uncorrelated components. 
The resulting estimates of the POS magnetic field strength are reported in Table~\ref{tab:mgf}. The uncertainties are estimated from, first, propagation of the standard error between the data and the fit, and, second, from Monte-Carlo simulations.
We find that the strength is around 17 $\mu$G, 5 $\mu$G and 6 $\mu$G in the North-Main, North-Western, and North-Eastern regions respectively, or globally of the order of 10 $\mu$G. 
This is in agreement with the global tendency in molecular clouds at the observed densities \citep{crutcher2010}. 
Determination of the magnetic field strength using the DCF method is generally subject to several assumptions, as well as the approach adopted here.
Supposing that errors are similar in different regions, a relative analysis of the results allows us to conclude that the North-Western region has a larger turbulent component ($\langle B_t^2 \rangle / \langle B_0^2 \rangle \simeq 0.67$ compared to $\simeq 0.14$ and $0. 11$) with the large-scale magnetic field strength two times weaker than in {\mn} filament. The values of $B_{0,POS}$ reported here are tentative and a possible range will be discussed in Sect.~\ref{sec:discussion}. 
As far as we know, there are no Zeeman line splitting observations in G202.3+2.5 (the Northern part). In the NGC 2264 protocluster located half-a-degree south to the {\mn} region, Zeeman line splitting observations were studied by \cite{maury2012}, but in that region we observe a clearly different geometry of the POS magnetic field.  The magnetic field component projected onto the line of sight (LOS) was estimated to 600 $\mu$G at largest, using IRAM 30-metre observations of CN(1-0) emission line at the resolution of $23''$. The Zeeman measurement with IRAM traces the small-scale magnetic field strength in the NGC 2264, while the {\planck} + DCF-derived magnetic field strength applies to the large-scale magnetic field of G202.3+2.5 and cannot be directly compared.

We compare our results with the results derived using a method that does not account for the beam dilution. \cite{hildebrand2009} proposed to use the two-point angular structure function within the assumption of the independence between the large-scale ordered and turbulent components that introduces no assumptions on the geometry of the large-scale field.
The details of the calculations are reported in App.~\ref{app:comparison}, and the magnetic field strength estimation is reported in the fourth column of Table~\ref{tab:mgf}. This method yields values about two orders of magnitude larger for the strength of the ordered component and it proves the necessity of accounting for the beam dilution to any data. It was also noted by \citet{planck2015-XXXV} that the method proposed by \cite{hildebrand2009} should be applied with caution to the {\planck} data.
 
\begin{table}[htp]
\begin{center}
\begin{threeparttable}
\caption{Results of the fit parameter $\langle B_t^2 \rangle / \langle B_0^2 \rangle$ from the Eq.~\ref{eq:houde53}, the derived magnetic field strength $B_{0,POS}$, the magnetic field strength derived using the angular structure function (see App.~\ref{app:comparison}), and the estimated mass-to-magnetic flux ratio parameter $\lambda$ in the three detected VCSs in the Northern part.}
\label{tab:mgf}
\begin{tabular}{|c|c|c|c|c|}
\hline \hline
\small Region & $\frac{ \langle B_t^2 \rangle } {\langle B_0^2 \rangle}$ & $B_{0,POS}$ & $B_{0,POS}^1$	& $\lambda^2$   \\
    &               & \small($\mu G$)	& \small($\mu G$) &   \\
\hline
\small {\mn} & $0.14$	& $16.0 \pm 2.5$ & $385 \pm 17$  & $0.95 \pm 0.41$   \\ 
\small North-Western &	$0.67$	& $5.1 \pm 0.9$       	& $362 \pm 14$		& $2.55 \pm 0.57$		\\ 
\small North-Eastern & 	$0.11$		& $6.5 \pm 1.1$					& $85 \pm 2$ 		&	$1.40 \pm 0.38$	\\
\hline
\end{tabular}
\begin{tablenotes}
      \small
      \item $^1$ calculated according to the method of \cite{hildebrand2009}, described in App.~\ref{app:comparison}
      \item $^2$ calculated using Eq.~\ref{eq:lambd}
    \end{tablenotes}
  \end{threeparttable}
\end{center}
\end{table}

The magnetic field may support the filamentary clouds against gravitational fragmentation. 
To quantify this, we evaluate the ratio $\lambda$ between the actual and critical mass-to-magnetic flux ratios $M/\Phi$ according to \cite{crutcher2004}:
\begin{equation}
\lambda = 7.6 \times 10^{-21} N_{\mathrm{H}_2} / B_{tot}\, ,
\label{eq:lambd}
\end{equation}
where $B_{tot}$ is the total magnetic field strength in $\mu G$ and $N(\mathrm{H}_2)$ is the column density of molecular hydrogen in molecules per cm$^2$.
We then determine the total magnetic field strength out of our estimation of $B_{POS}$:
\cite{heiles2005} argued that statistically $B_{POS}$ accounts for 0.79 of $B_{tot}$ on average, and we adopt this value. If $\lambda < 1$, the cloud is subcritical which means here that the cloud is not prone to collapse, and  $\lambda > 1$ the gravitational energy is larger than the  support provided by the magnetic field.
We average the column density data in the three VCS regions and obtain $\lambda = 0.95 \pm 0.41,\, 2.55 \pm 0.57,$ and $1.40 \pm 0.38$  for the North-Main, North-Western, and North-Eastern regions respectively.
This means that these regions are mostly supercritical, where the magnetic field alone does not provide sufficient support against collapse. As we shall see in Sect.~\ref{sec:gradients}, the VGs do not show significant signs of collapse, and turbulence may be the most likely factor providing support.

\subsection{Coupling between magnetic field structure and the cloud's dynamics}
\label{sec:gradients}
According to \cite{hu2020}, the relative orientation between the magnetic field as inferred from dust polarimetric observations and the velocity and intensity gradients of the gas emission can be used as an indicator of undergoing dynamical processes:
\begin{itemize}
\item \textit{i)} if both the VGs and IGs  are perpendicular to the magnetic field, this is an indication that the gravity is not a dominant factor in the region;
\item \textit{ii)} if both the VGs and IGs  are parallel to the magnetic field, this is an indication of a gravitational collapse;
\item \textit{iii)} if the VGs are perpendicular to both the magnetic field and the IGs, this is an indication of a shock.
\end{itemize}

\begin{figure}
    \centering
    \includegraphics[width = 0.45\textwidth]{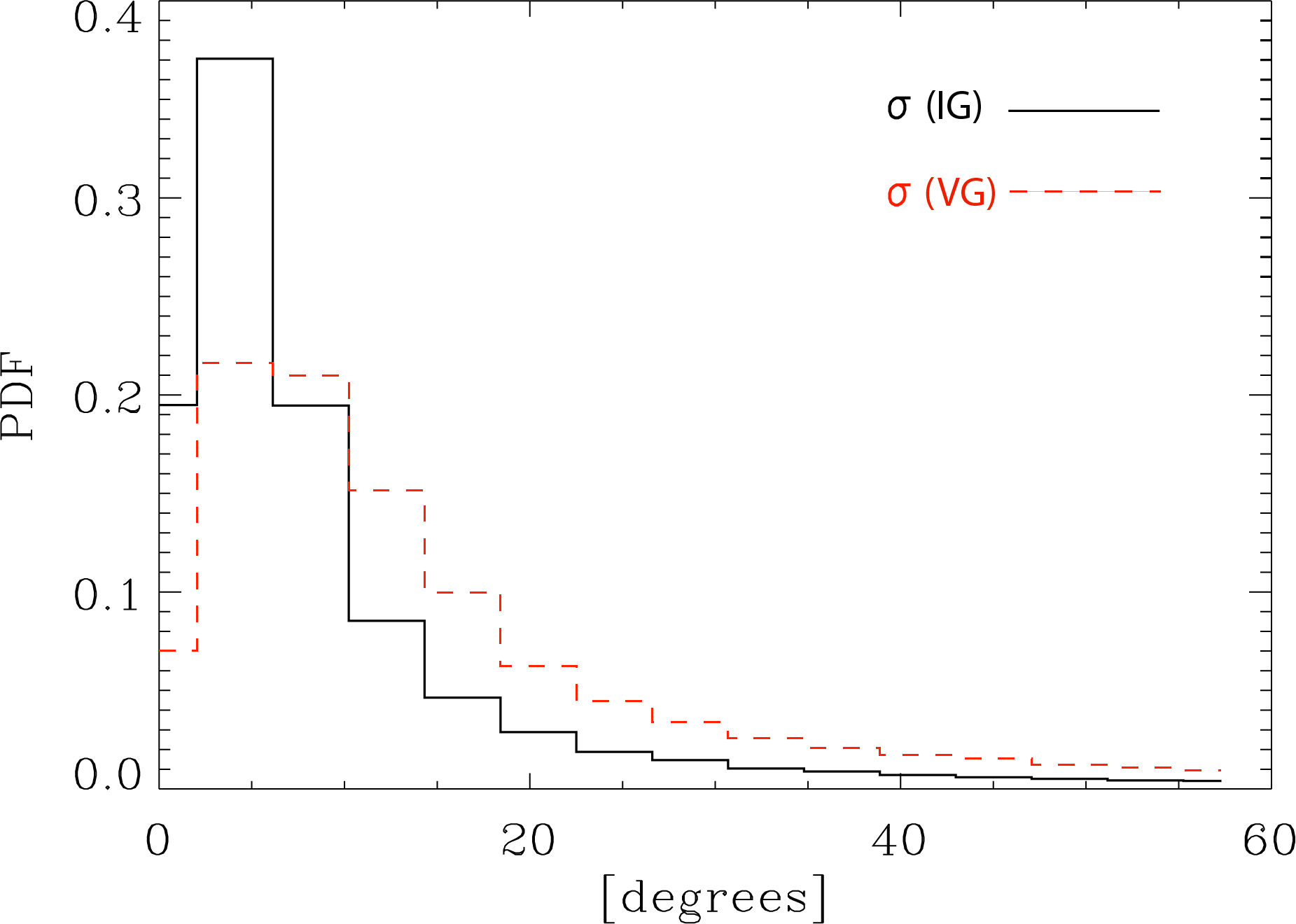} \\
    \includegraphics[width = 0.45\textwidth]{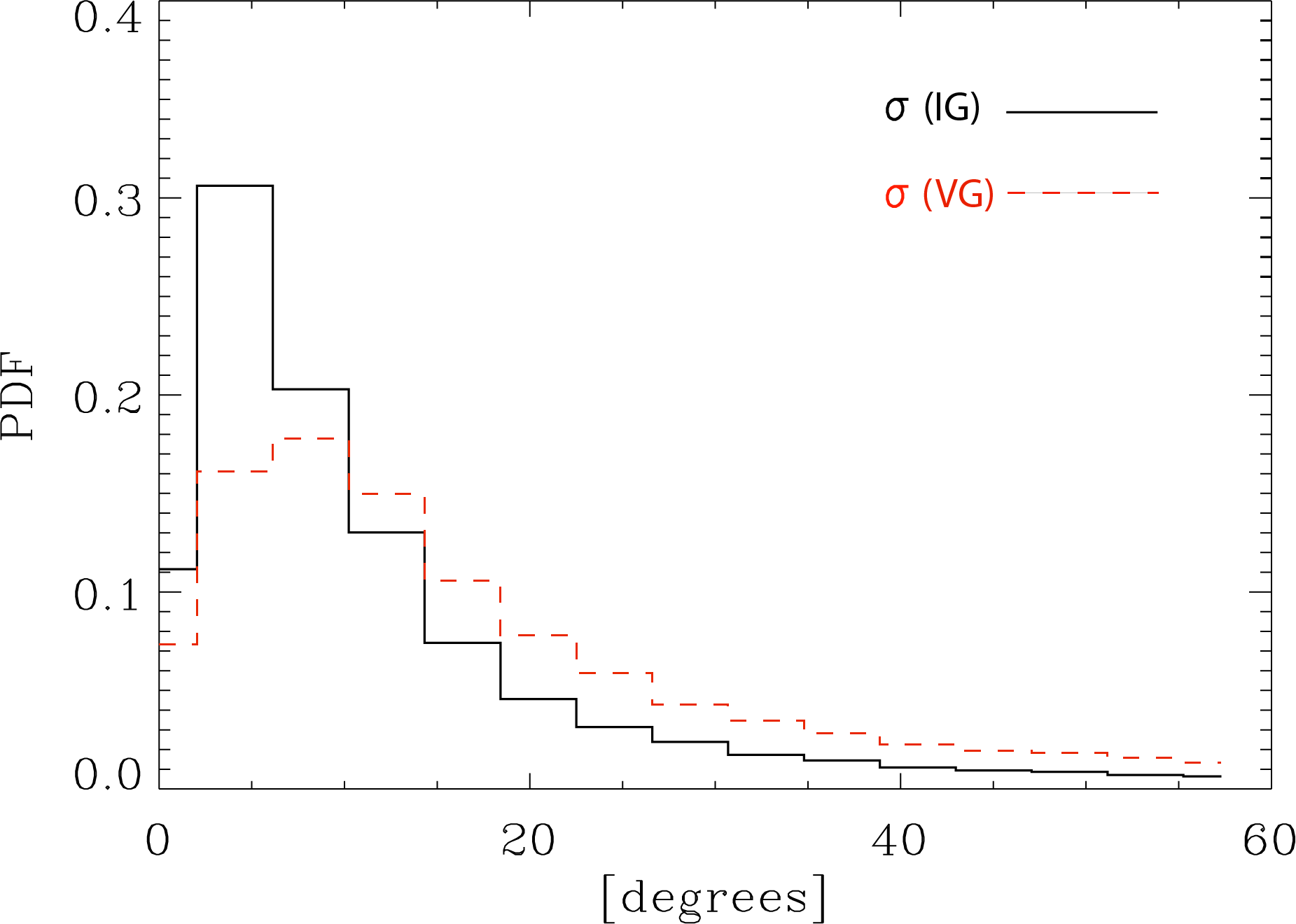}
    \caption{Uncertainties of the IGs (black plain curve) and VGs (red dashed curve) orientations in the Monoceros OB-1 East cloud.  \textbf{Top}: {\co}, \textbf{bottom}: {\coo}.}
    \label{fig:gradients_noise}
\end{figure}

We compute the IGs, VGs and VChGs in the two parts of the cloud and compare the derived directions between them and to the direction of the POS magnetic field. 
The uncertainty on the determination of the gradient angles can come from the systematic error in the map and from the sub-block averaging as described in Sect.~\ref{sec:vgt}. In the sub-block averaging process, only the statistically crucial angle, which corresponds to the peak of the Gaussian fit to the histogram over the considered pixels, is considered,  and the uncertainty of the sub-block averaging is taken as the error of the Gaussian fitting at two-sigma level.
Figure~\ref{fig:gradients_noise} shows the histograms of the gradient uncertainties. The PDFs peak between $5$ and $10$ degrees. We only consider the gradients with uncertainties lower than $30^{\circ}$.
Figure~\ref{fig:b_vs_grads_north} illustrates that the data in the Northern region has a sufficient signal-to-noise ratio, so that the gradients are defined at every data point. In contrast, Fig.~\ref{fig:b_vs_grads_south} illustrates that in faint regions there are no gradients that satisfy our selection criteria.
We set thresholds on the considered area at the {\co} integrated intensity level of $3$ K km s$^{-1}$ in the Southern part and $6$ K km s$^{-1}$ in the Northern part. This globally corresponds to column densities above $\sim 1.8 \times 10^{21}$ cm$^{-2}$ as discussed earlier. The comparison between spectroscopic and polarimetric results is performed above these thresholds. 

\subsubsection*{Northern part}
\begin{figure*}[htbp]
\centering
\begin{tabular}[t]{cc}
\begin{tabular}{c}
\includegraphics[height = 12 cm]{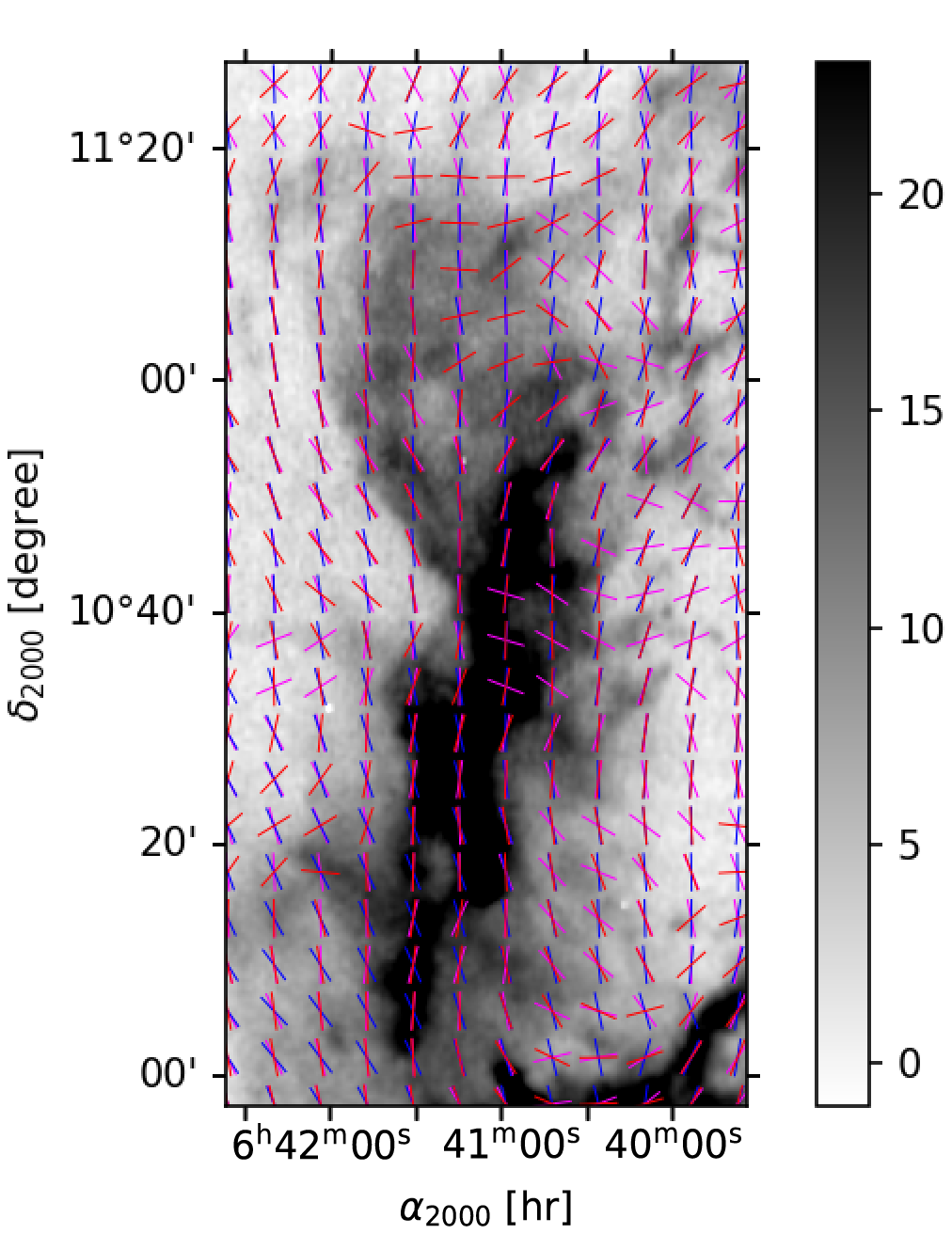}
\put(-200,360) {\tiny a) $^{12}$CO, smoothed}
\end{tabular}
&        
\begin{tabular}{cc}
\subfloat{\label{figur:1}\includegraphics[height = 6 cm]{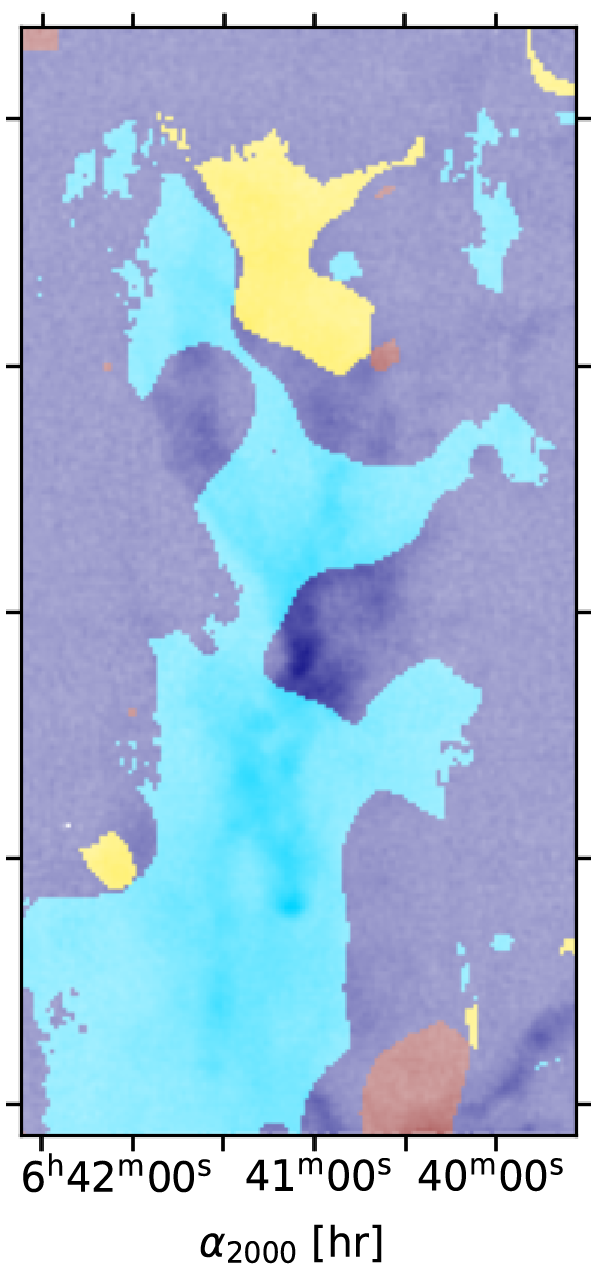}}
\put(-95,175) {\hspace{0.55 cm} \tiny b) $^{12}$CO, smoothed}
&
\subfloat{\label{figur:2}\includegraphics[height = 6 cm]{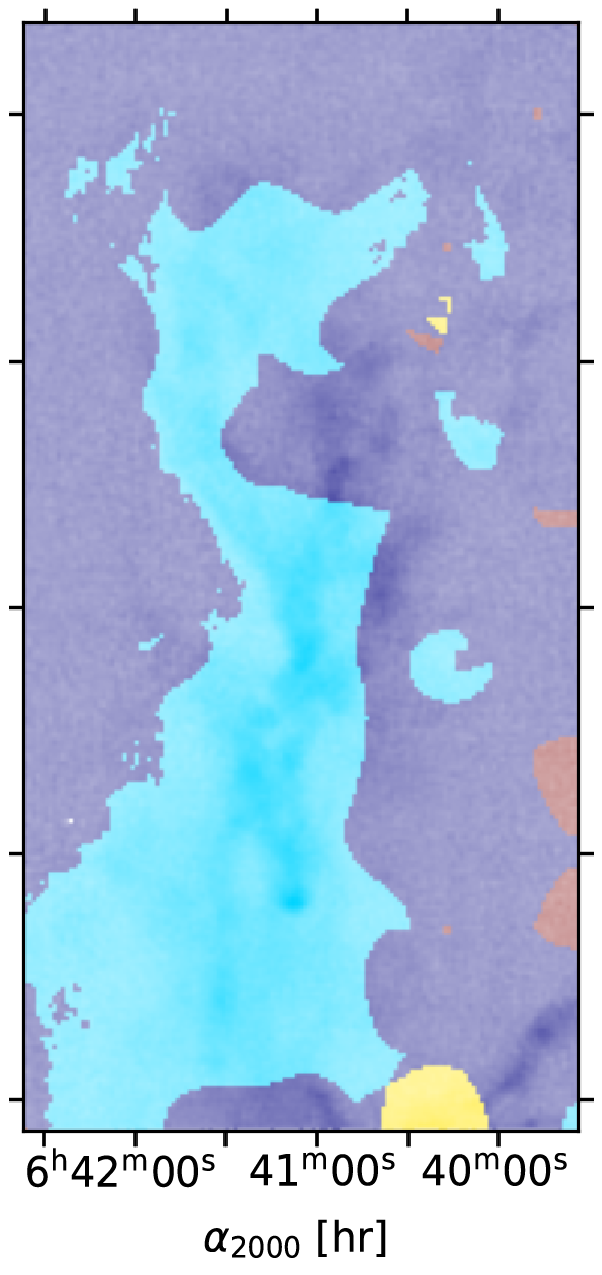}}
\put(-95,175) {\hspace{0.55 cm} \tiny c) $^{13}$CO, smoothed}
\\
\subfloat{\label{figur:3}\includegraphics[height = 6 cm]{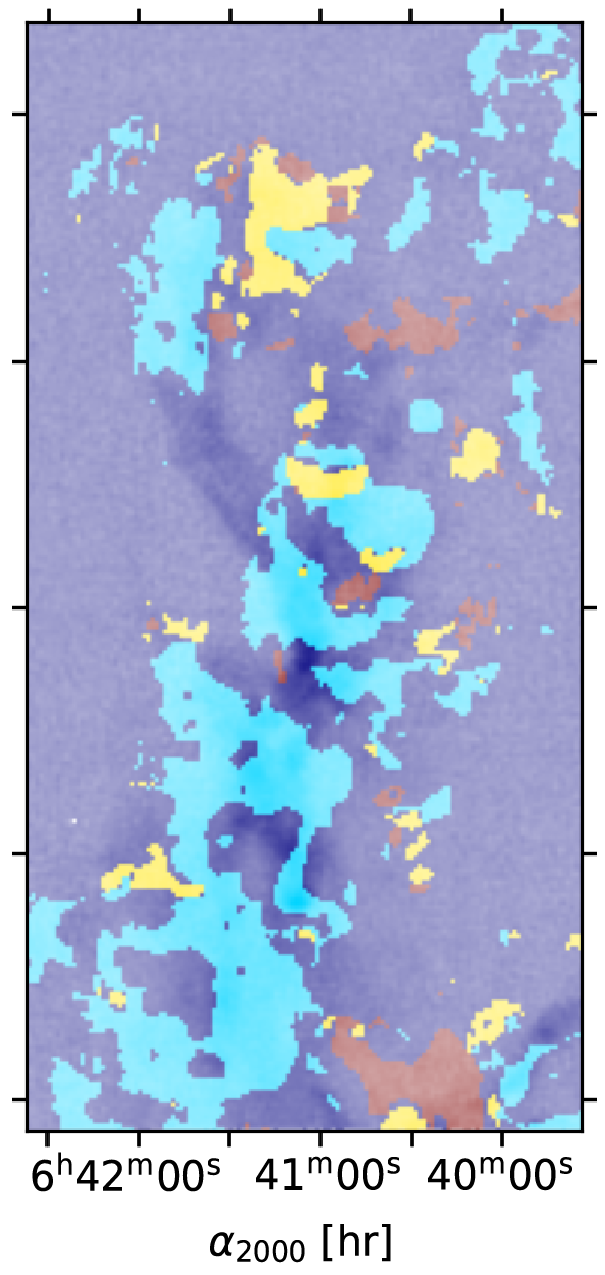}}
\put(-95,175) {\hspace{0.55 cm} \tiny d) $^{12}$CO, unsmoothed}
&
\subfloat{\label{figur:4}\includegraphics[height = 6 cm]{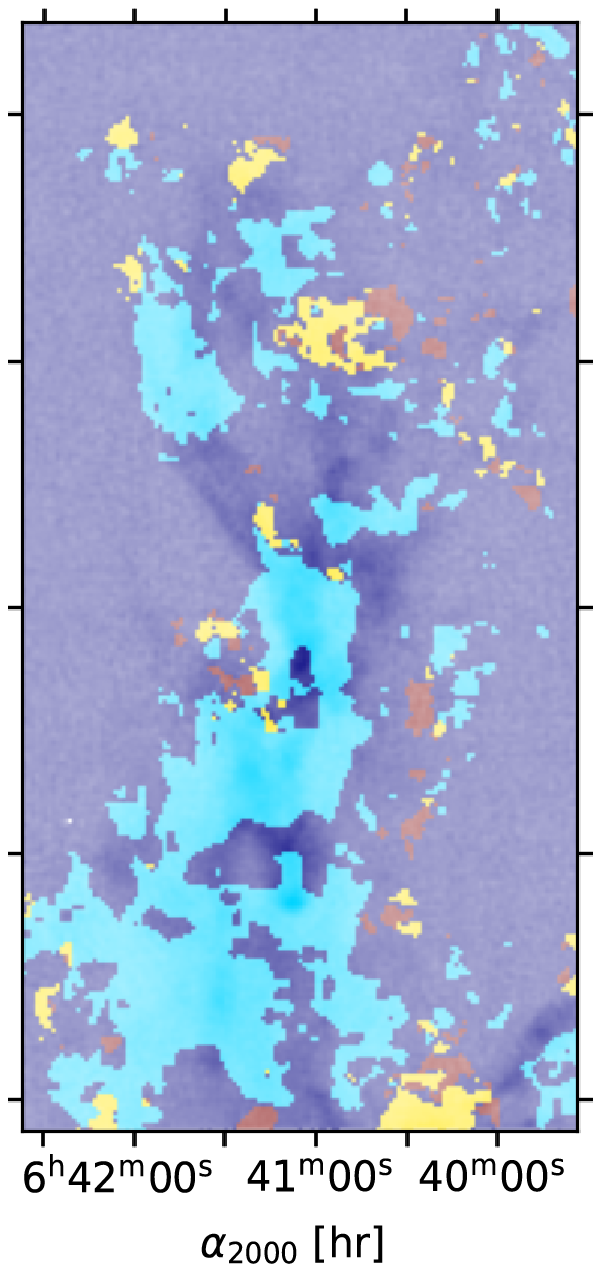}}
\put(-95,175) {\hspace{0.55 cm} \tiny e) $^{13}$CO, unsmoothed}
\end{tabular}
\end{tabular}
\caption{a) Northern part of Mon OB-1 East (corresponding to the yellow rectangle in Fig.~\ref{fig:pol_map}) with the POS magnetic field orientation (blue segments), the orientations of the rotated velocity gradients (magenta segments), and intensity gradients (red segments) overlaid on the TRAO $^{12}$CO integrated intensity map (between $v_{\rm{lsr}}=-3$ to $17$ km s$^{-1}$).\\
b-e) Same region, where the blue shaded pattern corresponds to the regions where gravity is not dominant (case \textit{i}), the red shaded pattern corresponds to the candidate collapse (case \textit{ii}), the yellow shaded pattern corresponds to the candidate shock regions (case \textit{iii}), overlaid on the $^{13}$CO integrated intensity grey scale map: b) based on $^{12}$CO TRAO data smoothed to $7\arcmin$ resolution, c) based on $^{13}$CO TRAO data smoothed to $7\arcmin$ resolution, d) based on $^{12}$CO TRAO data at the resolution of $47\arcsec$, e) based on $^{13}$CO TRAO data at the resolution of $47\arcsec$.}
\label{fig:b_vs_grads_north}
\end{figure*}

\begin{figure}
    \centering
    \includegraphics[width=0.45\textwidth]{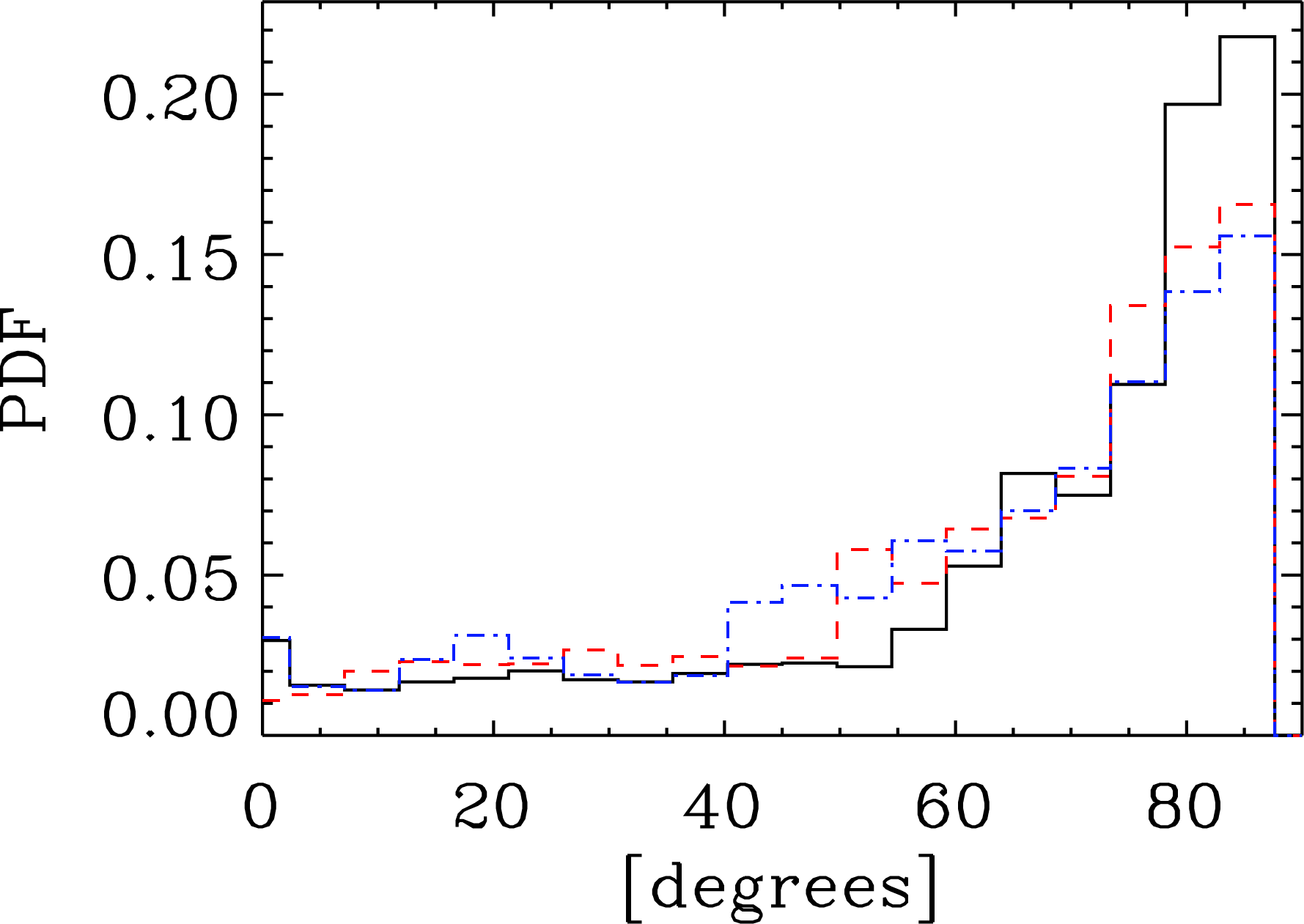} \\
    \includegraphics[width=0.45\textwidth]{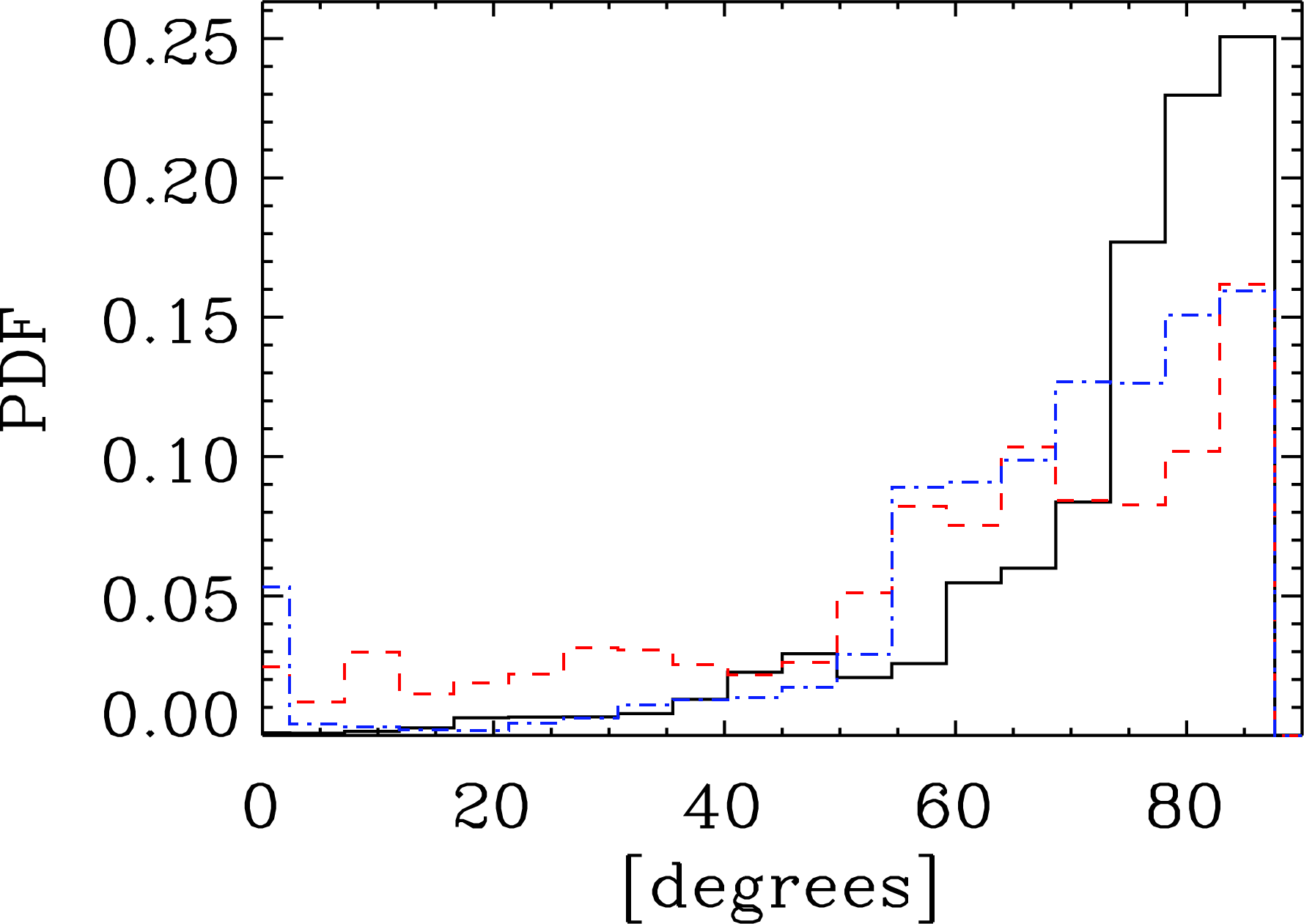}
    \caption{PDFs of the absolute differences between the POS magnetic field derived from the {\planck} data and the IGs in black (plain curve), VGs in red (dashed curve), VChGs in blue (dash-dotted curve) for the Northern part. \textbf{Top:} based on $^{12}$CO data. \textbf{Bottom:} based on $^{13}$CO data.}
    \label{fig:hists}
\end{figure}

We represent in the panel (\textit{a}) of Fig.~\ref{fig:b_vs_grads_north} the POS magnetic field traced by the interstellar dust polarised emission and the VGs and IGs rotated by $90^{\circ}$, calculated on the $^{12}$CO TRAO data in the Northern part of Mon OB1 East association, for illustration purpose.
We compute the angle differences between magnetic fields and VGs and IGs and represent the PDFs in Fig.~\ref{fig:hists}. Globally, the PDFs peak at 90$^{\circ}$, and a 30$^{\circ}$ range from the peak corresponds to more than $1 \, \sigma$ level. In what follows, we adopt this margin as a confident one. 
In panels \textit{b}-\textit{e} of Fig.~\ref{fig:b_vs_grads_north} we overlay the three dynamical cases on the grey scale image of the $^{13}$CO integrated intensity map.
There, the quiescent regions (case \textit{i}) are represented as blue shaded areas, the shock regions (case \textit{iii}) are represented as yellow shaded areas, and the sites of possible collapse (case \textit{ii}) are shown as red shaded areas.
We identify two main types of dynamical regions in the Northern part: the "quiescent" regions where the IGs and VGs are both perpendicular to the magnetic field orientation within $30^{\circ}$ (case \textit{i}) and the candidate "shock" regions where the VGs are perpendicular to and the IGs are parallel to the magnetic field within $30^{\circ}$ (case \textit{iii}). The former means that the magnetic field orientation derived from the gradient orientations (obtained by rotation of $90^{\circ}$) would give the same orientation of the magnetic field as the one derived from dust polarisation observations. 
In the {\mn} and North-Eastern filaments the $^{12}$CO and $^{13}$CO intensity and velocity gradients both agree with polarimetric data in estimating the magnetic field orientation. 
It means that, first, apparently there is no source of foreground contamination in the polarised emission map, and both polarimetric and spectroscopic data trace the same media.
Second, the turbulent motions in the gas and the magnetic field support are larger than the gravitational forces (case \textit{i}), as expected from theory on MHD turbulence that forms the basis of the gradient technique used in this study. 
Interestingly, the comparison of the  $^{12}$CO VGs and IGs with the dust-derived POS magnetic field in panel \textit{b} traces a shock in-between the North-Eastern and North-Western filaments, while we observe no such dynamics in $^{13}$CO (panel \textit{c}).

It is worth noting that the VG and IG maps were smoothed to the {\planck} resolution in order to be compared to the polarimetric data. 
However, these maps bring information on the magnetic field structure and the dynamics of the gas at higher resolution. 
We over-sample the {\planck} map and represent in the last two panels (\textit{d} and \textit{e}) of Fig.~\ref{fig:b_vs_grads_north} the dynamically active regions identified using the unsmoothed map of $^{12}$CO and $^{13}$CO. The corresponding histograms are reported in App.~\ref{app:figures}, Fig.~\ref{fig:north_unsm}.
As expected, the shock is also observed when using unsmoothed {\co} gradients between the two Northern filaments (panel \textit{d}). Moreover, there are signs of collapse around the shock region and in the North-Western filament.
In denser parts, traced by $^{13}$CO, the comparison between the unsmoothed gradients and the over-sampled {\planck} magnetic fields reveals a small shock region between the two filaments which disappears if the smoothing is applied.
In the Northern part of the junction region, in {\coo} (panel \textit{e}), we do not observe any obvious trend. 
However, if the smoothing is applied, this region shows a coherence between the magnetic field orientation derived from the spectroscopic data and the magnetic field orientation derived from polarisation data. 
This can be due to the confusion of multiple sub-structures of Monoceros OB-1 along the LOS  \citep{montillaud2019b}.
The coherence between magnetic field orientation derived using polarimetric and spectroscopic data is also lost at the southernmost end of the junction region, where a signature of merging of the North-Eastern  and {\mn} filaments was detected using the IRAM 30-metre observations \citep{montillaud2019b}. Comparison with higher resolution polarimetric data is necessary in order to analyse the magnetic field in this region.
The velocity channel gradients (VChGs) show a better agreement with the {\planck} data in the determination of the magnetic field orientation: their PDFs show narrower peak toward 90$^{\circ}$, while the VG's PDFs are flatter, especially for the $^{13}$CO data (bottom panel of Fig.~\ref{fig:hists}). However, the global behavior is very similar between the VGs and the VChGs. For this reason, we do not add the corresponding plots.

\subsubsection*{Southern part}

\begin{figure}
    \centering
    \includegraphics[height = 5.2 cm]{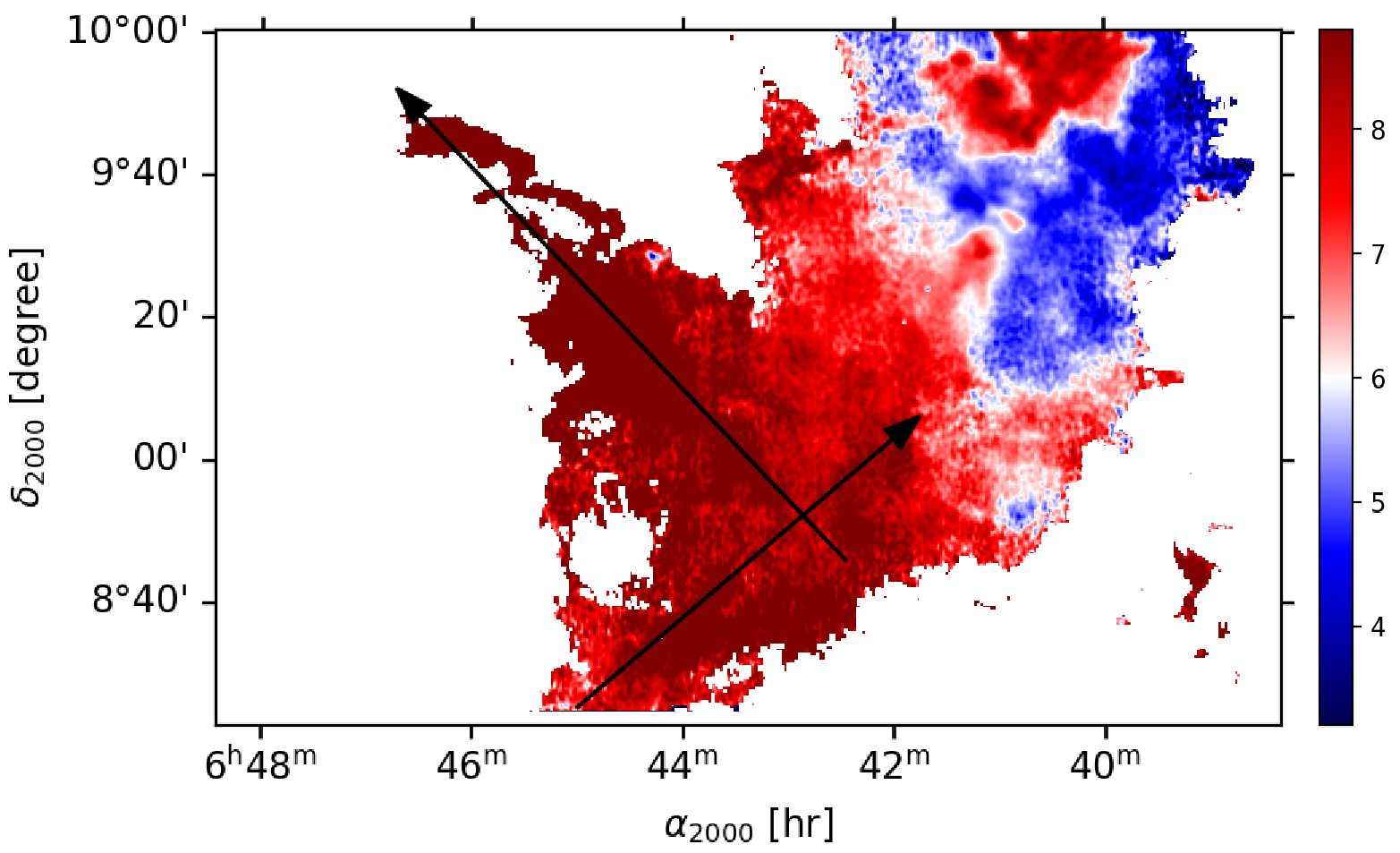}
    \caption{First moment map of $^{12}$CO emission, in km\,s$ ^{-1}$. The arrows show the cuts used for the position-velocity diagrams of Fig.~\ref{fig:PV_cut}.}
    \label{fig:mom1_south}
\end{figure}

The magnetic field traced by interstellar dust emission polarisation in the Southern part of the cloud seems to have two major parts. A uniform magnetic field orientated North-East to South-West  extends from the Eastern edge of the cloud to the densest part, even in regions with no significant CO detections, and is roughly perpendicular to the Galactic plane (see left panel of Fig.~\ref{fig:pol_map}). 
Figure~\ref{fig:mom1_south} shows the first moment map of {\co} emission, in which we observe that an elongated structure that joins the main cloud at the Eastern border and follows this magnetic field direction, is detected at a $v_{\rm lsr}$ from $8$ to $8.4$ km\,s$^{-1}$ in both $^{12}$CO and $^{13}$CO, whereas the rest of the cloud has velocities in the range $3$ to $7$ km\,s$^{-1}$.
In the southernmost part the {\planck} magnetic field is orientated South-North.

\begin{figure*}[htbp]
\center
\begin{tabular}{cc}
\includegraphics[height=5.3 cm]{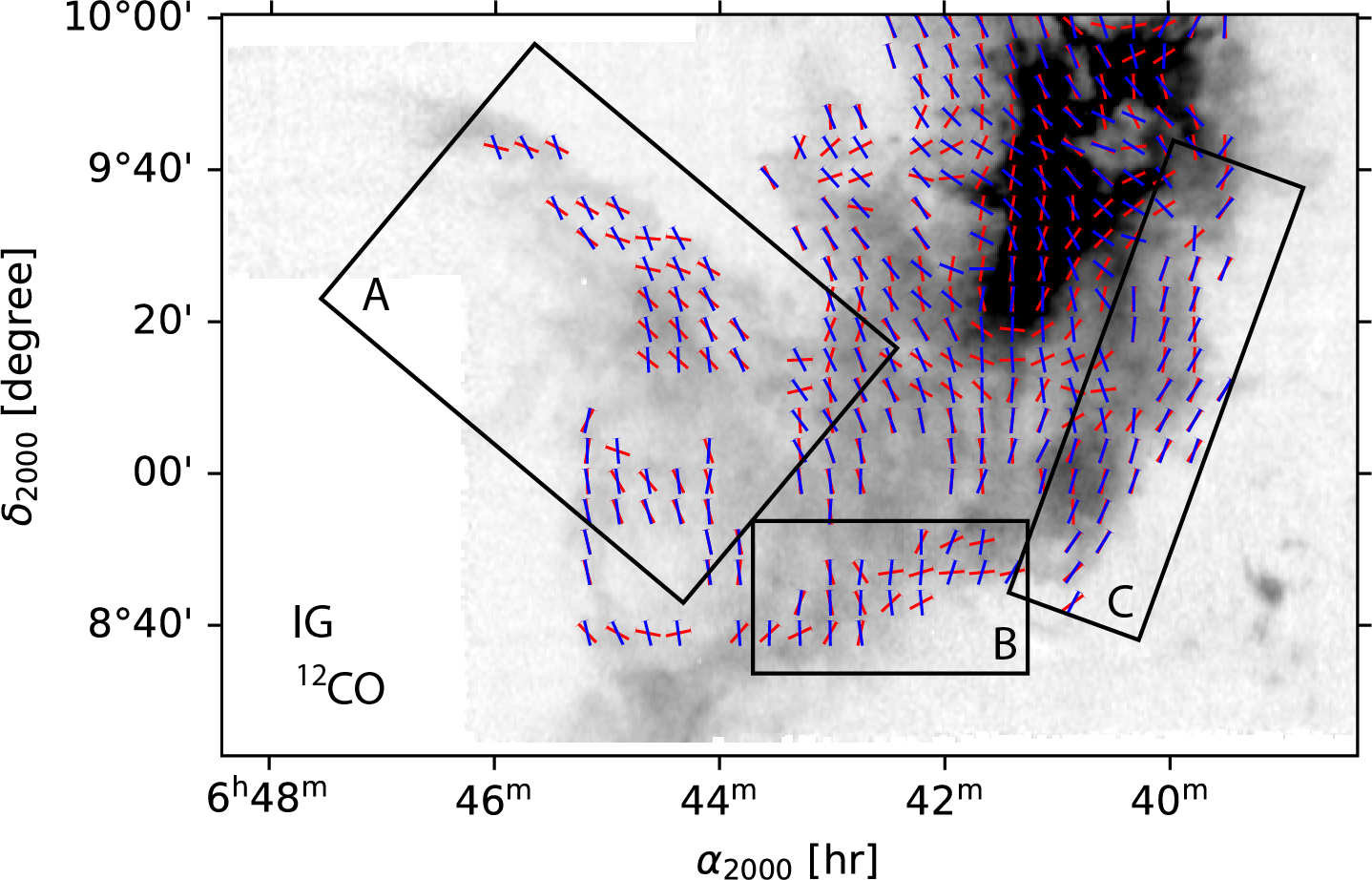} 
\put(-194,42) {\tiny a) }
&
\includegraphics[height=5.3 cm]{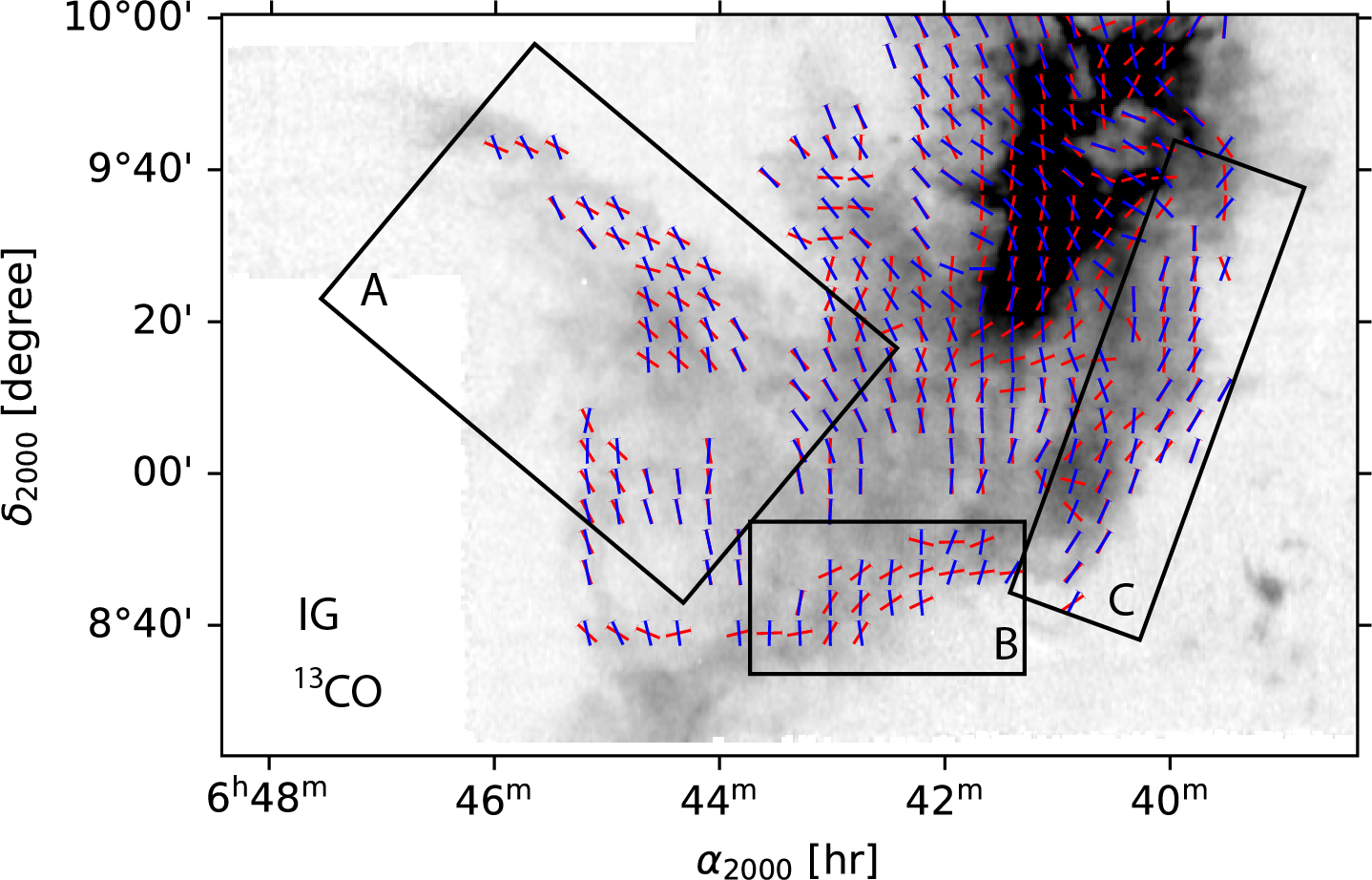}
\put(-194,42) {\tiny b) }
\\
\includegraphics[height=5.3 cm]{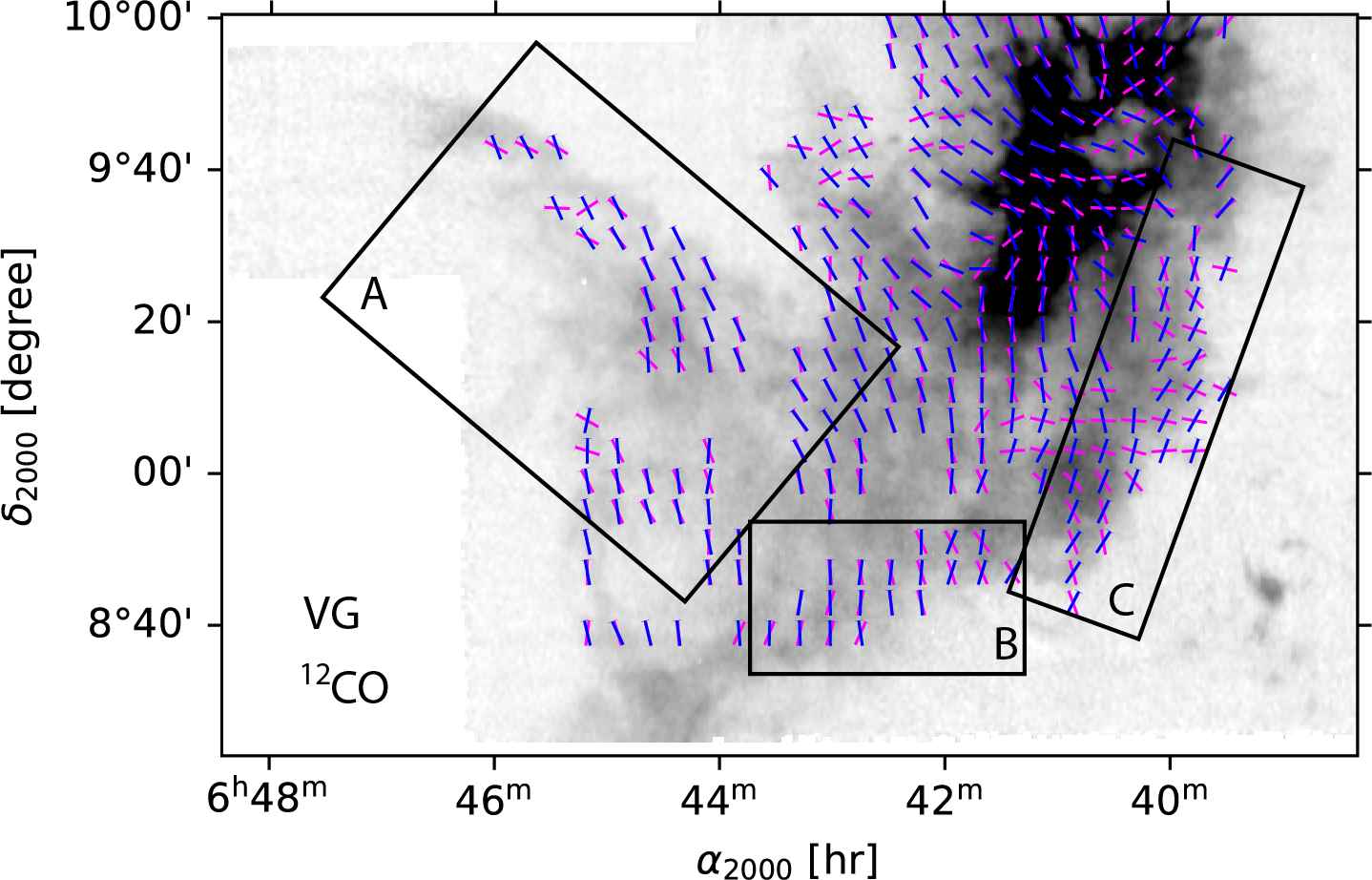} 
\put(-194,45) {\tiny c) }
&
\includegraphics[height=5.3 cm]{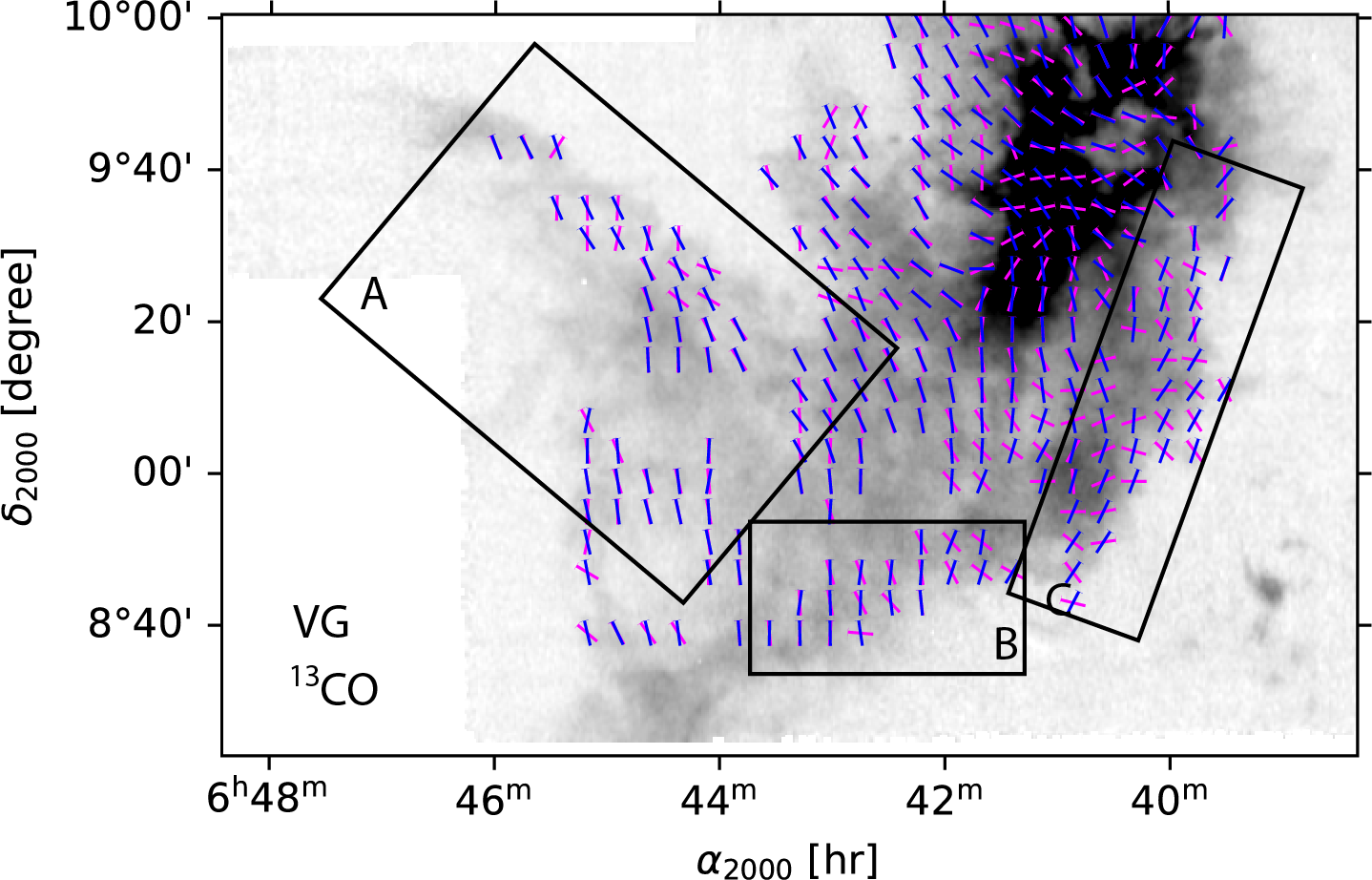}
\put(-194,42) {\tiny d) }
\end{tabular}
\caption{Southern part of Mon OB-1 East (corresponding to the red box in Fig.~\ref{fig:pol_map}) with the POS magnetic field orientation represented by the blue segments, overlaid on the TRAO $^{12}$CO integrated intensity gray-scale map. \\
\textbf{Top row:} Red segments show the orientation of the rotated intensity gradients based on the {\co} and {\coo} tracers (panels \textit{a} and \textit{b} respectively). \\
\textbf{Bottom row:}  Magenta segments show the orientation of the rotated velocity gradients based on the {\co} and {\coo} tracers (\textit{c} and \textit{d} respectively).}
\label{fig:b_vs_grads_south}
\end{figure*}

Figure \ref{fig:b_vs_grads_south} shows the POS magnetic field from the {\planck} data and the rotated velocity and intensity gradients, to illustrate the derived local magnetic field direction.
Contrary to the Northern part, the directions of the velocity and intensity gradients differ in more extended regions.
We schematically define three regions (rectangles A, B, and C in Fig. \ref{fig:b_vs_grads_south}) to identify the special cases of the relative orientation between the {\planck} magnetic field and the gradients rotated by $90^{\circ}$, one-by-one. We note that \citep{gonzales-casanova2017} found the VGs to be a more robust tool than the IGs, as they are stable to different regimes (subsonic or supersonic).
We observe a better agreement between the {\planck} magnetic field and the VG magnetic field than the IG magnetic field in the rectangle A and in NGC 2264 for both gas tracers.
In addition, the alignment is tighter in {\co} than in {\coo}. This was expected because {\co} traces the more extended envelope which might contribute more to the polarised emission detected by {\planck} in that LOS.
In the southernmost part (rectangle B), VGs and IGs are almost perpendicular to each other and the magnetic field derived from the VGs agrees with the {\planck} magnetic field direction, which indicates a candidate shock region.
Interestingly, at the Western edge (rectangle C), the magnetic field derived from the IGs is aligned with the {\planck} magnetic field while it is perpendicular to the rotated VGs. This does not correspond to any of the three dynamical cases. It might be due to the VGs and IGs being dominated by different regions in the LOS. 

The  histograms of the absolute difference between the orientation of the magnetic field derived from the polarimetric data and the orientation of the IGs and VGs are reported in Fig.~\ref{fig:south_hist} of App.~\ref{app:figures}. The distributions are also peaked at 90$^{\circ}$ with a spread which is larger than in the Northern part.

\begin{figure*}[htbp]
\center
\begin{tabular}{cc}
\includegraphics[height=5.3 cm]{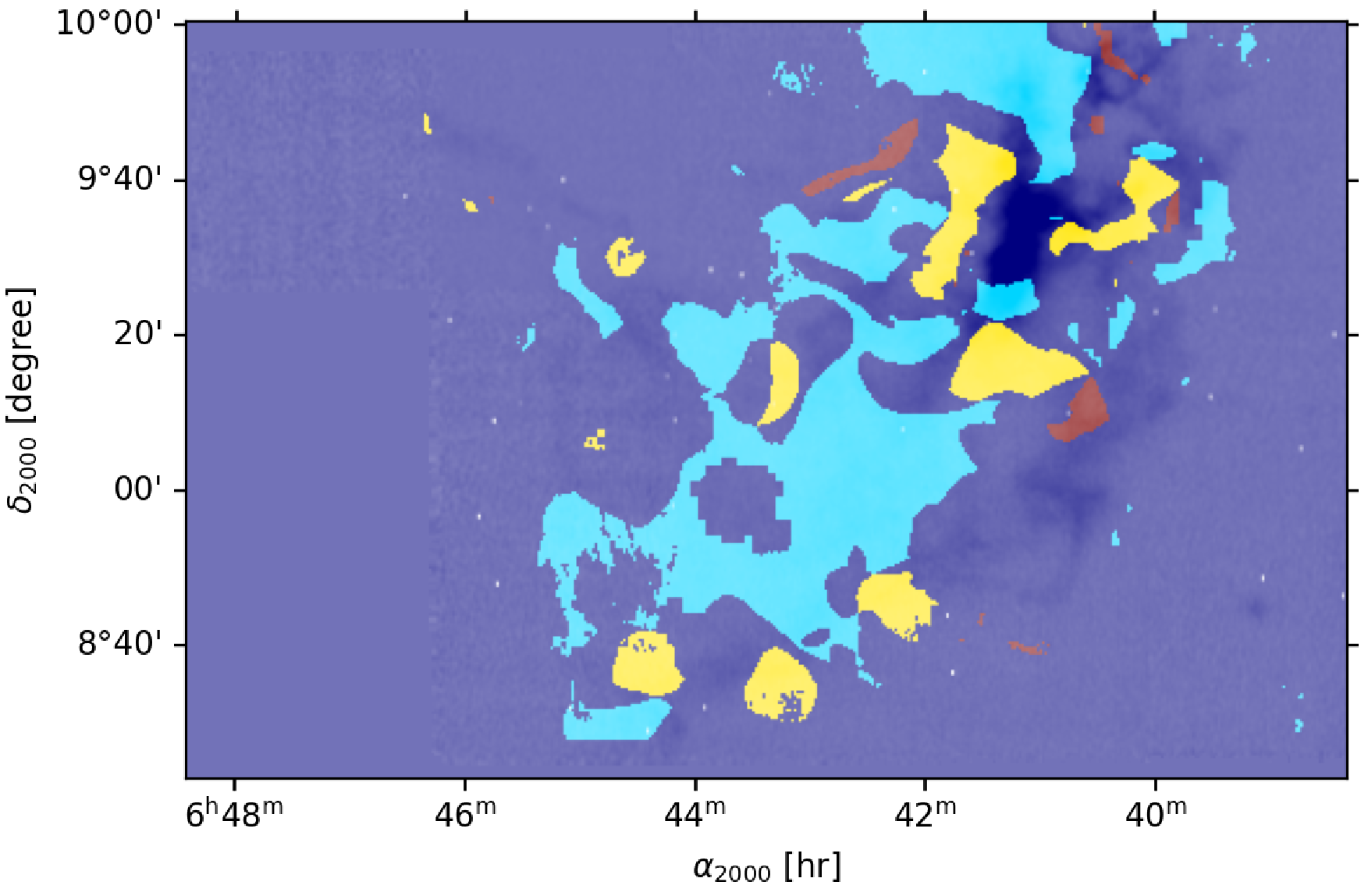} 
\put(-198,28) {\tiny a) {\co}}
&
\includegraphics[height=5.3 cm]{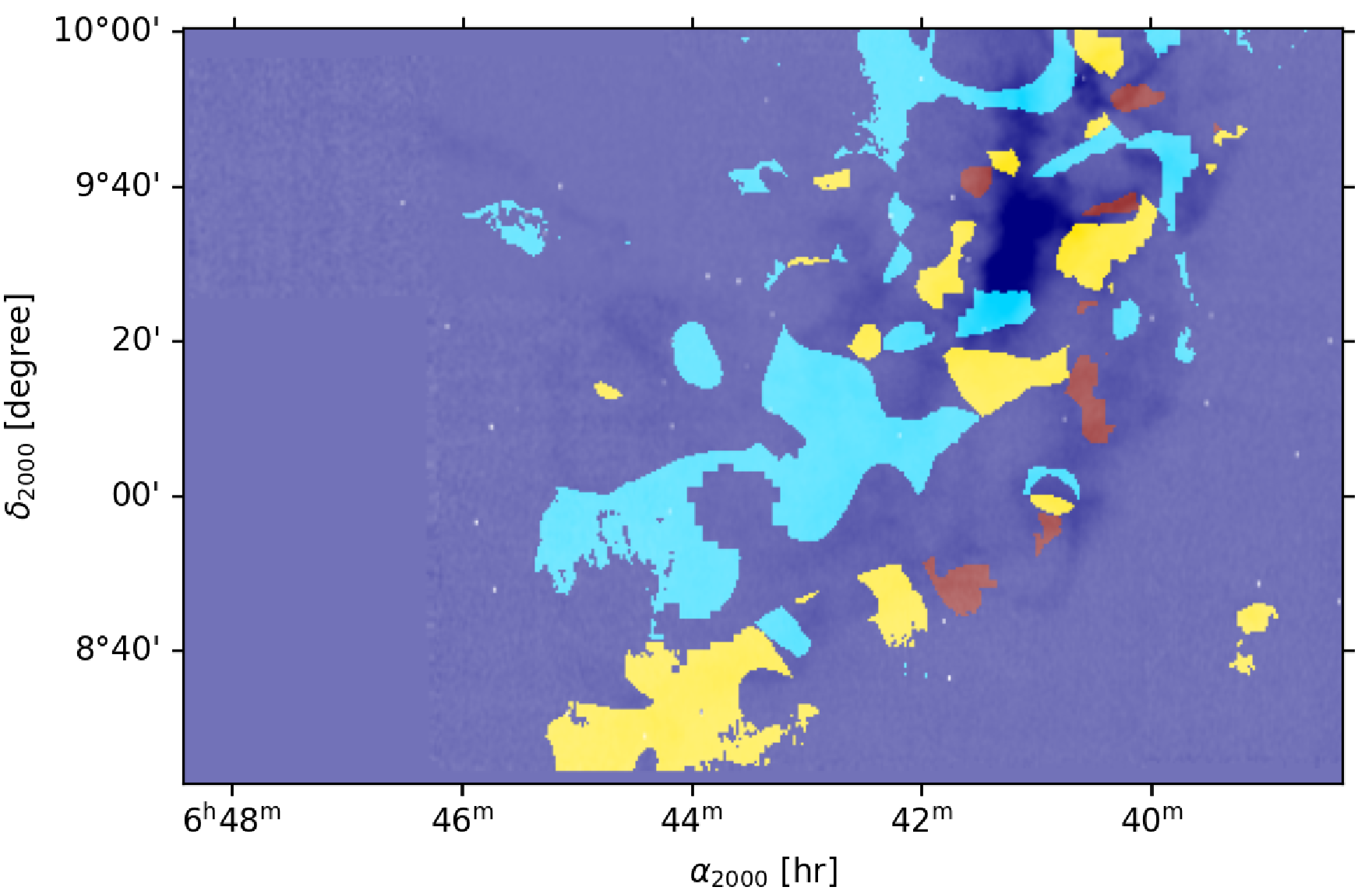}
\put(-198,28) {\tiny b) {\coo}}
\\
\includegraphics[height=5.3 cm]{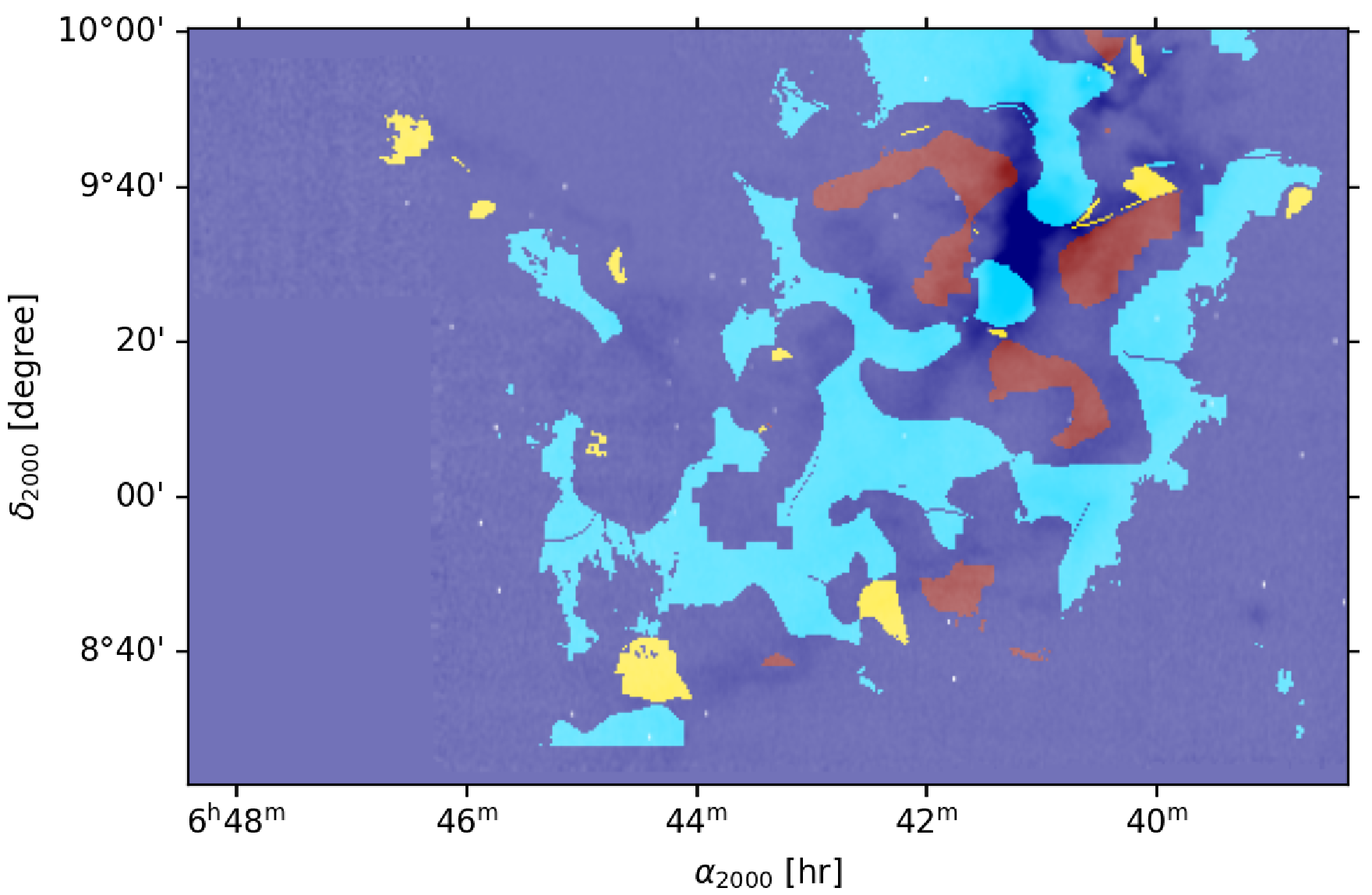} 
\put(-198,28) {\tiny c) {\co}, VChG}
&
\includegraphics[height=5.3 cm]{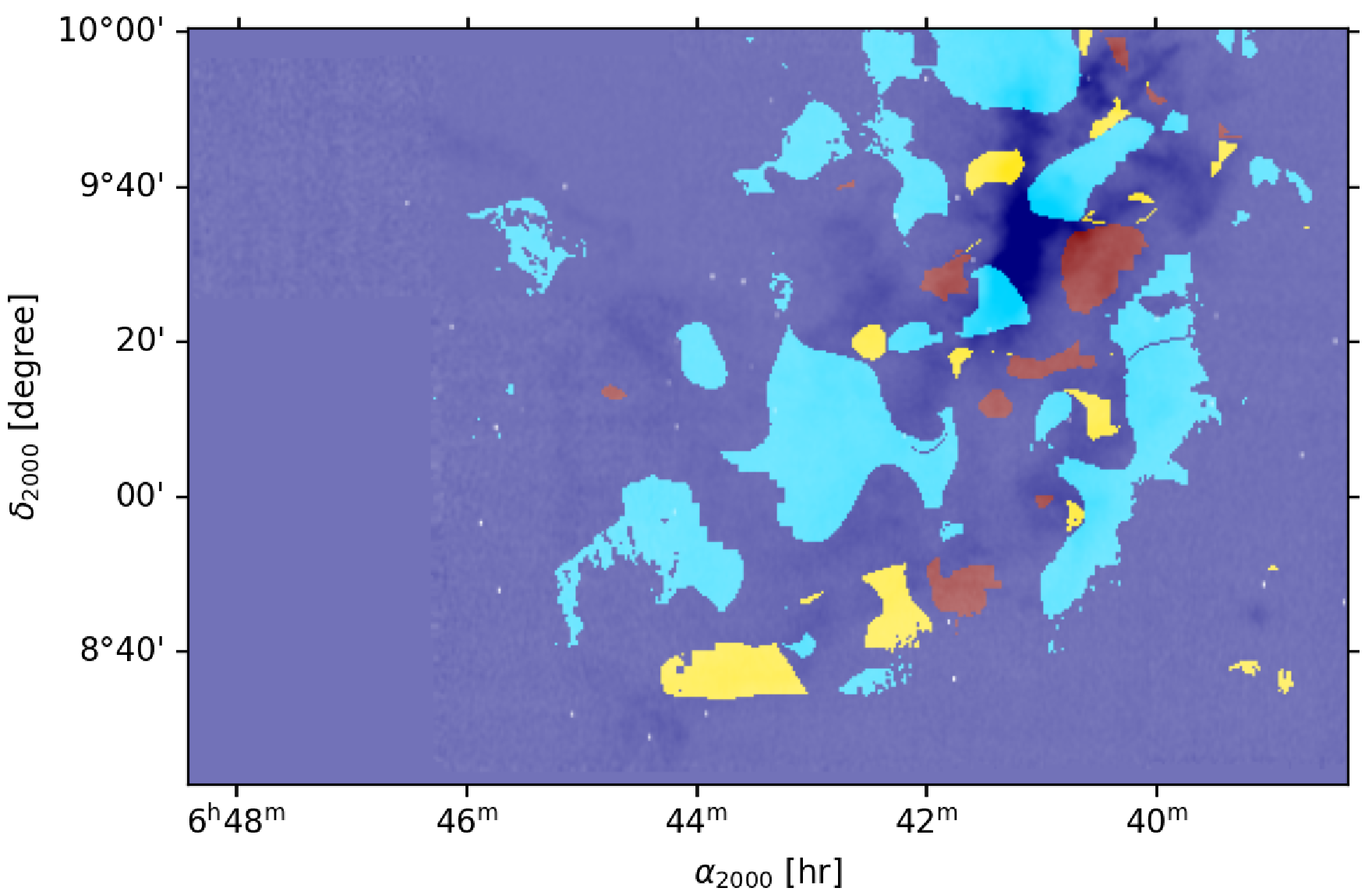}
\put(-198,28) {\tiny d) {\co}, VChG}
\end{tabular}

\caption{Southern part of Mon OB-1 East. The blue shaded patterns correspond to the regions where gravity is not dominant (case \textit{i}), the red shaded patterns correspond to the candidate collapsing regions (case \textit{ii}), and the yellow shaded patterns correspond to the candidate shock regions (case \textit{iii}), overlaid on the $^{13}$CO integrated intensity grey scale map. \\
\textbf{Top row:} based on the comparison between the velocity gradients, intensity gradients and the POS magnetic field direction derived from {\planck}: a) based on {\co} data; b) based on {\coo} data. \\
\textbf{Bottom row:} based on the comparison between the velocity channel gradients, intensity gradients and the POS magnetic field direction derived from {\planck}: a) based on {\co} data; b) based on {\coo} data.}
\label{fig:dynamics_south}
\end{figure*}

We present in the upper panels of Fig.~\ref{fig:dynamics_south} the three cases: quiescent (case \textit{i}) in blue, candidate collapsing (case \textit{ii}) in red, and candidate shock (case \textit{iii}) in yellow. The lower panels show the results when using VChGs instead of VGs.
The IGs and VGs globally show the same magnetic field orientation as the polarimetric data. 
There are signs of collapse around NGC 2264 and at the South-Western edge.
The {\co} VGs indicate a shock-like dynamics near the densest part of the cloud.
However, it disappears when considering the velocity channel gradients (VChGs). This probably means that several structures are present along the LOS and a combination of higher resolution polarimetric and spectroscopic data would be necessary.
Other candidate shock regions are detected in the Southern part and at the outer end of the Eastern elongated structure, both in VGs and VChGs.

\section{Discussion}
\label{sec:discussion}

The aim of this work is to study the large-scale magnetic field of the Monoceros OB-1 East cloud and analyze how it is connected with the cloud's dynamics. 
For this purpose, {\planck} polarimetric data and gradients of TRAO spectroscopic measurements are combined.
The resolution of the {\planck} data and the sub-block averaging of the IGs and VGs give access to scales larger than a parsec or larger than a few arcminutes in angular size. 
This raises questions about whether sub-resolution processes could impact our analysis, and how the data obtained with the large-scale, low-density tracer $^{12}$CO and the smaller scale, higher density tracer $^{13}$CO can be combined to characterize the large-scale magnetic field in this region. We first address these two questions, and then discuss separately the large-scale magnetic field in the Northern and Southern parts of the molecular complex.

The analysis of intensity (or column density) gradients of spectroscopic data for magnetic field studies within MHD turbulence is used by different authors \citep{soler2017, chen2016, heyer2020}. In the last sub-section we discuss our results based on the developments of \cite{hu2020} within the interpretations of the column density gradients behavior in the presence of magnetic fields and self-gravity proposed in the literature.

\subsection{ The effect of outflows on our large-scale study }
NGC 2264 is actively forming stars which is accompanied by outflows. The outflow activity affects the shape of the emission lines and may affect the calculation of the gradients. \cite{buckle2012} analysed the {\co} (J=3-2) emission and H$_2$ emission in the cluster and identified 46 outflows with angular sizes ranging from a few arcseconds to a few arcminutes in a 1 deg$^2$ region. They are clustered in a few groups where they often spatially overlap, and effectively cover a very minor area compared to the 6 deg$^2$ studied here. The authors concluded that the outflow activity is not significantly bringing energy and momentum and is not the dominant source of turbulence even at the scale of their study. In this analysis, the TRAO beam size (47{\arcsec}) is larger than most of the identified outflows, and the sub-block averaging of the gradients trims the contribution of the largest outflows.  Thus, the outflows have a negligible effect at the final resolution of the gradients. Also, the conclusions we draw using the comparison between the gradients and the polarization data, represented in Fig.~\ref{fig:dynamics_south} are barely connected to NGC 2264, but are rather related to less dense, extended structures.

\subsection{Toward a cloud tomography}
The sub-millimeter emission of the interstellar dust is optically thin, and the signal contains information gathered all along the LOS. The {\co} emission is generally optically thick in molecular clouds, and mostly traces the outer envelopes and less dense material within the cloud, while {\coo} is more optically thin and allows us to examine denser parts. We see this difference in the opacity of the two isotopologues as an advantage: when gradients from {\co} and {\coo} both agree with polarimetric data, the magnetic field structure is very probably coherent across different layers, from outer envelopes to inner and relatively denser parts. Conversely, when {\co} and {\coo} gradients bring different information, it suggests a complex magnetic field morphology through the thickness of the cloud. A similar idea was proposed by \cite{hu2019} who suggested that multiple gas tracers ({\co}, {\coo}, C$^{18}$O, CS, HNC, HCO$^{+}$, HCN) could be used to make a cloud's tomography. While we only have two gas tracers, the available high $\snr$ {\planck} data in the region allows us to draw a global picture of the interplay between the molecular cloud and the magnetic field. 
\cite{hsieh2019} showed that the VGs can safely be applied to both optically thin and optically thick gas tracers to infer the magnetic field morphology.
In the Northern part, the IGs and VGs derived from {\co} and {\coo} data (Fig.~\ref{fig:b_vs_grads_north}) globally provide the same direction of the magnetic field. 
Since {\coo} was found to be essentially optically thin in this region \citep{montillaud2019b}, this indicates that the large-scale magnetic field orientated South-North is permeating the filaments at least in gas layers of volume densities of the order of those probed by those transitions ($\sim 10^2 - 10^3$ cm$^{-3}$).
In the Southern part, the magnetic field direction in the cloud's envelope, traced by the IGs and VGs from the {\co} data is different from the magnetic field direction in the denser part of the cloud, traced by {\coo} (Fig.~\ref{fig:dynamics_south}). In the envelope and in the more diffuse gas the magnetic field is in the South-North and South-East to North-West direction while in the denser part it is in the North-East to South-West direction.
\cite{rapson2014} and \cite{park2002} showed that the cloud has a complex, continuous star formation history, with at least one past burst in the southern part of the cloud outside of NGC 2264 and one on-going inside the cluster. Both studies arrived to the conclusion that the cloud is generally quiescent, if not considering NGC 2264. Our analysis supports this statement as we observe a global agreement between the magnetic field directions derived from the spectroscopic gradients and from polarimetric data globally in the cloud.\\

In the two following sections, we discuss each part of the cloud separately.

\subsection{The Northern filaments}
The Northern part of the cloud seen in submillimeter and radio wavelengths has a filamentary shape which bifurcates into two filaments.
\cite{montillaud2019b} identified the structures as the {\mn}, North-Western, North-Eastern filaments, and the junction region (Fig.~\ref{fig:names}) and showed that the North-Western and North-Eastern filaments are moving towards each other, suggesting that they are experiencing a collision at their junction.
We first re-examine this scenario using our results, and then discuss the contribution of the magnetic field to the stability of the cloud substructures.

Our results suggest that the {\mn} and North-Eastern filaments originally formed one structure.
This idea is supported by the following findings:
\begin{itemize}
    \item The {\mn} and North-Eastern filaments have similar polarised intensities properties: their $Q$ and $U$ parameters shown in Fig.~\ref{fig:qu_smallmap} have globally the same polarities, negative and positive respectively, while the North-Western filament has an opposite polarity, both $Q$ and $U$ being negative.
    \item They have similar behaviours in intensity and velocity gradient orientations with respect to the observed POS magnetic field: the IGs and VGs both agree with the polarimetric measurements regarding the derived magnetic field orientation. The difference of the angles in the {\mn} and North-Eastern filaments, shown in the panels \textit{b} and \textit{c} of Fig.~\ref{fig:b_vs_grads_north}, do not exceed 30$^{\circ}$. This is true to a greater extent if we consider the Northern part of the North-Eastern filament, where polarimetric and spectroscopic data show the same orientation of the magnetic field in both $^{12}$CO and $^{13}$CO.
\end{itemize} 

We also detect a signature of a shock in the diffuse gas using the comparison of the IGs and VGs of {\co} with the polarimetric data. 
It is located in the northernmost end of the cloud between the two Northern filaments. The {\coo} data does not show such a trend. 
This may either mean that the dense structures were formed before the collision, or indicate that the dissipation of the shock in the diffuse gas takes longer time.
The shock detection in the combination of polarimetric and spectroscopic data supports the hypothesis of a collision of the filaments proposed by \cite{montillaud2019b}.

Our results shed a new light on the findings of \citet{montillaud2019b} who interpreted the {\mn} and North-Eastern filaments as two originally distinct structures in the process of merging at the level of the junction region, while the North-Western filament was understood as the continuity of the {\mn} filament. Our analysis suggests that the {\mn} and North-Eastern filaments were one and entered in a collision with the North-Western filament, while the magnetic field is dragged along with the matter during the evolution of the structures.

\subsubsection{Magnetic support and filament stability}

The Northern filaments have a large population of dense cores \citep{montillaud2019a,rapson2014}.
Analysis by \citet{montillaud2019b}, based on the {\herschel} data, suggested that the filaments are subject to fragmentation if the magnetic field does not provide support. 
In Sect.~\ref{sec:strength} we used the DCF method to estimate the strength of the magnetic field in the Northern filaments. 
We adopted an approach proposed by \cite{houde2009} to account for beam dilution. However, the DCF method, even with applied corrections, can be uncertain by a factor of two or more, as discussed in \citet{crutcher2012}. 
Also, there exists an uncertainty on the determination of $n_{\rm H_2}$ from observations of $^{13}$CO, which is generally assumed to be a tracer of gas at $n=10^3$ cm$^{-3}$ rather than the values adopted here ($n=10^2$ cm$^{-3}$)  and supported by recent studies \citep{evans2020}. 
With these caveats, it is clear that the absolute value of the B field remains quite uncertain. However, assuming that the errors committed on the measurements of the magnetic field strength are similar in the various regions, differences between their values may reflect genuine variations from region to region. In the following, we discuss the differences between the regions in the frame of this assumption.
Although the values for the magnetic field strength are similar (below $10\, \mu $G) in the North-Eastern and North-Western filaments, the ratio of the turbulent-to-ordered component is larger for the North-Western filament. 
The ratio between the actual and critical mass to magnetic flux ratio $\lambda$ for the North-Western filament ($\lambda \simeq 2.55$) is about two times larger than for the North-Eastern filament ($\lambda \simeq 1.4$) while those of the {\mn} and North-Eastern filaments have less discrepancies ($\lambda \simeq 0.95$). This confirms again that the North-Western filament is dynamically more active, 
and the magnetic field properties of the {\mn} and North-Eastern filaments are similar with low $\lambda$.
We note that if a larger, e.g. by one order of magnitude, gas volume density was taken in the calculation,
the strength of the magnetic field would be larger by a factor of $\simeq 3$ and the two Northern filaments would be considered as sub-critical (with $\lambda = 0.8$, and $0.4$ for the North-Western and North-Eastern), supported by the magnetic field. 
In addition, we do not detect signs of gravitational accretion, at the considered scales,
when comparing the VGs and IGs to the polarimetric data. This indicates that stability against gravity is provided by turbulence and/or magnetic fields. This is in line with studies aiming to explain the inefficiency of star formation in molecular clouds \citep{zuckerman1974}. Simulations \citep{clark2004,dobbs2011,bonnell2011} and observations \citep{barnes2016,nguyen-luong2016} showed that globally molecular clouds are gravitationally unbound entities with sites of very dense bound gas in which star formation occurs. 

The velocity dispersion used in this analysis is provided by the {\coo} data, while the polarisation data at the given resolution ($7\arcmin$ or $1.5$ pc) trace the cold dust emission of structures such as cold clumps or dense filaments. Thus, these results characterize the gravitational to magnetic balance at intermediate densities only (lower than $10^{3}$ cm$^{-3}$), and do not contradict potentially different behaviours at smaller scales in dense cores. The present study should be complemented with higher angular resolution data to tackle this question. For example, one could consider using the data from the Atacama Cosmology Telescope (ACT, \citealt{thornton2016}) at $1\arcmin$ resolution \citep{naess2020} to make the comparison between polarimetry- and spectroscopy-derived orientations of the POS magnetic field but also to estimate the strength using techniques involving polarization-gradient relation within ideal MHD framework from the polarimetric data only \citep{koch2012}.

\subsection{Southern part}

\begin{figure}
    \centering
    \includegraphics[width=0.4\textwidth]{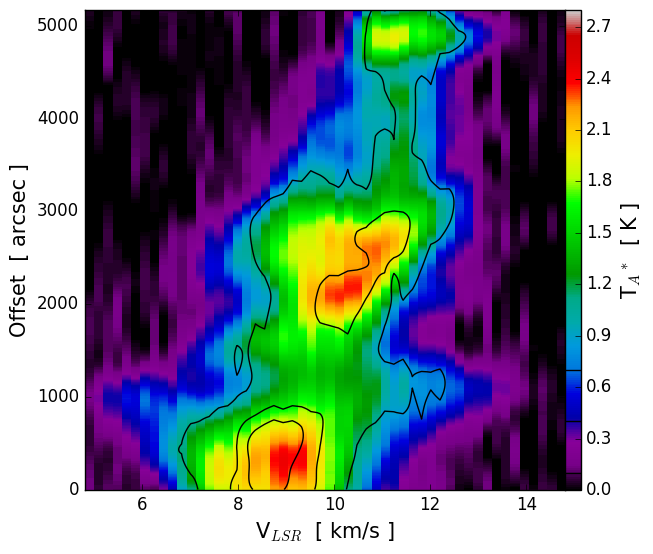}\\
    \includegraphics[width=0.4\textwidth]{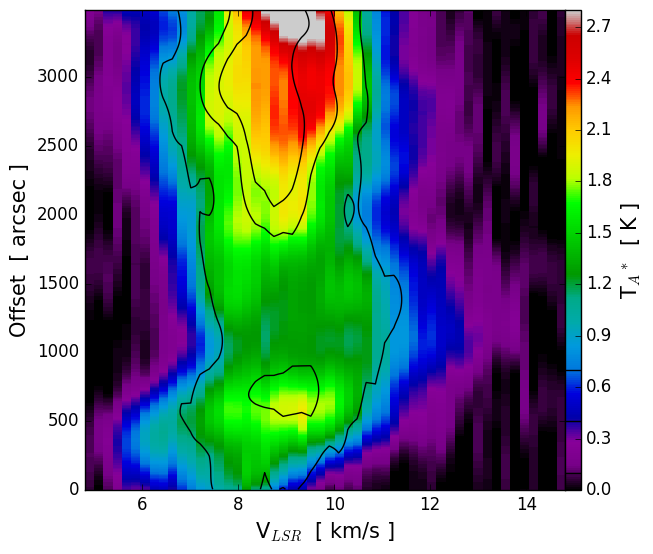}
    \caption{Position-velocity diagrams of the Eastern (top) and Southernmost (bottom) inflow candidates in the Southern part of the Monoceros OB-1 East. The colorscale shows {\co} emission, while the black lines show contours of {\coo} at $T_A^*=0.1$, 0.4, and 0.7 K. The location of the position-velocity cuts and their orientations are shown by the arrows in Fig.~\ref{fig:mom1_south}.}
    \label{fig:PV_cut}
\end{figure}

A large ordered magnetic field orientated roughly perpendicular to the Galactic plane is permeating the Southern part of the cloud across its densest part, and the whole region seems elongated perpendicular to this magnetic field. 
We discuss here two structures of interest in this region. The diffuse region framed by rectangle A in Fig.~\ref{fig:b_vs_grads_south}
is elongated alongside its POS magnetic field lines and presents a strong velocity gradient towards the bulk of the cloud (Fig.~\ref{fig:PV_cut}). This suggests an inflow, either channelled by the magnetic field, or dragging it. However, in this region, Fig.~\ref{fig:dynamics_south}
shows mostly blue patterns (case (i)), where gravity does not dominate the dynamics, and a few small and scattered yellow patches (shocks, case (iii)) leaving the reason for this possible inflow unclear. Its orientation, nearly perpendicular to the Galactic plane, may indicate that the gas follows the Galactic gravitational potential.

A second structure, at the southernmost end part of the cloud, below the rectangles A and B, covers a large velocity range. It appears as a red compact patch in Fig.~\ref{fig:mom1_south} ($\sim 8$ km\,s$^{-1}$), but its position-velocity profile (Fig.~\ref{fig:PV_cut}) shows that it spans velocities from $\sim 6$ to 12 km\,s$^{-1}$ over only $\sim$10\arcmin, suggesting that it corresponds to a structure greatly inclined with respect to the plane of the sky. Interestingly, the four frames of Fig.~\ref{fig:dynamics_south} show a large yellow pattern indicating a large zone of shock between this structure and the bulk of the cloud, suggesting that this second structure is also flowing into the cloud. This draws a picture where star formation in the main cloud is fueled by inflows of diffuse gas on scales as large as 10-15 pc.

We notice other intriguing observations in this region. Signs of shocks are present in NGC 2264 when using the velocity centroid maps, but they disappear when using velocity channel maps. Also, the densest area of the Southern part, corresponding to the location of the NGC 2264 open cluster, is located exactly at the confluence of two distinct magnetic field regions. These observations would request further investigation, but are less relevant for the present study focused on the large scale magnetic field. We delay them to a future study more focused on the area of the cluster.

\subsection{Comparison to alternative approaches}
Predictions of the magnetically-influenced filaments formation and evolution from numerical simulations as well as corresponding observational evidences boosted theoretical investigations of the underlying physical processes. This work employs the technique of intensity and velocity gradients \citep{yuen2017,lazarian2018vgt,2018MNRAS.480.1333H}.
There also exist alternative approaches that utilize gradients of column density. Specifically, \citet{chen2016} and \citet{soler2017} use the gradients of column density to study its relative orientation with respect to the magnetic field. 

  It is worth noting that HRO employed in \citet{chen2016} and \citet{soler2017} studies the relative orientation of the gradients of column density with respect to the magnetic field as revealed by polarization. On the contrary, the intensity and velocity gradients techniques obtain magnetic field direction. Moreover, the VGT and IGT can detect self-gravity effects without employing polarization data. It was shown by \citet{hu2020} that the results of identification of collapsing regions obtained without polarization measurements are consistent with the HRO results obtained with polarization measurements. Recently, using velocity gradients the regions of gravitation collapse were identified in Serpens G3-G6 molecular cloud \citep{hu2021}. Our present results for the Monoceros OB-1 East also show  the consistency of the velocity gradients and HRO predictions.
  Moreover, the combination of the velocity and intensity gradients of molecular line observations with polarimetric observations of interstellar dust allows us to investigate a possibility of other dynamical processes such as the shock or regions dominated by turbulence and magnetic field which are beyond the techniques that utilizes the column density gradients versus polarization data.

\section{Conclusion and perspectives}
\label{sec:conclusion}

We analysed the large-scale magnetic field structure of the Monoceros OB-1 East molecular cloud with its complex morphology of interconnected filaments in order to study the possible influence of the magnetic field on the evolution and formation of the cloud and on its active star formation.  
We used the {\planck} 353 GHz polarised channel to trace the plane-of-the-sky (POS) component of the magnetic field associated with the cold dust and the novel technique of estimation of intensity and velocity gradients (IG and VG). This approach allowed us to trace dynamically active regions such as shocks or turbulence and magnetic field dominated regions via comparison of the polarimetric observations with the spectroscopic data. 

The magnetic field in the Monoceros OB-1 East cloud has a greater influence on the dynamics in the Southern part of the cloud than in the Northern part. In the Northern part, the magnetic field follows the filamentary structure of the cloud, in the South - North direction, and this regular field seems to have the same direction in the envelope, as traced by {\co} emission, and in denser parts, as traced by {\coo} emission.
Our study suggests that the {\mn} and North-Eastern filament were once one whole structure, that collided with the North-Western filamentary cloud. 
In this scenario, the magnetic field was kept dragged during the evolution of the cloud, and is probably providing support against fragmentation at the scales larger than a parsec, at least in the {\mn} filament, in the direction, perpendicular to the magnetic field and the filament. It might also channel a flow of matter along the filament, but we do not detect signs of accretion in the Northern filaments at large scales. However, the comparison of unsmoothed gradients and polarization data show small-feature collapsing regions. We conclude that in the Northern part the magnetic fields and turbulence might be the two stabilizing factors, at scales larger than a parsec.
The Southern part, in contrast, shows a complex structure of the magnetic field. The data suggest an inflow of matter towards the dense parts of the cloud which is channelled along the magnetic field lines in the North-East to South-West direction. Alternatively, the magnetic field could be dragged by the gas motion towards the main cloud.  In more diffuse parts, the magnetic field is also generally orientated in the South - North direction.

Our study of Monoceros OB1-East molecular cloud supports the emerging paradigm: the molecular cloud filaments with higher densities than their environments are mostly elongated perpendicular to the magnetic field, as we see in the Southern part. The lower density-contrasted filaments tend to align with the magnetic field, as we observe in the Northern part, or in the diffuse inflow of the Southern part.

The studies of the relative orientation between filamentary molecular clouds and magnetic fields indicate on the change from parallel to perpendicular with increasing column density \citep{planck2014-XXXV,soler2019}, and the relative orientation also depends on the characteristics of the environment and on the scale such as cores or clumps \citep{zhang2014,hull2017,alina2019}. Recently \cite{pillai2020} reported a discovery of a change of the relative orientation from perpendicular to parallel with increasing density in the open cluster in the Serpens South cloud using Sofia HAWC+ data with sub-arcminute resolution. This is a step forward in the paradigm of the relative orientation between matter and magnetic fields in molecular clouds. A further analysis of the Monoceros OB-1 East cloud and the NGC2264 open cluster is of interest to investigate whether this trend is observed in different molecular clouds.

A detailed analysis of higher {$\snr$} data in velocity channels region by region, using higher angular resolution data, would improve our knowledge of the magnetic field properties in the cloud.
For example, \cite{naess2020} obtained a large data set of polarised emission at 90, 150, and 220 GHz from ACT. They developed an algorithm to combine {\planck} and ACT data. Their data include the Monoceros OB1-East cloud region and a detailed comparison with the TRAO data might bring new pieces of information regarding the large-scale magnetic field.
It might be also of interest to derive the strength of the magnetic field using other techniques such as the polarisation-intensity gradient based method by \citet{koch2012}.
Also, an analysis of spectroscopic data of dense gas tracers, such as N$_2$H$^+$, using the VGs and IGs techniques along with higher resolution polarimetric measurements would be necessary to investigate a possible connection between the magnetic field and the dynamics. 
This will be especially relevant in the complex structures such as the junction region or the NGC 2264 region.

\section*{Acknowledgements}

This research is funded by the Science Committee of the Ministry of Education and Science of the Republic of Kazakhstan (Grant No. AP08855858).
DA acknowledges the Nazarbayev University Faculty Development Competitive Research Grant Programme No110119FD4503. EA acknowledges support from the MES RK grant AP05135753 and the MES RK funded grant “Center of Excellence for Fundamental and Applied Physics” No BR05236454. AL and YH acknowledge the support from the NSF AST 1816234, NASA TCAN 144AAG1967 and NASA ATP AAH7546. Flatiron Institute is supported by the Simons Foundation.
This work was supported by the Programme National “Physique et Chimie
du Milieu Interstellaire” (PCMI) of CNRS/INSU with INC/INP co-funded by
CEA and CNES. TL acknowledges the supports from the international partnership program of Chinese academy of sciences through grant No.114231KYSB20200009, National Natural Science Foundation of China (NSFC) through grant NSFC No.12073061, and Shanghai Pujiang Program 20PJ1415500.

This research made use of JPBLib, an IDL library developed by Jean-Philippe Bernard, and Astropy, a community-developed core Python package for Astronomy. 
\appendix
\section{Relative orientation between the filaments and the magnetic fields}
\subsection*{Filament detection algorithm}
A number of algorithms for identification of filamentary patterns aroused with the discovery of the cosmic web structure of the galaxy clusters \citep{bond1996} and of the ubiquitous nature of the interstellar filaments as revealed by Herschel \citep{andre2014}.
Among many we use the Rolling Hough Transform (RHT) \citep{clark2014} because it allows us to identify structures regardless of their intensity relative to the maximum intensity over the map, which means that not only the densest filament is detected. Furthermore, it allows us to trace the structure extent which means that not only the crest is detected. The RHT is based on the Hough Transform \citep{hough1962}. It centers at each pixel a user-defined kernel of rectangular shape, computes the total intensity for every position angle in the parameter space and builds a histogram from which a final value is picked up. The method strongly depends on the input parameters, such as kernel size, length and threshold for determination of the final angle over the histogram, that are chosen for each map separately depending on the resolution and the angular size of the structures. Also, the rectangular kernel privileges linear structures and the detected pattern can be more linear than the real filament. However, choosing a relatively short aspect ratio for the kernel, that is between two and four, makes it possible to reflect the global curvature of the structure.
We apply RHT to the {\planck} column density map. 
The explored parameter ranges were between $15\arcmin$ and $31\arcmin$ for the length and $3\arcmin$ to $9\arcmin$ for the width, the smoothing kernel is set to be equal to the kernel width.
The final parameters are $21\arcmin$ and $6\arcmin$ for the kernel's length and width respectively, and the normalised histogram threshold for the identification of the maximum is set to 0.65.
It is worth noting that to detect more ramified structures, as, for example, in the {\herschel} map \citep{montillaud2019a}, we would suggest to use the optimized version which does not require to set up the histogram threshold \citep{carriere2019}.

\subsection*{Results}
\label{app:rht}
\begin{figure}
  \hfill \begin{minipage}[c]{0.5\linewidth}
    \includegraphics[width = 4 cm]{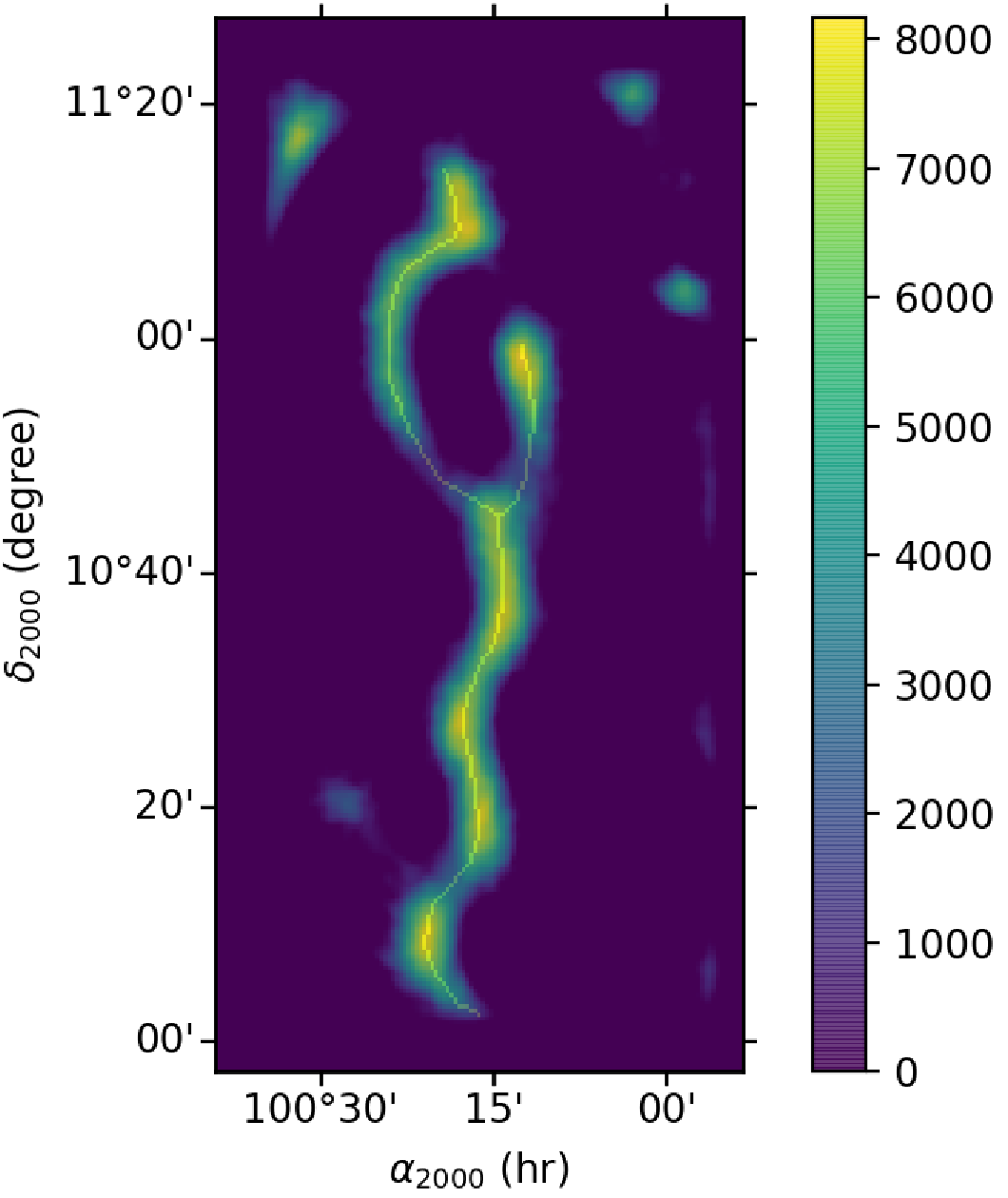}  
  \end{minipage}
  \begin{minipage}[c]{0.4\linewidth}
   \caption{Rolling Hough Transform (RHT) intensity calculated over the {\planck} column density map. The yellow curve shows the crest of the detected structure.}
    \label{fig:rht}
  \end{minipage}
\end{figure}

\begin{figure}[htbp]
\center
\begin{tabular}{cc}
\end{tabular}
\includegraphics[width = 4 cm]{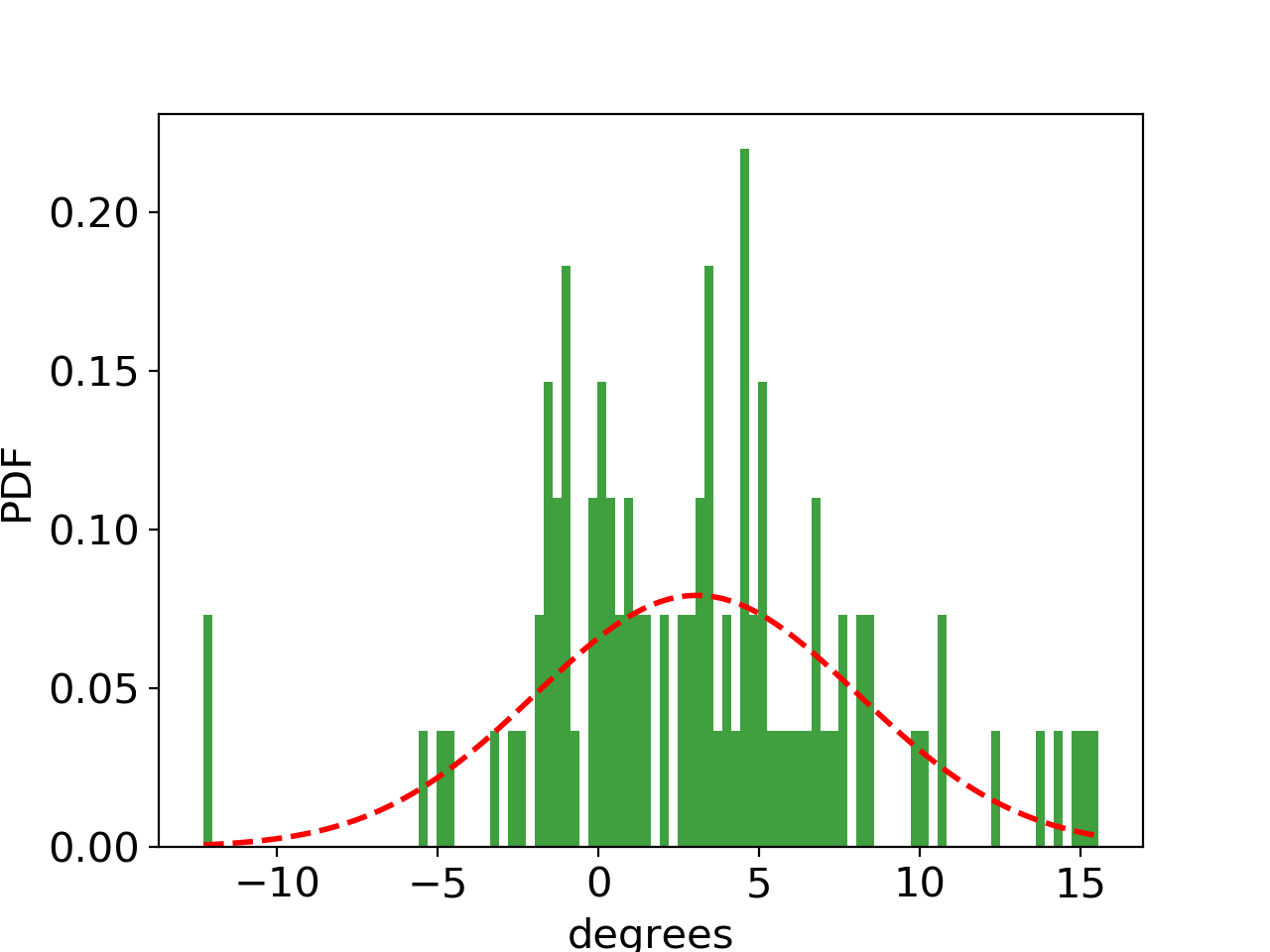} 
\includegraphics[width = 4 cm]{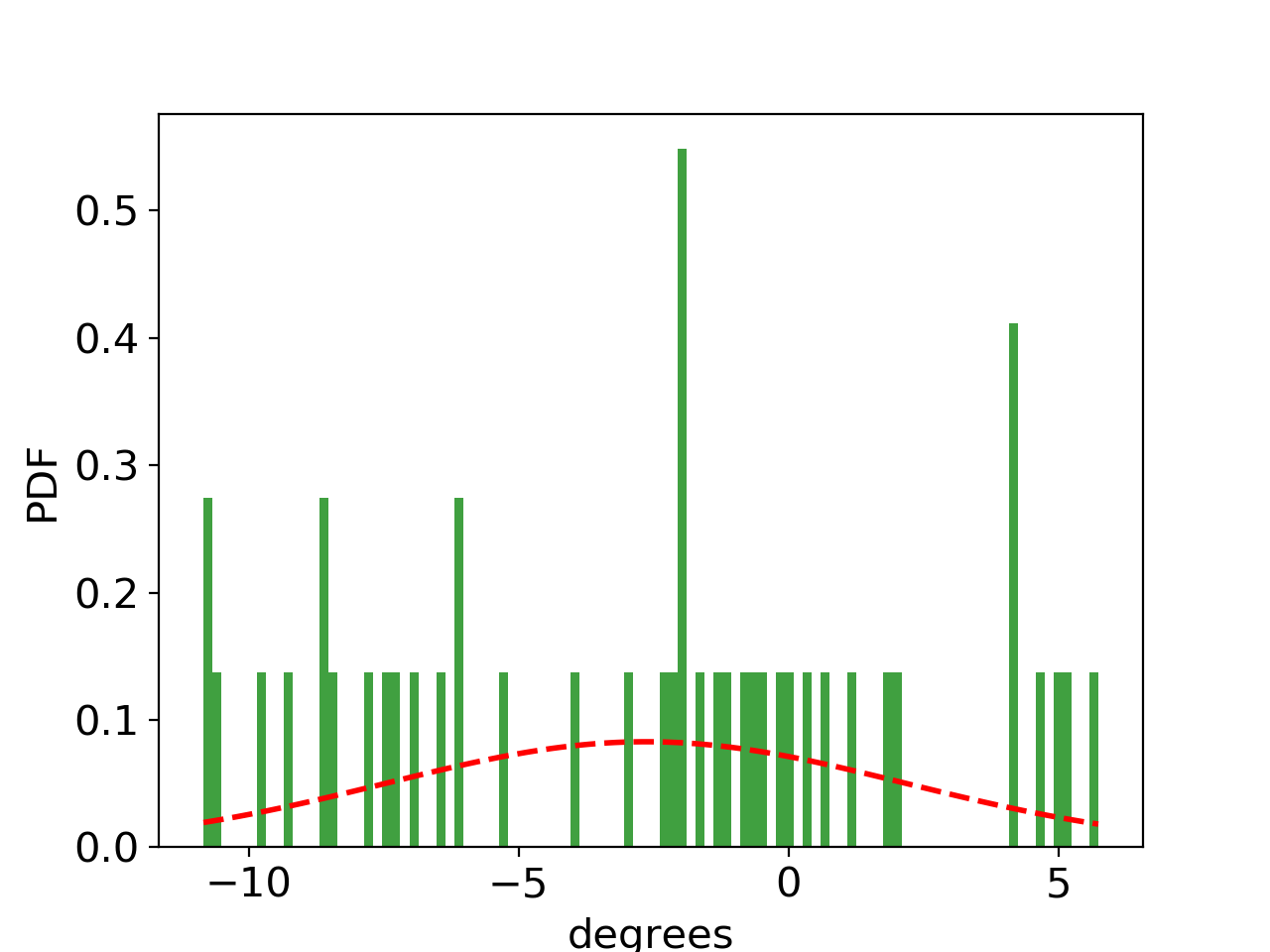}
\caption{PDFs of the POS magnetic field angles derived from the {\planck} 353\,GHz data in the North-Eastern (left panel) and the North-Western (right panel) filaments detected by the Rolling Hough Transform. The red dashed curve represents the Gaussian fits.}
\label{fig:bfils}
\end{figure}

\begin{figure}[htbp]
\center
\begin{tabular}{cc}
\includegraphics[width = 4 cm]{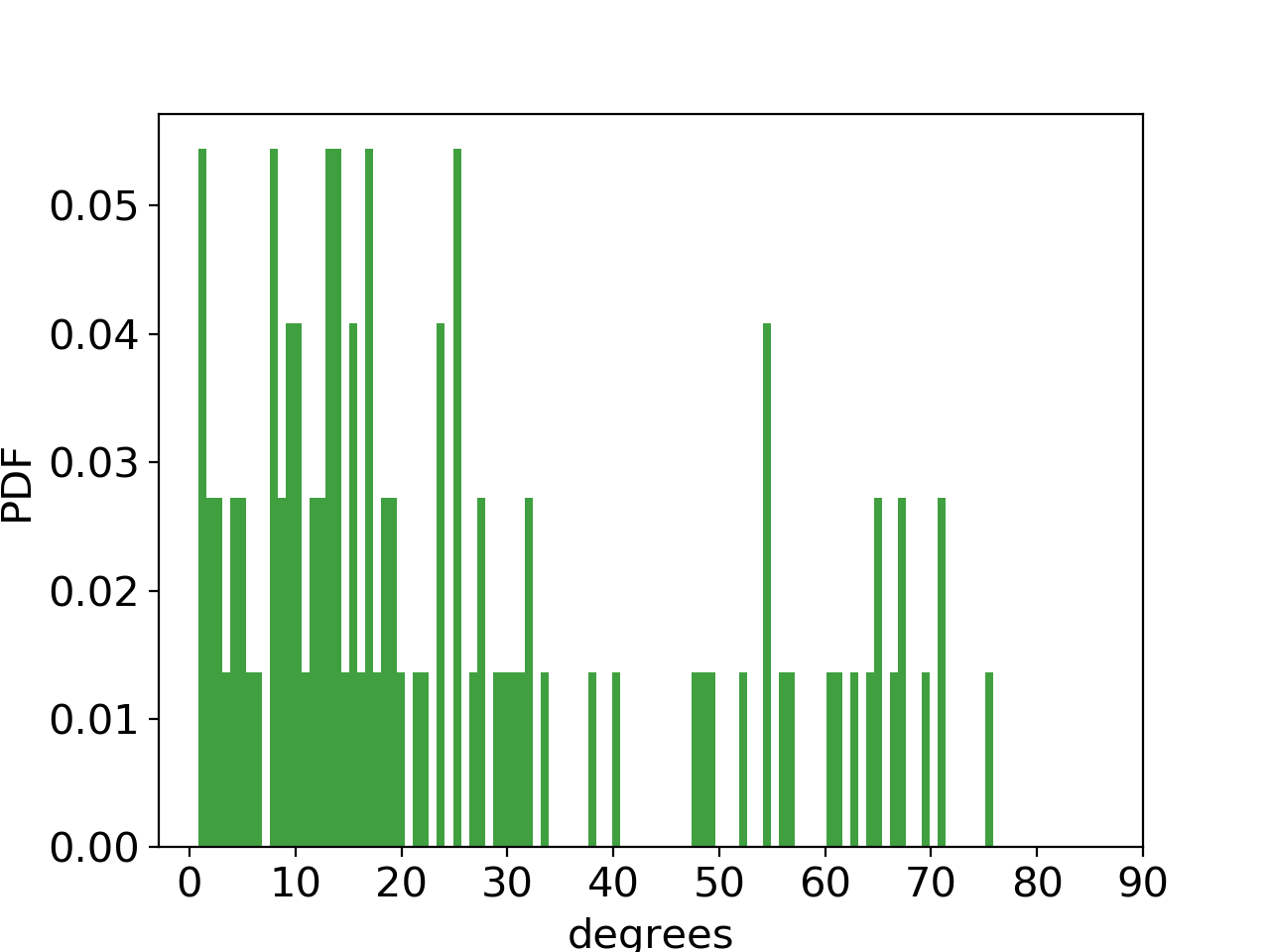}   
\includegraphics[width = 4cm]{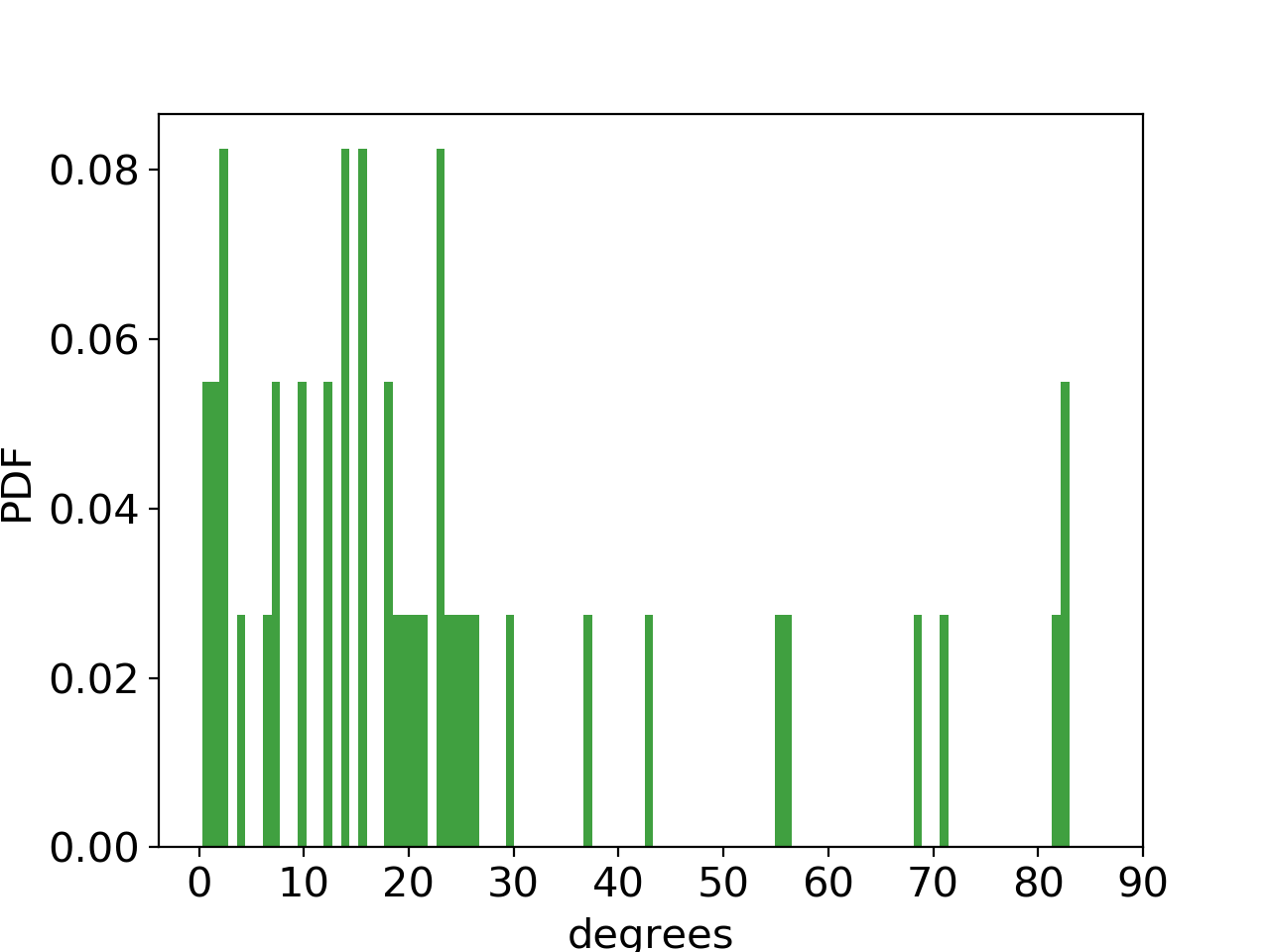}

\end{tabular}
\caption{PDFs of the absolute difference between the POS magnetic field angles and the orientation of the matter structures derived using the Rolling Hough Transform in the North-Eastern (left panel) and the North-Western (right panel) filaments.}
\label{fig:diffs}
\end{figure}

The RHT was applied to the {\planck} column density map in order to quantitatively define the two Northern filaments as seen in the continuum map and study the variations in the relative orientation of the filaments with respect to the magnetic field.
We represent in Fig.~\ref{fig:rht} the detected structure in the Northern part of the Monoceros OB-1 East cloud, where we clearly observe the three major filaments.
Figure~\ref{fig:bfils} shows the PDFs of the magnetic field angles in the North-Eastern and North-Western filaments. The red dashed lines represent the attempts of Gaussian fits to the data. In the North-Eastern filament the magnetic field angle spans from roughly $-10^{\circ}$ to $15^{\circ}$ with the maximum of the fit at $\valeast$, while in the  North-Western filament it is almost uniformly distributed between $-10^{\circ}$ and $5^{\circ}$ with a peak at $\valwest$. We determine the average dispersion along each filament as the mean value of the standard deviations of the angles taken perpendicular to the crest, and we obtain $\stdeast$ and $\stdwest$ for the North-Eastern and North-Western filaments respectively. 
Figure~\ref{fig:diffs} shows the PDFs of the absolute difference between the magnetic field angles and the filament orientations. We observe that the magnetic field and the matter are mostly aligned with respect to each other in both filaments. 

\section{Determination of the magnetic field strength using angular structure function}
\label{app:comparison}
In this section, we briefly summarise the method described in \cite{hildebrand2009}
The two-point polarisation-angle structure function is defined as a root mean square of the angle difference between a pair of points located at a distance $\mathit{l}$ from each other.
In terms of the Stokes parameters the polarisation-angle structure function has the following expression \citep{planck2014-xix, alina2016}:
\begin{eqnarray}
S(\mathit{l}) = \Big[ \frac{1}{N(l)}\sum_{i=1}^{N(l)} \Big(\dfrac{1}{2} \arctan \lbrack U(\mx)Q(\mxl)-Q(\mx)U(\mxl),  \nonumber\\
Q(\mx)Q(\mxl)+U(\mx)U(\mxl) \rbrack \Big)^{2} \Big]^{1/2} \, .
\label{eq:struct_fun}
\end{eqnarray}

\begin{figure}
    \centering
    \includegraphics[width = 0.3\textwidth]{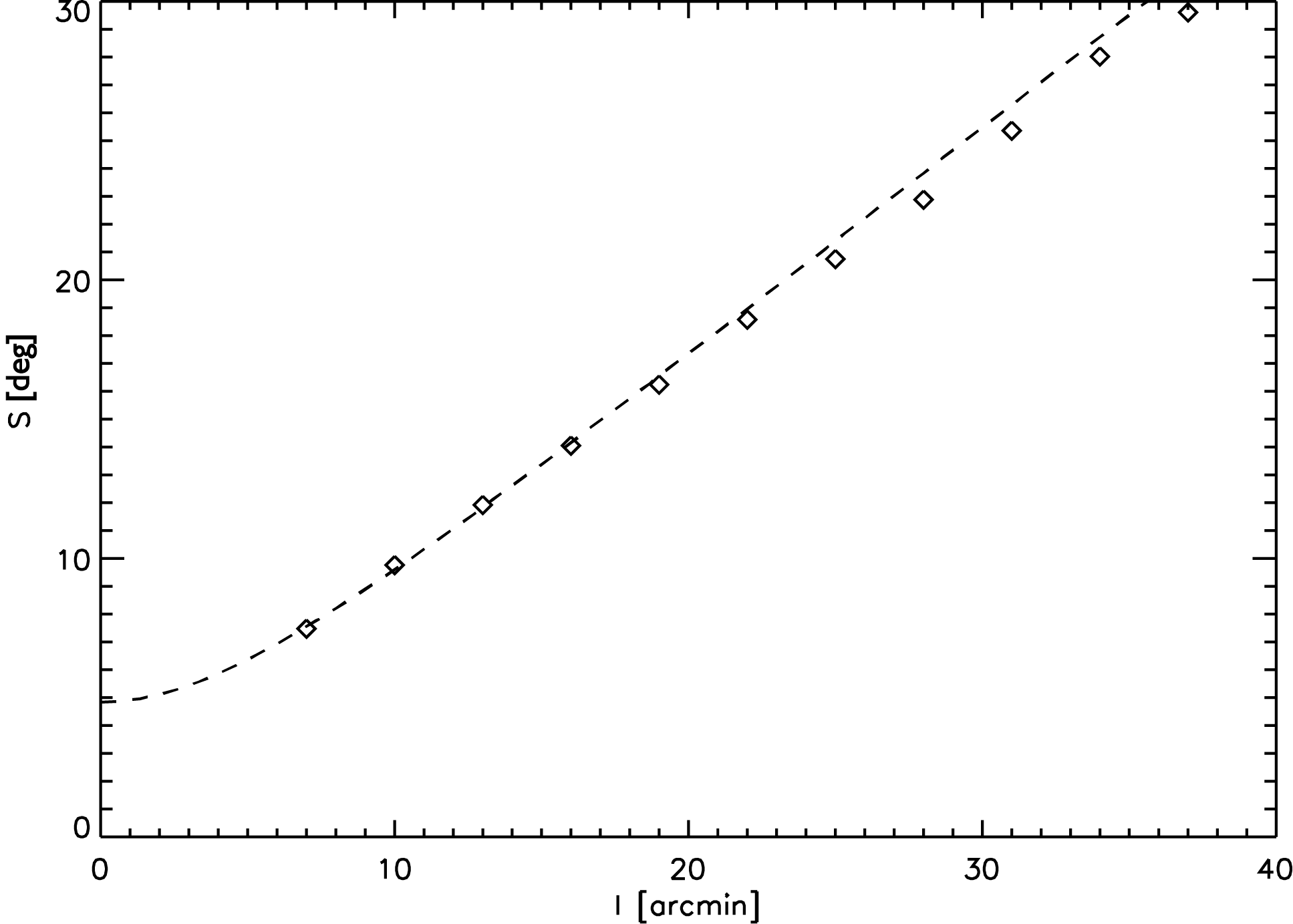}\\
    \includegraphics[width = 0.3\textwidth]{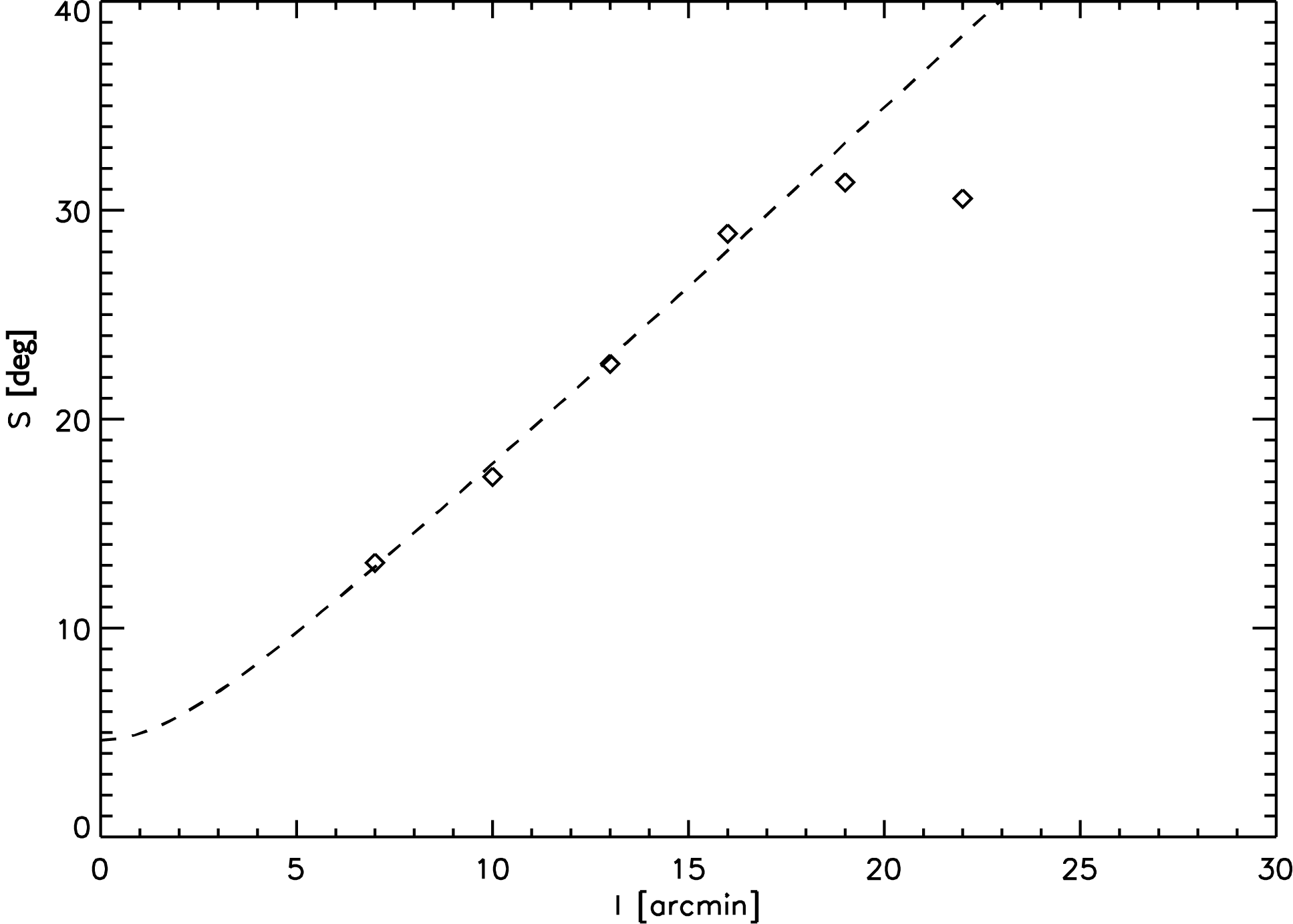}
    \includegraphics[width = 0.3\textwidth]{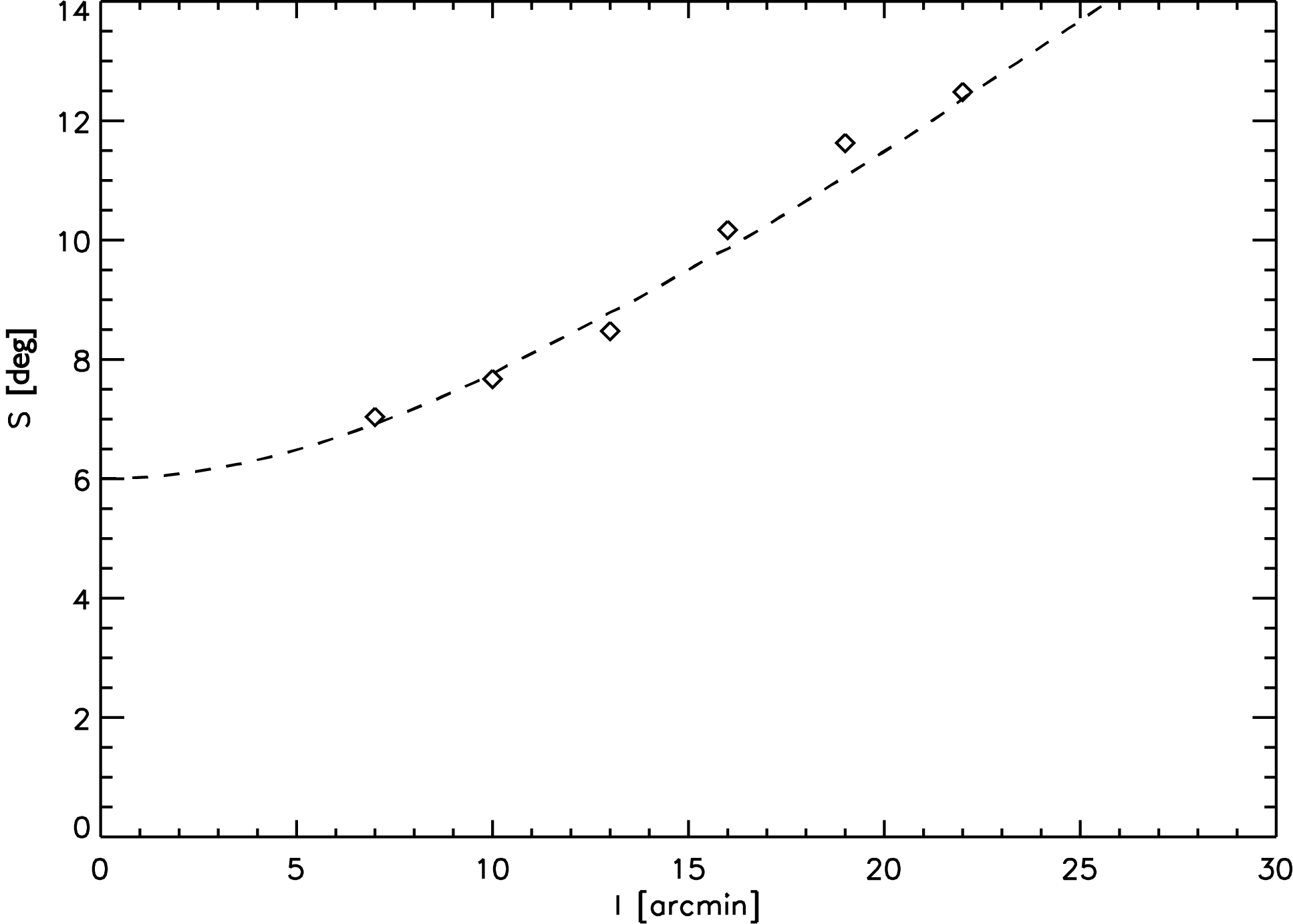}
    \caption{The structure function calculated according to Eq.~\ref{eq:struct_fun} for the {\mn}, North-Western and North-Eastern filaments (diamonds) for different lags $\mathit{l}$ and the fit to the data following Eq.~\ref{eq:sfit} (dashed curve).}
    \label{fig:structure_function}
\end{figure}

\cite{hildebrand2009} approach was the following. Within the assumption that the contribution of the large-scale structured magnetic field and the turbulent component of the magnetic field are independent, both contribute quadratically along with the measurement uncertainty $\sigma_M$ to the total angle structure function:
\begin{equation}
S \simeq \sqrt{ b^2 + m^2 \mathit{l}^2 + \sigma_M(\mathit{l})^2} \, ,
\label{eq:sfit}
\end{equation}
where $m \mathit{l}$ defines the large-scale component, and $b$ corresponds to the turbulent dispersion of the large-scale magnetic field.
In practice, $b$ can be determined from the zero-intercept of the measured polarisation angle structure function. This allows us to calculate the large-scale magnetic field component using the DCF method:
\begin{equation}
B_{0} \simeq \sqrt{8 \pi \rho } \frac{\sigma (v)}{b}
\end{equation}
We used the same lags for the calculation of $\mathit{S}$ as in Sect.~\ref{sec:dcf}. The resulting curves and the fits are represented in Fig.~\ref{fig:structure_function}. The uncertainties are calculated by propagation of the standard error between the fit and the data.

It is worth noting that \cite{lazarian2020} proposed the following expression for 
the structure function of polarization angle to be more appropriate in the frame of the MHD turbulence:
\begin{equation}
SF\{V\}(R) = q R^{1+m} + p R^2 
\label{structure_f}
\end{equation}
where $q$ and $p$ are constants, $V$ is the velocity centroid and $R$ is the distance separation. In other words, at the scales less than the injection scale of the turbulence one measures the scaling that reflects the power-law scaling of 3D turbulent motions. Using the representation given by Eq.~\ref{structure_f}, \cite{lazarian2020} proposed their new technique of using the structure functions that may be used in studies of the magnetic fields in molecular clouds. 

\section{Additional figures}
\label{app:figures}
\begin{figure}[htpb]
    \centering
    \begin{tabular}{cc}
            \includegraphics[width = 4.1 cm]{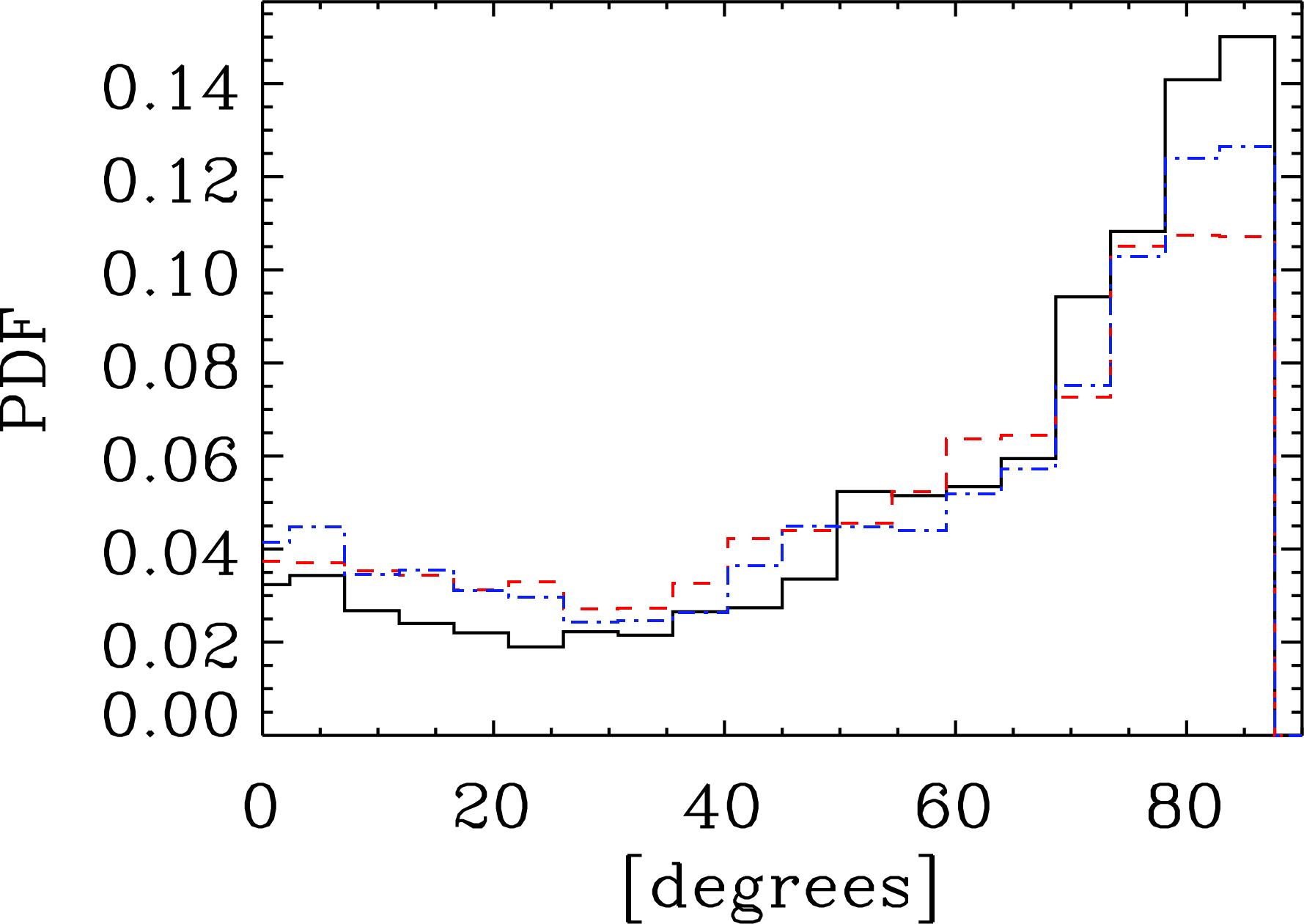} 
         & 
            \includegraphics[width = 4.1 cm]{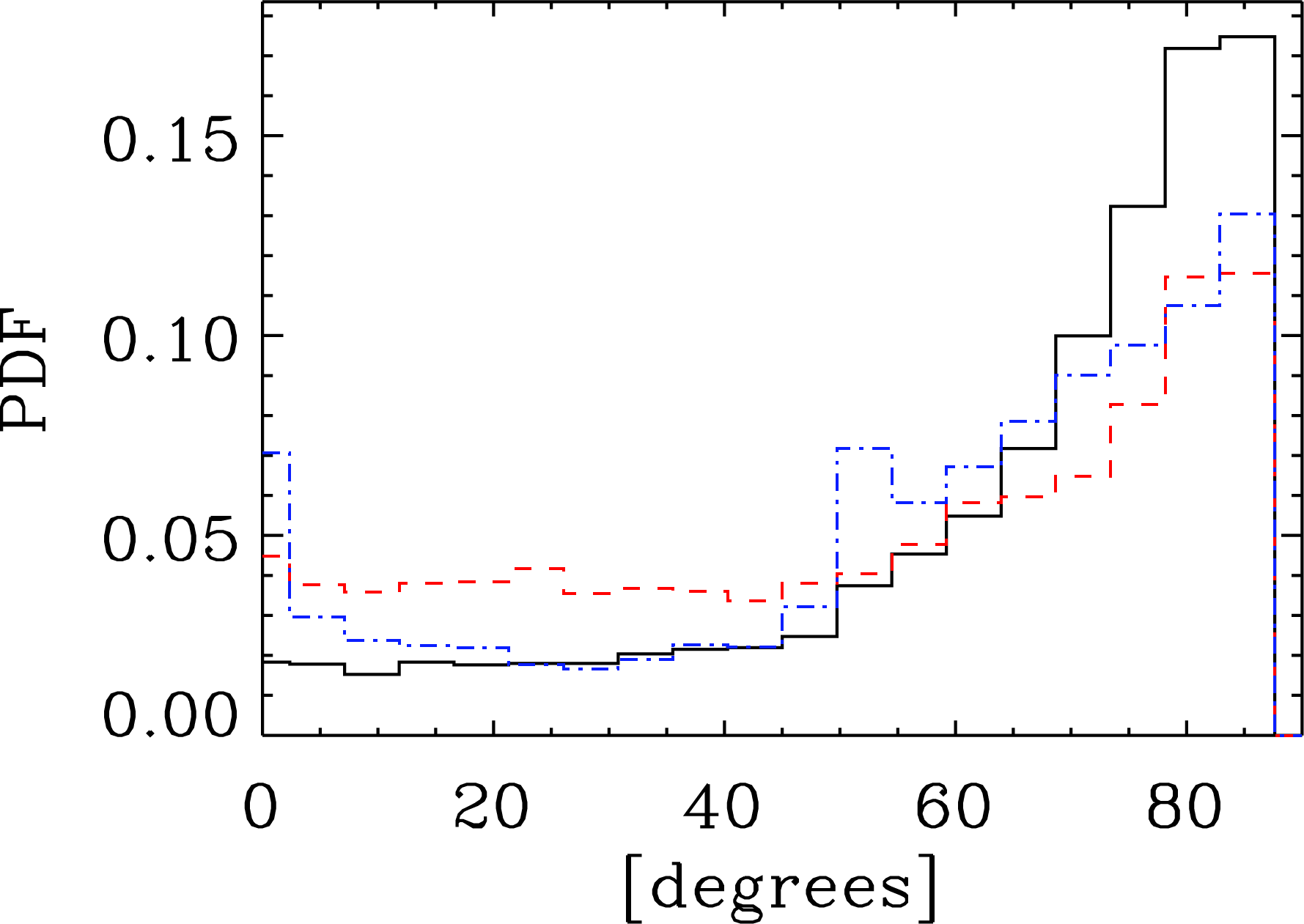} 
    \end{tabular}
    \caption{PDFs of the absolute differences between the POS magnetic field derived from the {\planck} data and the unsmoothed ($47\arcmin$ resolution) IGs in black (plain curve), VGs in red (dashed curve), VChGs in blue (dash-dotted curve) for the Northern part. Left: based on $^{12}$CO data, right: based on $^{13}$CO data.}
    \label{fig:north_unsm}
\end{figure}

\begin{figure}[htpb]
    \centering
    \begin{tabular}{cc}
            \includegraphics[width = 4.1 cm]{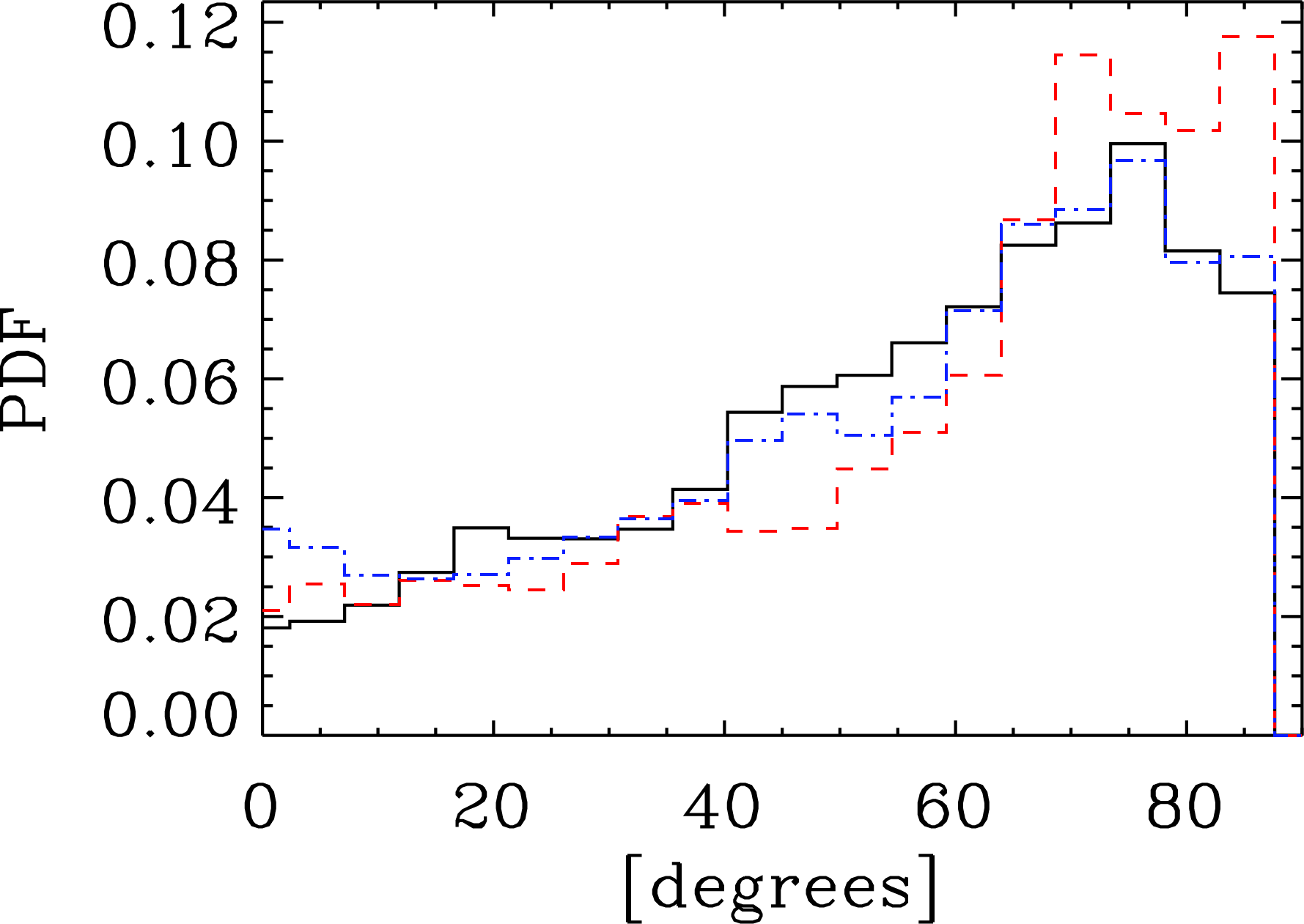} 
         & 
            \includegraphics[width = 4.1 cm]{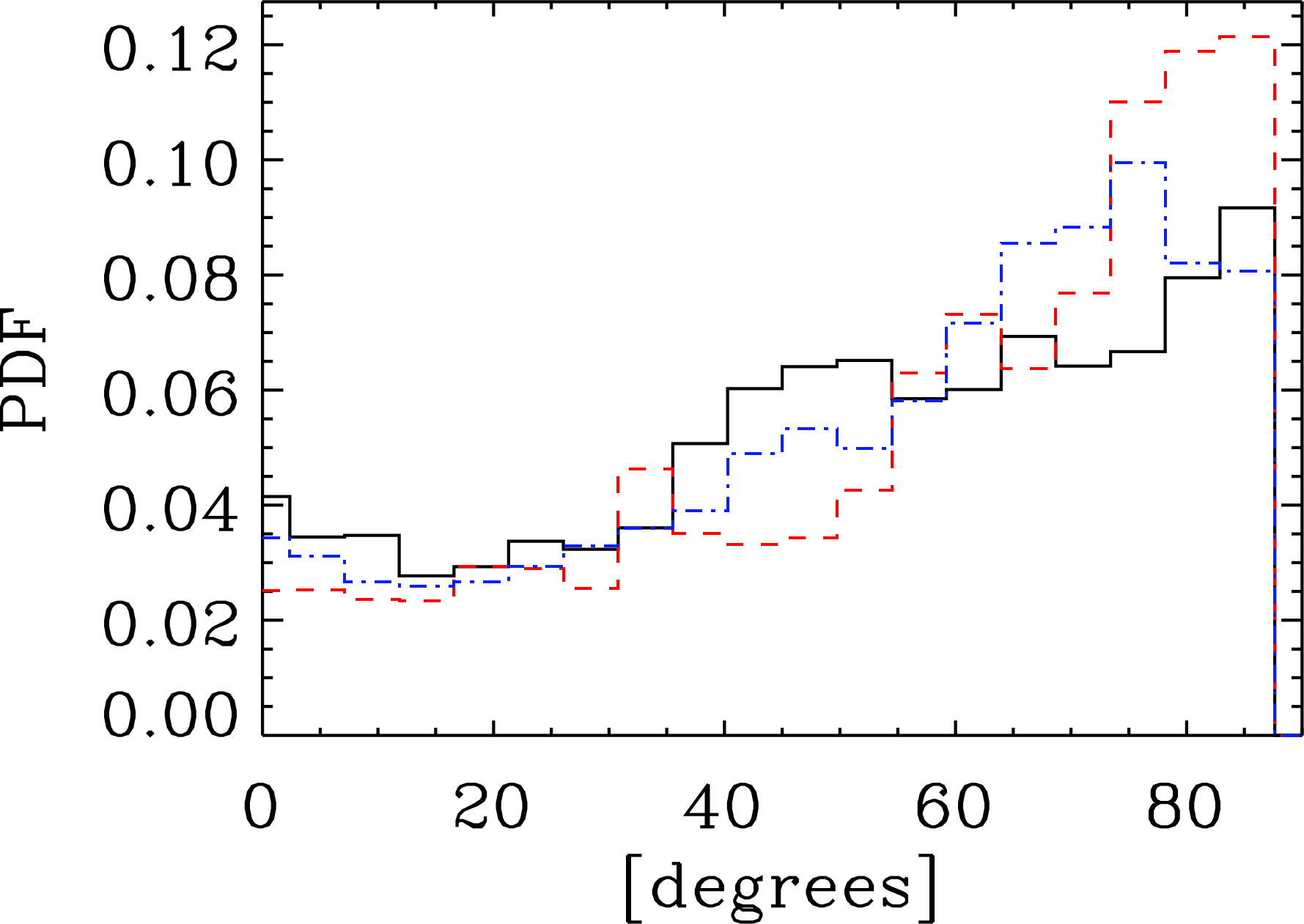} 
    \end{tabular}
    \caption{PDFs of the absolute differences between the POS magnetic field derived from the {\planck} data and the IGs in black (plain curve), VGs in red (dashed curve), VChGs in blue (dash-dotted curve) for the Southern part. Left: based on $^{12}$CO data, right: based on $^{13}$CO data.}
    \label{fig:south_hist}
\end{figure}

Here we represent:
\begin{itemize}
\item the uncertainty on $\polang$ in the area represented by contours in Fig.~\ref{fig:snrp}, excluding pixels with $\snr(p) \leq 2$ and $\sigma_{\polang} > 10^{\circ}$. The \snr($\polang$) in the {\mn}, North-Eastern and North-Western VCSs are shown in green (dash-dotted), red (dotted) and blue (dashed) respectively and generally do not exceed 5$^{\circ}$;
\item the rms noise map of the TRAO 14-metre {\coo} data in greyscale with the same contours of the {\co} integrated intensity at 2 K km s$^{-1}$ as in Fig.~\ref{fig:snrp};
\item the PDFs of the absolute difference between the magnetic field orientation derived from the {\planck} 353 GHz polarised channel data and the orientation of the IGs and VGs derived from the TRAO 14-metre telescope CO data. The unsmoothed gradients are used to produce the PDF shown in Fig.~\ref{fig:north_unsm}, while the PDF in Fig.~\ref{fig:south_hist} is built using the CO data smoothed to the resolution of $7\arcmin$.
\end{itemize}

\begin{figure}
    \centering
    \includegraphics[height = 4 cm]{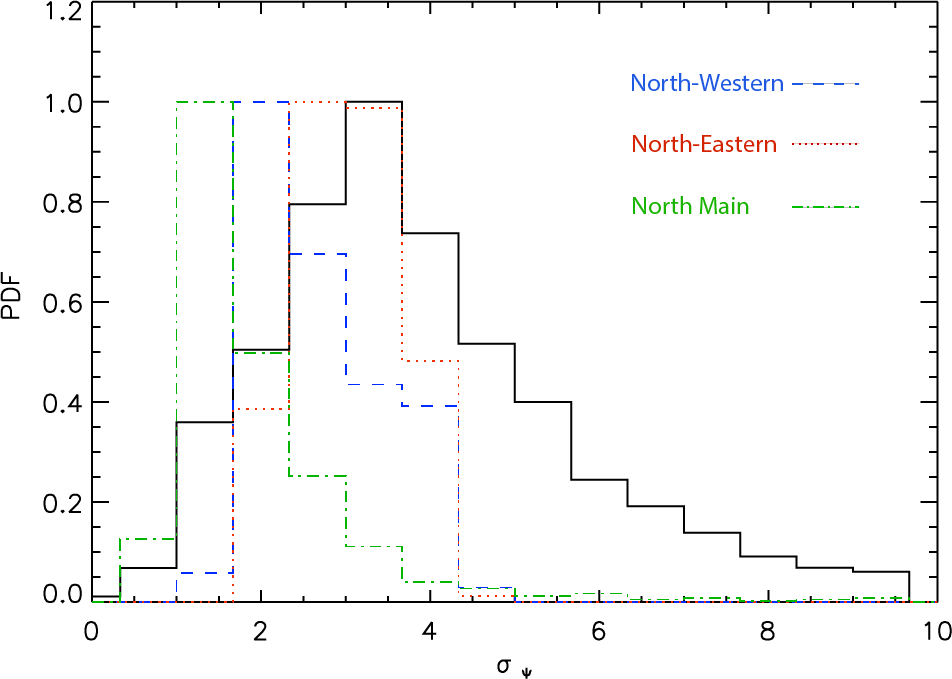}
    \caption{PDFs of the uncertainty on polarization angle, $\sigma_{\polang}$, computed using Bayesian analysis \citep{Montier2} inside the area delimited by the contour in Fig.~\ref{fig:snrp} (plain black) and inside the {\mn}(green dash-dotted line), North-Eastern (red dotted line) and North-Western (blue dashed line) VCSs.}
    \label{fig:sigpsi}
\end{figure}
\begin{figure}
    \centering
    \includegraphics[height = 4 cm]{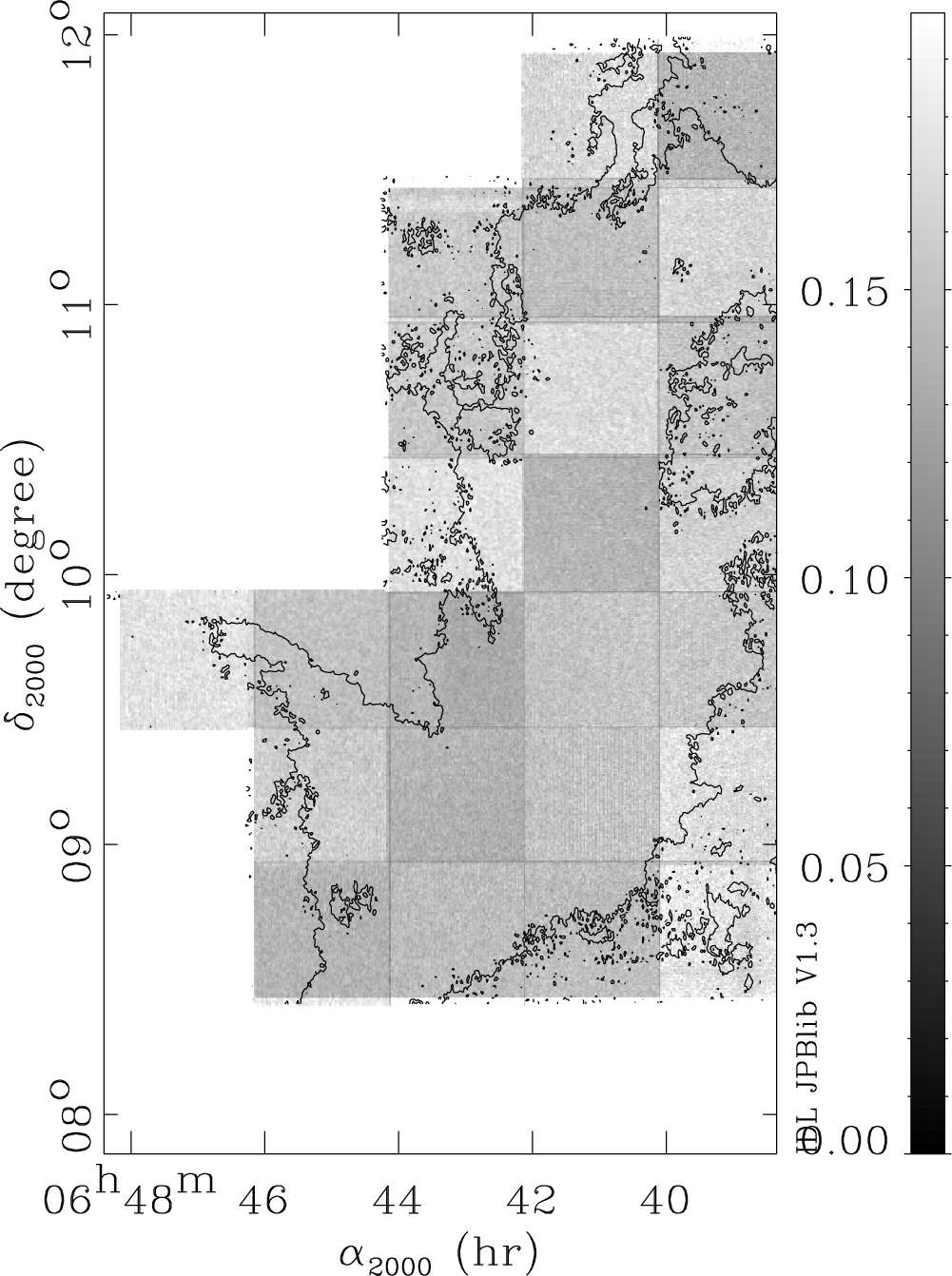}
    \caption{Rms noise map of the {\coo} data, in K. The contours are given by {\co} integrated intensity at 2 K km s$^{-1}$, same as in Fig.~\ref{fig:snrp}}
    \label{fig:rms}
\end{figure}

\section{MHD turbulence, reconnection and Gradient Technique}

The Gradient Technique (GT) is rooted in the advanced MHD turbulence theory and the theory of turbulent magnetic reconnection. The Velocity Gradient Technique (VGT) is a branch of the GT and the Intensity Gradient Technique (IGT) can be viewed as an outgrowth of the VGT.  Here we briefly explain the basic elements of the theory that are essential for understanding the VGT and IGT. An in-depth discussion of the properties of MHD turbulence and its relation to the turbulent reconnection can be found in the monograph by \citet{BL19}.

The current understanding of MHD turbulence theory is related to the pioneering study in \citet{GS95} (hereafter GS95). There, the concept of scale-dependent anisotropy increasing with the decrease of scale was first introduced.
However, this GS95 anisotropy scaling is derived in the global magnetic field reference frame, in which the predicted scaling is not observable. In fact, the scale-dependent anisotropy is present in the so-called {\it local} system of reference. This concept of local system of reference is frequently a point of confusion for many researchers and thus will be discussed below. Here we would like to stress that for the GT the idea of the local system of reference is central and crucial.

The local system of reference naturally arises in the theory of turbulence that is based on the turbulent reconnection in theory \citet{LV99} (hereafter LV99). 
There, it was demonstrated that MHD turbulence can be presented as a collection of eddies with their rotation axis aligned with magnetic field. This physical picture is possible as the time scale for the turbulent reconnection coincides with the eddy turnover time. The turbulent reconnection is an essential part of the dynamics of turbulent eddies, which enables the mixing of magnetic field lines perpendicular to the magnetic field direction. In this direction, turbulent eddy motions do not experience the magnetic back-reaction. In other words, if eddy is rotating around its surrounding magnetic field , the eddy evolves in a hydrodynamic type manner. The evolution of this type of eddies with their rotation axis parallel to the magnetic field direction presents the path of the minimal resistance for the energy cascade. Therefore, most of the energy is channeled through such anisotropic cascade. 

The motions perpendicular to the local magnetic field are Kolmogorov types due to the absence of the back-reaction of magnetic field. Therefore the Kolmogorov relations are valid for the eddy size $l_\bot$, perpendicular to the magnetic field, and for the eddy perpendicular velocity $v_l$, that is, $v_l\sim l_\bot^{1/3}$.

As eddies mix up magnetic field, this causes Alfv\'{e}n waves propagating along the magnetic field at velocity $V_A$. The period of the Alfv\'{e}n wave generated this way $\sim l_{\|}/V_A$ is equal to the turnover time of the magnetic eddies, the latter being $\sim l_{\bot}/v_l$. The relation
$l_{\|}/V_A\sim l_{\bot}/v_l$ corresponds to the critical balance in GS95 theory. However, in the latter theory the scales $l_\bot$ and $l_\|$ are measured in respect to the mean, rather than to the local magnetic field direction. 

The considerations above reflect the spirit of arguments that was used in LV99 to derive the following relation between the parallel and perpendicular scales of the eddies in the local reference frame of the eddies:
\begin{equation}
\label{eq.lv99}
 l_\parallel\simeq L_{inj}(\frac{l_\bot}{L_{inj}})^{\frac{2}{3}}M_A^{-4/3}
\end{equation}
where $M_A$ is the ratio of the injection velocity $v_{L}$ to the Alfv\'{e}n speed $V_A$ and $L_{inj}$ is the injection scale of turbulence.
This universal scale-dependent anisotropy of Alfv\'{e}nic turbulence in the local magnetic field reference frame has been demonstrated in \citet{2000ApJ...539..273C},\citet{2002ApJ...564..291C}, and \citet{2001ApJ...554.1175M}.

Combining Eq.~\ref{eq.lv99} and the "critical balance" expressed in the local reference frame, i.e., $l_\bot V_A\sim l_\parallel v_l$, one can get the scaling relation for velocity fluctuations (see LV99):
\begin{equation}
\label{eq:KS}
    v_{l}\simeq v_L\left(\frac{l_{\perp}}{L_{inj}}\right)^{\frac{1}{3}}M_A^{\frac{1}{3}} 
\end{equation}
where $v_L$ is the injection velocity of turbulence. 

From the point of view of the GT, the eddies parallel to the local direction of magnetic field present a remarkable possibility of determining the direction of magnetic field by measuring the gradients of the absolute value of velocity. Indeed, the latter are perpendicular to the local magnetic field direction (see \citealt{hu2020} for pictorial illustration). Thus the velocity gradients can trace magnetic fields. In Alfv\'{e}nic turbulence, the magnetic field amplitude fluctuations are proportional to the velocity fluctuations.\footnote{From the mathematical point of view the symmetry of magnetic and velocity fluctuations in Alfvenic turbulence is evident if the description of MHD turbulence with Elsasser variables is employed. (see \citet{BL19}). On the intuitive level, this symmetry is related to the Alfven expression that defines $\delta v_l \sim \delta b_l$.}  
 Therefore magnetic amplitude gradients can also be used to trace magnetic field direction. It is important that the velocity gradient's amplitude for Kolmogorov-type eddies is increasing with the decrease of the eddy size, i.e. $v_l/l_\bot \sim l_\bot^{-2/3}$. As a consequence, the smallest resolved eddies corresponding to the telescope beam size contribute most to the observed gradient measure. These smallest eddies trace well the local magnetic field around them.

The relation between the turbulent velocity  fluctuations and density fluctuations is not so direct as between the velocities and magnetic field (see Kowal et al. 2007). However, for a range of parameters the density acts as a so-called "passive scalar" and is advected by velocity fluctuations. In this case the statistics of velocity is imprinted on the density statistics and the application of the gradient technique to tracing magnetic field is justified. 

Explicitly, since the anisotropic relation indicates $l_\bot \ll l_\parallel$, the velocity gradient $\nabla v_l$ and density gradient $\nabla \rho_l$ scale as \citep{hu2020}:
\begin{equation}
\label{eq.grad}
\begin{aligned}       
        \nabla \rho_l&\propto \frac{\rho_{l}}{l_\bot}\simeq\frac{\rho_0}{c}{F}^{-1}(|\hat{k}\cdot\hat{\zeta}|)\nabla v_l\\
        \nabla v_l&\propto\frac{v_{l}}{l_\bot}\simeq \frac{v_L}{L_{inj}}\left(\frac{l_{\perp}}{L_{inj}}\right)^{-\frac{2}{3}}M_A^{\frac{1}{3}}\\
\end{aligned}
\end{equation}
where $\rho_0$ is the mean density, $\hat{\zeta}$ is the unit vector for the Alfv\'{e}nic mode (fast mode or slow mode), $c$ is the  propagation speed of the corresponding mode, and $v_l$ is the turbulence velocity at scale $l$. ${F}^{-1}$ denotes the inverse Fourier transformation. The direction of density gradient and velocity gradient is perpendicular to the local direction of the local magnetic field. This consideration is at the core of VGT and IGT.

\bibliographystyle{aa}
\bibliography{thebibliography_all.bib}

\end{document}